\documentclass[a4paper,12pt]{article}
\pdfoutput=1
\usepackage{epsfig}
\usepackage{amssymb,subfigure}
\usepackage{amsfonts}
\usepackage{amsmath}
\usepackage{euscript}
\usepackage{verbatim}
\usepackage{latexsym}
\usepackage{graphicx,color}
\usepackage{caption}
\usepackage{float}
\usepackage{slashed}
\usepackage{graphicx}
\graphicspath{ {images/} }
\usepackage{ytableau}

\usepackage{multirow}
\usepackage{relsize}
\usepackage{lscape}

\usepackage{mathtools}

\usepackage[colorinlistoftodos]{todonotes}
\usepackage[colorlinks=true, allcolors=blue]{hyperref}

\usepackage{wrapfig}
	\usepackage[T1]{fontenc}
	\usepackage{tikz}
	\usetikzlibrary{decorations.pathmorphing}
\usepackage{tikz}
\usetikzlibrary{shapes.geometric, arrows,patterns,snakes}
\tikzstyle{ellip} = [ellipse, minimum width=3cm, minimum height=1cm,text centered, draw=black]

\newskip\humongous \humongous=0pt plus 1000pt minus 1000pt

\newif\ifdtup

\def  \bcen   {\begin{center}}
\def  \ecen   {\end{center}}
\def  \beqa   {\begin{eqnarray}}
\def  \eeqa   {\end{eqnarray}}
\def  \nn     {\nonumber }

\allowdisplaybreaks[1]

\catcode`\@=11

\@addtoreset{equation}{section}

\def\@normalsize{\@setsize\normalsize{15pt}\xiipt\@xiipt
\abovedisplayskip 14pt plus3pt minus3pt%
\belowdisplayskip \abovedisplayskip
\abovedisplayshortskip \z@ plus3pt%
\belowdisplayshortskip 7pt plus3.5pt minus0pt}

\def\small{\@setsize\small{13.6pt}\xipt\@xipt
\abovedisplayskip 13pt plus3pt minus3pt%
\belowdisplayskip \abovedisplayskip
\abovedisplayshortskip \z@ plus3pt%
\belowdisplayshortskip 7pt plus3.5pt minus0pt
\def\@listi{\parsep 4.5pt plus 2pt minus 1pt
     \itemsep \parsep
     \topsep 9pt plus 3pt minus 3pt}}

\relax

\catcode`@=12

\catcode`\@=11

\def\section{\@startsection{section}{1}{\z@}{3.5ex plus 1ex minus
   .2ex}{2.3ex plus .2ex}{\large\bf}}

\def\SymBoxes#1#2#3#4{\newdimen\un@t \un@t#3%
\raisebox{#1}{\rule{#2\un@t}{#4}\hskip-#2\un@t
\@tempdimb\un@t \advance\@tempdimb by-#4\@tempcntb#2\relax%
\@whilenum{\@tempcntb>0}\do{
\rule{#4}{\un@t}\hskip\@tempdimb \advance\@tempcntb by\m@ne}%
\hskip-#2\un@t \rule[\un@t]{#2\un@t}{#4}%
\rule[\un@t]{#4}{#4}\hskip-#4
\rule{#4}{\un@t}}\hskip-#4}                

\begin{document}

\newcommand{\beq}{\begin{equation}}
\newcommand{\eeq}{\end{equation}}
\newcommand{\bea}{\begin{eqnarray}}
\newcommand{\eea}{\end{eqnarray}}
\newcommand{\beas}{\begin{eqnarray*}}
\newcommand{\eeas}{\end{eqnarray*}}
\newcommand{\defi}{\stackrel{\rm def}{=}}
\newcommand{\non}{\nonumber}
\newcommand{\bquo}{\begin{quote}}
\newcommand{\enqu}{\end{quote}}
\renewcommand{\(}{\begin{equation}}
\renewcommand{\)}{\end{equation}}
\def \eqn#1#2{\begin{equation}#2\label{#1}\end{equation}}
\def\IZ{{\mathbb Z}}
\def\IR{{\mathbb R}}
\def\IC{{\mathbb C}}
\def\IQ{{\mathbb Q}}
\def\de{\partial}
\def\Tr{ \hbox{\rm Tr}}
\def\H{ \hbox{\rm H}}
\def\HE{ \hbox{$\rm H^{even}$}}
\def\HO{ \hbox{$\rm H^{odd}$}}
\def\K{ \hbox{\rm K}}
\def\Im{ \hbox{\rm Im}}
\def\Ker{ \hbox{\rm Ker}}
\def\const{\hbox {\rm const.}}
\def\o{\over}
\def\im{\hbox{\rm Im}}
\def\re{\hbox{\rm Re}}
\def\bra{\langle}\def\ket{\rangle}
\def\Arg{\hbox {\rm Arg}}
\def\Re{\hbox {\rm Re}}
\def\Im{\hbox {\rm Im}}
\def\exo{\hbox {\rm exp}}
\def\diag{\hbox{\rm diag}}
\def\longvert{{\rule[-2mm]{0.1mm}{7mm}}\,}
\def\a{\alpha}
\def\dag{{}^{\dagger}}
\def\tq{{\widetilde q}}
\def\p{{}^{\prime}}
\def\W{W}
\def\N{{\cal N}}
\def\hsp{,\hspace{.7cm}}

\def\br{\nonumber\\}
\def\IZ{{\mathbb Z}}
\def\IR{{\mathbb R}}
\def\IC{{\mathbb C}}
\def\IQ{{\mathbb Q}}
\def\IP{{\mathbb P}}
\def \eqn#1#2{\begin{equation}#2\label{#1}\end{equation}}

\newcommand{\sgm}[1]{\sigma_{#1}}
\newcommand{\idd}{\mathbf{1}}

\newcommand{\C}{\ensuremath{\mathbb C}}
\newcommand{\Z}{\ensuremath{\mathbb Z}}
\newcommand{\R}{\ensuremath{\mathbb R}}
\newcommand{\rp}{\ensuremath{\mathbb {RP}}}
\newcommand{\cp}{\ensuremath{\mathbb {CP}}}
\newcommand{\vac}{\ensuremath{|0\rangle}}
\newcommand{\vact}{\ensuremath{|00\rangle}                    }
\newcommand{\oc}{\ensuremath{\overline{c}}}
\begin{titlepage}
\bigskip
\def\thefootnote{\fnsymbol{footnote}}

\begin{center}
{\Large
{\bf Model independent Analysis of Dirac CP Violating Phase for some well known mixing scenarios
\vspace{0.2in}
}
}
\end{center}

\bigskip
\begin{center}
{Sumit K. GARG$^a$,\footnote{\texttt{sumit.k@cmr.edu.in}} \vspace{0.15in} \\ }
\vspace{0.1in}

\end{center}

\renewcommand{\thefootnote}{\arabic{footnote}}

\begin{center}
$^a$ {Department of Physics, \\
CMR University, Bengaluru 562149, India }

\end{center}

\noindent
\begin{center} {\bf Abstract} \end{center}
We present a model independent analysis of Leptonic CP violation for some well known mixing scenarios.
In particular, we considered modified schemes for Bimaximal(BM), Democratic(DC), Hexagonal(HG) and Tribimaixmal(TBM) mixing 
for our numerical investigation. These model independent corrections to mixing matrices are parameterized in terms of complex rotation matrices ($U$) with 
related modified PMNS matrix of the forms \big($U_{ij}^l\cdot V_{M},~V_{M}\cdot U_{ij}^r$ \big ) where $U_{ij}^{l, r}$ is complex rotation in ij sector and  
$V_{M}$ is unperturbed mixing scheme. We present generic formulae for mixing angles, Dirac CP phase($\delta_{CP}$) and Jarkslog Invariant($J_{CP}$) 
in terms of correction parameters. The parameter space of each modified mixing case is scanned for fitting neutrino mixing angles using $\chi^2$ approach
and the corresponding predictions for Leptonic CP Phase($\delta_{CP}$) and Jarkslog Invariant($J_{CP}$) has been evaluated 
from allowed parameter space. The obtained ranges are reported for all viable cases.




\vspace{1.6 cm}
\vfill

\end{titlepage}

\setcounter{page}{2}

\setcounter{footnote}{0}



\section{Introduction}

The observation of neutrino oscillations~\cite{neutroscill} is a major milestone in particle physics over the last few decades.
Solar and atmospheric neutrino studies~\cite{solar, atmospheric} provided the first reliable evidence of neutrino flavor change when these subatomic particles
travels through vacuum and matter. These observations undoubtedly confirmed the existence of physics beyond the domain of standard model. 
With  arrival of reactor and accelerator experiments~\cite{Dayabay, T2K, Doublechooz, Minos, RENO} in neutrino physics arena, neutrino physics 
entered into precision era with the determination 
of oscillation parameters with much greater accuracy. It remarkably helped for improving our understanding about neutrino oscillation physics. 

In a 3 flavor scenario, neutrino mixing is described by $3\times 3$ unitary matrix which can 
parametrized in terms of  3 mixing angles and 6 phases. However 5 phases are redundant and thus
can be rotated away leaving behind only 1 physical phase. Thus light neutrino mixing is given in standard form as~\cite{upmns}
\begin{eqnarray}
U &=& \left( \begin{array}{ccc} 1 & 0 & 0 \\ 0 & c^{}_{23}  & s^{}_{23} \\
0 & -s^{}_{23} & c^{}_{23} \end{array} \right)
\left( \begin{array}{ccc} c^{}_{13} & 0 & s^{}_{13} e^{-i\delta_{CP}} \\ 0 & 1 & 0 \\
- s^{}_{13} e^{i\delta_{CP}} & 0 & c^{}_{13} \end{array} \right)
\left( \begin{array}{ccc} c^{}_{12} & s^{}_{12}
 & 0 \\ -s^{}_{12}  & c^{}_{12}  & 0 \\
0 & 0 & 1 \end{array} \right) \left( \begin{array}{ccc} 1 & 0
 & 0 \\ 0  & e^{i\rho}  & 0 \\
0 & 0 & e^{i\sigma} \end{array} \right)
,\label{standpara}\nonumber
\end{eqnarray}
where $c_{ij}\equiv \cos\theta_{ij}$, $s_{ij}\equiv \sin\theta_{ij}$ and $\delta_{CP}$ is the Dirac CP violating phase.
Here two additional phases $\rho$ and $\sigma$ known as Majorana phases are not
relevant as they don't affect the neutrino oscillations. Thus we safely assumed their values to be zero in this study. 
The so far interesting picture which emerged in this scenario is that two mixing angles seems to be large while third one 
remains small. As far as leptonic CP phase is concerned, the situation is not much clear as still wide range of 
values~\cite{globalfit11,globalfit12, globalfit21,globalfit22, globalfit3} can be accommodated at $3\sigma$ level of confidence. Moreover data from long-baseline accelerator, 
solar and KamLAND is also consistent~\cite{globalfit3} at $2\sigma$ or less in CP conserving limit for NH as well as IH. However on the other hand,
some initial hints also emerged from experiments like T2K~\cite{T2KCPViol} and NOvA~\cite{NovACPViol} refering towards maximal CP violation in 
this sector. Thus neutrino physics is in interesting phase and is suppose to reveal many secrets in forthcoming years.

Before the start of exciting era of non zero $\theta_{13}$, many mixing schemes like tribimaximal~\cite{scott1,scott2,scott3,scott4,scott5}, bimaximal~\cite{BM1,BM2,BM3,BM4,BM5,BM6,BM7,BM8}, 
democratic~\cite{DM1,DM2,DM3} and Hexagonal~\cite{HG} were proposed which offered to explain neutrino mixing data with a common novel prediction of zero
reactor mixing angle i.e. $\theta_{13}=0$.
The atmospheric mixing angle($\theta_{23}$) is maximal for BM, TBM and HG while it takes a value of
of $54.7^{\circ}$ in DC case. The solar mixing angle($\theta_{12}$) is maximal for BM and DC scenario while it predicts a
lower value of $35.3^{\circ}$ and $30^{\circ}$ in TBM and HG case respectively. 
However, nuclear reactor based Daya Bay~\cite{Dayabay} experiment in China which basically looks for disappearance of $\bar{\nu}_e$ 
gave first conclusive result about the fate of $\theta_{13}$. This collaboration reported the value of 1-3 mixing angle consistent with data at $5.2\sigma$
significance level in the range $\sin^2 2\theta_{13}= 0.092\pm 0.016(stat)\pm 0.05(syst)$ under a three flavor scenario. 
Earlier reactor based Japanese experiment, T2K~\cite{T2K} which is a long baseline neutrino oscillation experiment observed similar hints of 
non zero $\theta_{13}$ corresponding to $\nu_{\mu}\rightarrow \nu_e$ transition in a three flavor scenario. The  value of 1-3 mixing angle consistent with data 
at 90\% CL was reported to be  in the range $ 5^\circ(5.8^\circ) < \theta_{13} < 16^\circ(17.8^\circ)$ for Normal (Inverted) neutrino mass hierarchy. 
These observations are also being verified by other oscillation experiments like Double Chooz~\cite{Doublechooz}, Minos~\cite{Minos} and RENO~\cite{RENO}. 

The discovery of non zero $\theta_{13}$ is an important turning point in neutrino physics which provided a very crucial input for model building.
Thus with these findings along with inputs from recent global fits~\cite{globalfit11,globalfit12, globalfit21,globalfit22, globalfit3} for 
neutrino masses and mixing angles (given in Table~\ref{Table1}), it is quite evident that these mixing scenarios can only provide the 
main structure of the consistent neutrino matrix at leading order. Hence all such mixing schemes should be tested for possible modifications~\cite{1largeth13,2largeth13,3largeth13,4largeth13,5largeth13,6largeth13,7largeth13,8largeth13,9largeth13,10largeth13,11largeth13,12largeth13,13largeth13,14largeth13,15largeth13,16largeth13,17largeth13,18largeth13,19largeth13,20largeth13,21largeth13,22largeth13,23largeth13,24largeth13,25largeth13,26largeth13,27largeth13,28largeth13,29largeth13,30largeth13,models1,models2,models3,models4,models5,models6,models7,models8,models9,models10,models11,models12, pertbsinangles1,pertbsinangles2,pertbsinangles3,pertbsinangles4,pertbsinangles5,pertbsinangles6, S41,S42,S43,S44,S45, A41,A42,A43,A44,A45,A46,A47,A48,A49} to check their viability with current oscillation data. 
In literature, these corrections 
are often being parametrized in terms of complex rotation matrices~\cite{pertbsinangles1,pertbsinangles2,pertbsinangles3,pertbsinangles4,pertbsinangles5,pertbsinangles6, skg2} which acts on 12, 23 or 13 sector of these special matrices. 
This simpler way of parameterizing the corrections is quite helpful to understand the nature of corrections which a particular sector of these special matrices
should get in order to be consistent with neutrino mixing data.

In this study, we addressed the role of possible corrections~\cite{HGPertub1,HGPertub2,skgetal1,skg1,skg2} which are parameterized by 
one complex rotation matrix~\cite{skg2} for these mixing schemes. Thus modified PMNS matrix will be of the forms  \big($U_{ij}\cdot V_{M},~V_{M}\cdot U_{ij}$\big), where $V_M$ is unmodified mixing matrix and $U_{ij}$ is rotation in $ij$ sector of complex plane. As we know from theoretical point of view, 
the form of PMNS matrix is given by $U_{PMNS} = U_l^{\dagger} U_\nu$ so 
these modifications might originate from charged lepton~\cite{chrgdleptcrrs1,chrgdleptcrrs2,chrgdleptcrrs3,chrgdleptcrrs4,chrgdleptcrrs5,chrgdleptcrrs6,chrgdleptcrrs7,chrgdleptcrrs8,chrgdleptcrrs9,chrgdleptcrrs10,chrgdleptcrrs11} and neutrino~\cite{neutrinocrrs1,neutrinocrrs2,neutrinocrrs3,neutrinocrrs4,neutrinocrrs5,neutrinocrrs6,neutrinocrrs7,neutrinocrrs8} sector. 
Here we performed numerical analysis by scanning parameter space for each case of these mixing schemes. The main characteristics of 
our detailed numerical investigation are:\\ 
{\bf{(i)}} The latest global fit results~\cite{globalfit3} have been used for our model independent analysis. These results are obtained
by taking into account latest data from long-baseline accelerator, 
solar and Kam-LAND, short-baseline reactor, and atmospheric neutrino experiments.\\
{\bf{(ii)}} We invoked $\chi^2$ approach~\cite{skgetal1,skg1,skg2} for studying the situation of mixing 
angle fitting in parameter space. This will give essential information about magnitude and sign of  correction parameters. It will also help in 
comparing level of fitting achieved for various cases under different modification schemes.\\
{\bf{(iii)}} The correlations among neutrino 
mixing angles are studied by varying all mixing angles in their $3\sigma$ limits. It is different from the approach where 
one out of three mixing angles is fixed at a particular value for discussing the 
correlation between remaining two mixing angles. This in our view show a complete picture and thus we present our numerical findings in terms of  
2 dimensional scatter plots instead of going for line plots. \\ 
{\bf{(iv)}} Finally the limits on leptonic CP phase($\delta_{CP}$) and Jarkslog invaraint($J_{CP}$) has been derived from the
parameter space which is consistent with 3 mixing angles global fit data. The obtained ranges will thus
act as a prediction of that mixing case.

These results can be helpful for understanding the structure of corrections that these well known mixing schemes require in order to be 
consistent with neutrino mixing data. This model independent investigation can also be useful in filtering out viable models from vast number of possibilities 
in neutrino model building physics. Moreover all allowed cases have  clear prediction 
of CP Dirac phase which can be easily tested from current/planned neutrino experiments. However mapping these results from model
dependent prospective is deferred  for future consideration.

The paper is organized as follows. In section 2, we give general discussion about methodology of our work. In sections 3-5, we present our 
numerical results for various possible correction cases under different mixing schemes. Finally in section 6, we give the summary and 
conclusions of our study.


\section{General Setup}

The form of mixing matrix for mixing scenarios under consideration is given as follows:
\begin{eqnarray} \nn
U_{\rm TBM} &=& \left ( \begin{array}{rrr}
\sqrt{2\over 3}&\sqrt{1\over 3}&0\\
-\sqrt{{1\over 6}}&\sqrt{{1\over 3}}&\sqrt{1\over 2}\\
-\sqrt{{1\over 6}}&\sqrt{{1\over 3}}&-\sqrt{{1\over 2}}
\end{array}
\right )\; , \hspace{0.2cm}
U_{\rm BM}=\left(
\begin{array}{rrr}
\sqrt{1\over 2 } & \sqrt{1\over 2 } & 0 \\
-{1 \over 2 }& {1 \over 2 } & \sqrt{1\over 2 } \\
{1 \over 2 } &  -{ 1 \over 2 } &\sqrt{1\over 2 }
\end{array}\right) \; , \\
U_{\rm DC} &=& \left ( \begin{array}{rrr}
\sqrt{\frac{1}{2}}&\sqrt{\frac{1}{2}}&0\\
\sqrt{\frac{1}{6}}&-\sqrt{\frac{1}{6}}&-\sqrt{\frac{2}{3}}\\
-\sqrt{\frac{1}{3}}&\sqrt{\frac{1}{3}}&-\sqrt{\frac{1}{3}}
\end{array}
\right )\;, \hspace{0.2cm}
U_{\rm HM}=\left(
\begin{array}{rrr}
\sqrt{3}\over 2 & 1\over 2 & 0 \\
-{1 \over {2\sqrt{2}} }&~~ ~{\sqrt{3} \over {2\sqrt{2}} } & ~~-\sqrt{1\over 2 } \\
-{1 \over {2\sqrt{2}} } &  {\sqrt{3} \over {2\sqrt{2}} } &\sqrt{1\over 2 }
\end{array}\right) \;. \label{vtri}
\end{eqnarray}

\begin{table} 
\begin{center}
\begin{tabular}{lccccc}

\hline

\hline

Mixing Angle & BM mixing & DC mixing & TBM mixing & HG mixing\\

\hline

$\theta_{23}^\circ$ & $45$ & 54.7 & 45 & 45\\

\hline

$\theta_{12}^\circ$ & $45$ & 45 & 35.3 & 30\\

\hline

$\theta_{13}^\circ$  & $0$  & 0 & 0 & 0\\

\hline

\label{mixingangles} 

\end{tabular}
\end{center}
\vspace{-1cm}
\caption{\it{Mixing angle values from unpertrubed special matrices.  All angles are in degrees($\theta^\circ$).}}

\label{mixingangles} 
\end{table}

\begin{center}
\begin{tabular}{ |p{3.2cm}||p{2.0cm}|p{2.5cm}|p{2.5cm}|p{2.5cm}|  }
 \hline
 \multicolumn{5}{|c|}{} \\
 \hline
 Normal Hierarchy  & Best fit & $1\sigma$ range& $2\sigma$ range & $3\sigma$ range\\
 \hline
 $\sin^2\theta_{12}/10^{-1}$   & $3.04$   & $2.91-3.18$ &   $2.78-3.32$ & $2.65-3.46$\\
  \hline
  $\sin^2\theta_{13}/10^{-2}$   & $2.14$   & $2.07-2.23$ &   $1.98-2.31$ & $1.90-2.39$\\
  \hline
    $\sin^2\theta_{23}/10^{-1}$   & $5.51$   & $4.81-5.70$ &   $4.48-5.88$ & $4.30-6.02$\\
      \hline
    \hline
   Inverted Hierarchy &&&&\\
   \hline
 $\sin^2\theta_{12}/10^{-1}$   & $3.03$   & $2.90-3.17$ &   $2.77-3.31$ & $2.64-3.45$\\
  \hline
  $\sin^2\theta_{13}/10^{-2}$   & $2.18$   & $2.11-2.26$ &   $2.02-2.35$ & $1.95-2.43$\\
  \hline
    $\sin^2\theta_{23}/10^{-1}$   & $5.57$   & $5.33-5.74$ &   $4.86-5.89$ & $4.44-6.03$\\
   \hline
\end{tabular}\captionof{table}{\it{Three-flavor oscillation neutrino mixing angles from fit to global data~\cite{globalfit3}.}}\label{Table1} 
\end{center}


The resulting value of mixing angles from above mixing schemes are given in Table~\ref{mixingangles}.
They share a common prediction of vanishing value of reactor mixing angle i.e. $\theta_{13}=0^{\circ}$. 
The atmospheric mixing angle($\theta_{23}$) is $45^{\circ}$ for BM, TBM and HG case while it takes a higher
value of $54.7^{\circ}$ in DC case. However solar mixing 
angle($\theta_{12}$) is $45^{\circ}$ for BM and DC scenario while it takes a value of $35.3^{\circ}$ and $30^{\circ}$ in TBM and HG case respectively. 
Thus these mixing angles are in conflict with recent experimental observations which provide best fit values at 
 $\theta_{13} \sim 8^{\circ}$, $\theta_{12} \sim 33^{\circ}$ and $\theta_{23} \sim 47^{\circ}$.
Hence these mixing schemes should be probed for possible corrections in order to investigate  their consistency with
current neutrino oscillation data.

In this study, we investigated different cases pertaining to PMNS matrix of the 
forms  $U_X \cdot V_M$ and $V_M \cdot U_X$ where $U_X$ denotes a complex rotation in $ij$ sector and $V_M$ is any one of these special matrices. 
The correction matrix $U_X$ can be expressed in terms of mixing matrix
as $R_X= \{ R^{}_{23}, R^{}_{13}, R^{}_{12} \}$ in general with a single phase parameter($\sigma$) as follows
\begin{eqnarray} \nn
&&U^{}_{12} = \left (\begin{array}{ccc}
\cos \alpha & \sin \alpha~e^{-i\sigma} &0\\
-\sin \alpha ~e^{i\sigma}&\cos \alpha &0\\
0&0&1
\end{array}
\right )\;,  U^{}_{23} = \left (\begin{array}{ccc}
1&0&0\\
0&\cos \beta  &\sin \beta~e^{-i\sigma}  \\
0&-\sin \beta~e^{i\sigma} & \cos \beta
\end{array}\right )\;, \\ && U^{}_{13} = \left ( \begin{array}{ccc}
\cos \gamma &0&\sin \gamma~e^{-i\sigma}  \\
0&1&0\\
-\sin \gamma~e^{i\sigma}  &0& \cos \gamma
\end{array}
\right )\; \label{vb}
\end{eqnarray}
Here $R^{}_{12}$, $R^{}_{23}$ and $R^{}_{13}$ represent the rotations in 12, 23 and 13 sector with corresponding rotation angle
$\alpha$, $\beta$, $\gamma$ respectively. The related PMNS matrix for single rotation case is given by:
\begin{eqnarray}
&& V^{L}_{\rm ij}= U_{ij}^{l}  \cdot V_{M}^{}  \; , \label{p1}\\
&& V^{R}_{\rm ij}= V_{M}^{} \cdot U_{ij}^{r}    \; ,\label{p2} 
\end{eqnarray}
where  $(ij) =(12), (13), (23)$ respectively. 

The effect of these corrections is studied by implementing 
$\chi^2$ function which is a measure of overall departure of obtained values of mixing angles 
in parameter space from the values that are given from global fits~\cite{globalfit11,globalfit12, globalfit21,globalfit22, globalfit3}. It is given by the expression
\begin{equation}
   \chi^2 = \mathlarger{\mathlarger{‎‎\sum}}_{i=1}^{3‎} \{ \frac{\theta_i(P)-\theta_i^{exp}}{\delta \theta_i^{exp}} \}^2
\end{equation}
with $\theta_i(P)$ are the values of mixing angles obtained from mixing scheme in parameter space which is a function
of one of correction parameter ($\alpha,\beta, \gamma$) along with a phase parameter $\sigma$.
$\theta_i^{exp}$ are the best fitted value of neutrino mixing angles obtained from latest global fit data with corresponding $1\sigma$ deviation $\delta \theta_i$. 
The unperturbed values of $\chi^2$ for all considered mixing schemes with Normal and Inverted Hierarchy are given in Table~\ref{chisqunperturb}.\\

\begin{tabular}{ |p{2.2cm}||p{2.2cm}|p{2.2cm}|p{2.2cm}|p{2.2cm}|  }
 \hline
 \multicolumn{5}{|c|}{$\chi^2$ value with uncorrected Mixing schemes} \\
 \hline
 Hierarchy & BM &DC&TBM & HG\\
 \hline
 Normal   & $927.6$    & $933.1$ &   $721.5$ & $732.8$\\
  \hline
  Inverted   & $1065.5$    & $1086.4$ &   $857.6$ & $868.0$\\
  \hline
\end{tabular}\label{chisqunperturb} \\
 
As evident from above table, NH is little favorable as its $\chi^2$ value is lower as compared to its IH counterpart. 
 Here we investigated the role of various possible mixing cases for bringing $\chi^2$ further down in parameter space and thus reaching closer
 to experimental best fit. A good numerical fit should produce  low $\chi^2$ value in parameter region.

\section{Numerical Results}
In this section,  we present and discuss numerical findings  of our investigation for Normal and Inverted Hierarchy. We
studied the role of these corrected mixing schemes in producing large $\theta_{13}$~\cite{1largeth13,2largeth13,3largeth13,4largeth13,5largeth13,6largeth13,7largeth13,8largeth13,9largeth13,10largeth13,11largeth13,12largeth13,13largeth13,14largeth13,15largeth13,16largeth13,17largeth13,18largeth13,19largeth13,20largeth13,21largeth13,22largeth13,23largeth13,24largeth13,25largeth13,26largeth13,27largeth13,28largeth13,29largeth13,30largeth13} and fitting 
other mixing angles. The resulting value of $\delta_{CP}$  and $J_{CP}$ from allowed parameter space is treated as prediction of that
mixing case. All these well known mixing scenarios have common prediction of $\theta_{13}=0$ since 
13 element is zero. Thus all such mixing schemes can be put in following generic form
\begin{eqnarray} \nn
V_{\rm Mix}=\left(
\begin{array}{rrr}
a_{11} & a_{12} & 0 \\
a_{21}&~~ ~a_{22} & ~~a_{23} \\
a_{31} &  a_{32} & a_{33 }
\end{array}\right) \;. \label{vtri}
\end{eqnarray}

where all matrix elements are real for our considered cases.
Here we present our formulae for mixing angles, $\delta_{CP}$ and $J_{CP}$ in terms of correction parameters using above
generic form. The same formuale can also be used in other situations which have similar prediction of $\theta_{13}=0$.

The scanning of parameter space is performed by randomly varying $\alpha$, $\beta$ and $\gamma$ in the
range [-0.5,~0.5] whereas the phase parameter($\sigma$) is chosen from the interval $[-\pi, ~\pi]$. A good numerical fit to global fitting data
should produce low $\chi^2$ and thus plotting data points are selected by invoking the condition $\chi^2 < \chi^2_{i}$($i$= BM, DC, TBM and HG)
in our scanning subroutines. The discussion on obtained results is presented from following plots:\\
{\bf{(i)}} A 2-dimensional projection of $\chi^2$ over correction parameters revealing overall situation of fitting in parameter space.
 The reported value of $\chi^2_{min}$ in these plots corresponds to best level of fitting for all three 
 mixing angles. \\
{\bf{(ii)}} Scattered plot of $\theta_{13}$ over $\theta_{23}-\theta_{12}$ plane for getting information about best level of fitting and 
ranges of mixing angles that can be achieved in parameter space.\\
{\bf{(iii)}} The  scattered plots of  $\delta_{CP}$ and $J_{CP}$ vs mixing angles within allowed region for determining
the range of these quantities.\\
In figures of $\chi^2$ over modification parameters ($\theta_1, \theta_2$) red, blue and light green color 
regions pertains to $\chi^2$ value in the interval $[0, 3]$,  $[3, 10]$  and  $ > 10$ respectively. The white region of plot corresponds to
completely disallowed part of $\chi^2 > \chi^2_{unperturbed}$.

In neutrino mixing angles figures, green band refers to $1\sigma$ and full colored band to $3\sigma$ values
of $\theta_{13}$. Also `$\times$' refers to the case which is unable to fit mixing angles even at $3\sigma$ level while `-' pertains 
to the situation where $\theta_{13}$ remains unchanged i.e. $\theta_{13}=0$. For showing  the mapping between left and right figures, 
we marked the $\chi^2 < 3, [3, 10]$ regions in mixing angle plots using color codings. The white region corresponds to $ 3 < \chi^2 < 10 $ 
whereas yellow region refers to 
$\chi^2 < 3$. Horizontal and vertical dashed black, dashed pink and thick black lines corresponds to $1\sigma$, $2\sigma$ and $3\sigma$ ranges 
of the other two mixing angles. Now we will take up all considered cases one by one.

\section{Rotations-$U_{ij}^l.V_{M}$}

Here we first consider the corrections for which the form of modified PMNS matrix is given by $U_{PMNS} = U_{ij}^l.V_{M}$.
It will introduce changes in $i^{\text{th}}$ and $j^{\text{th}}$ row of unperturbed matrix. In subsequent subsections, we investigate
the role of this mixing scheme in fitting neutrino mixing angles and its prediction for $\delta_{CP}$ and $J_{CP}$.

\subsection{12 Rotation}

This mixing scheme pertains to complex rotation in 12 sector of these special matrices. Here rotation matrix operates
from left side and thus impart changes in first two rows of unperturbed mixing matrix. The expressions for neutrino mixing angles 
in this scheme are given as

\beqa
 \sin^2\theta_{13} &=&  a_{23}^2 \sin^2\alpha,\\
 \sin^2\theta_{23} &=& \frac{a_{23}^2\cos^2\alpha}{\cos^2\theta_{13}},\\
  \sin^2\theta_{12} &=& \frac{a_{12}^2\cos^2\alpha + a_{22}^2\sin^2\alpha + a_{12}a_{22}\sin 2\alpha \cos\sigma}{\cos^2\theta_{13}}
\eeqa

The Jarsklog invariant and CP Dirac phase is given by the expressions 
\beqa
 \sin^2\delta_{CP} &=& C_{12L}^2 \left(\frac{p_{1\alpha}}{p_{2\alpha\sigma} p_{3\alpha\sigma}}\right)\sin^2\sigma,\\
J_{CP} &=& J_{12L} \sin2\alpha \sin\sigma
\eeqa

where 

\beqa
 J_{12L} &=& \frac{1}{2} a_{23}^2 \sqrt{1-a_{23}^2}~ C_{12L},\\
 C_{12L} &=& \frac{(a_{11}a_{22}-a_{12}a_{21})(a_{11}a_{12}+a_{21}a_{22})}{a_{23}^2 \sqrt{1-a_{23}^2}},\\
 p_{1\alpha} &=& 1+a_{23}^4\sin^4\alpha-2 a_{23}^2\sin^2\alpha,\\
 p_{2\alpha\sigma} &=& 1-a_{12}^2\cos^2\alpha - (a_{22}^2 +a_{23}^2)\sin^2\alpha - a_{12}a_{22}\cos\sigma \sin 2\alpha,\\
 p_{3\alpha\sigma} &=& a_{12}^2\cos^2\alpha + a_{22}^2\sin^2\alpha + a_{12}a_{22}\cos\sigma \sin 2\alpha 
\eeqa

\begin{figure}[!t]\centering
\begin{tabular}{c c} 
\hspace{-5mm}
\includegraphics[angle=0,width=80mm]{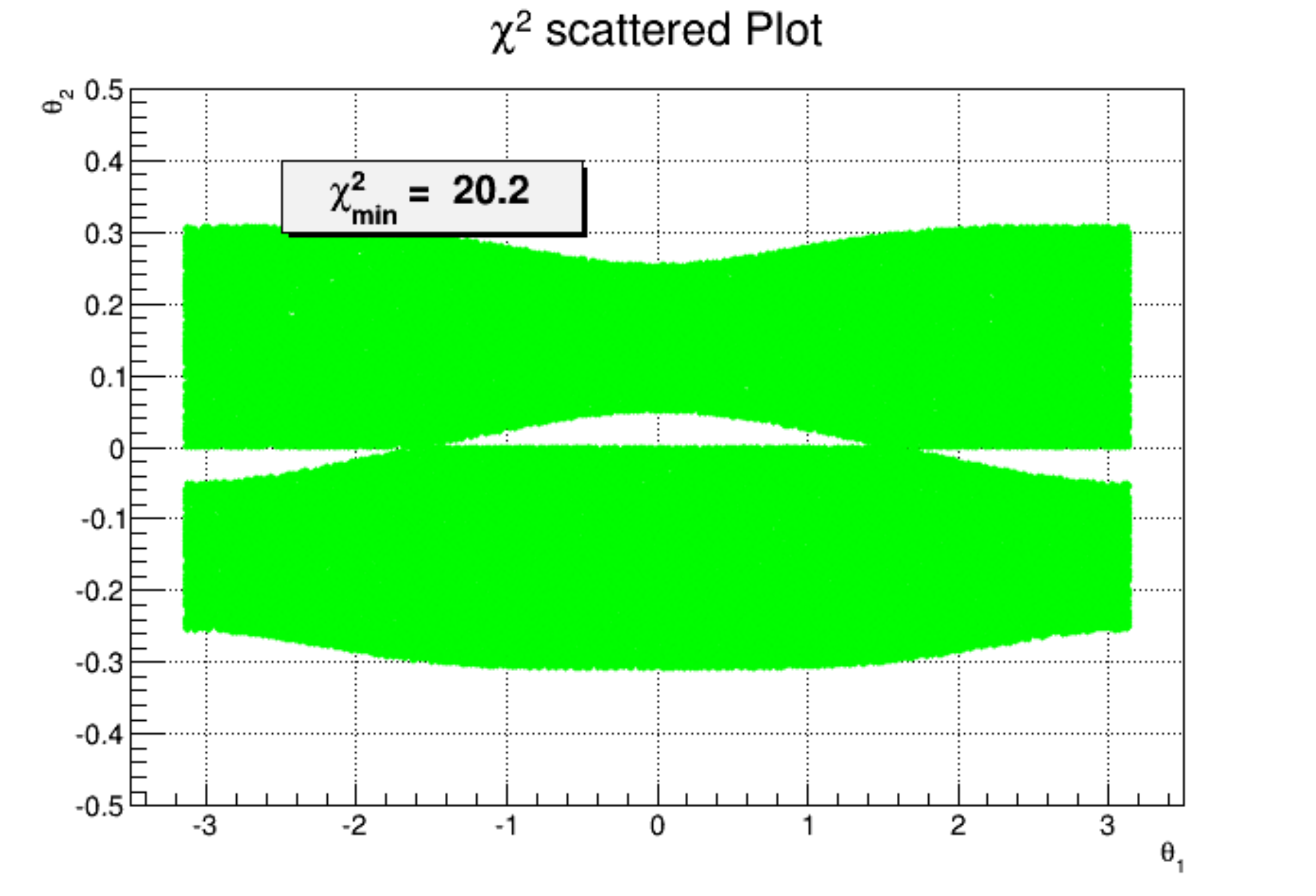} &
\includegraphics[angle=0,width=80mm]{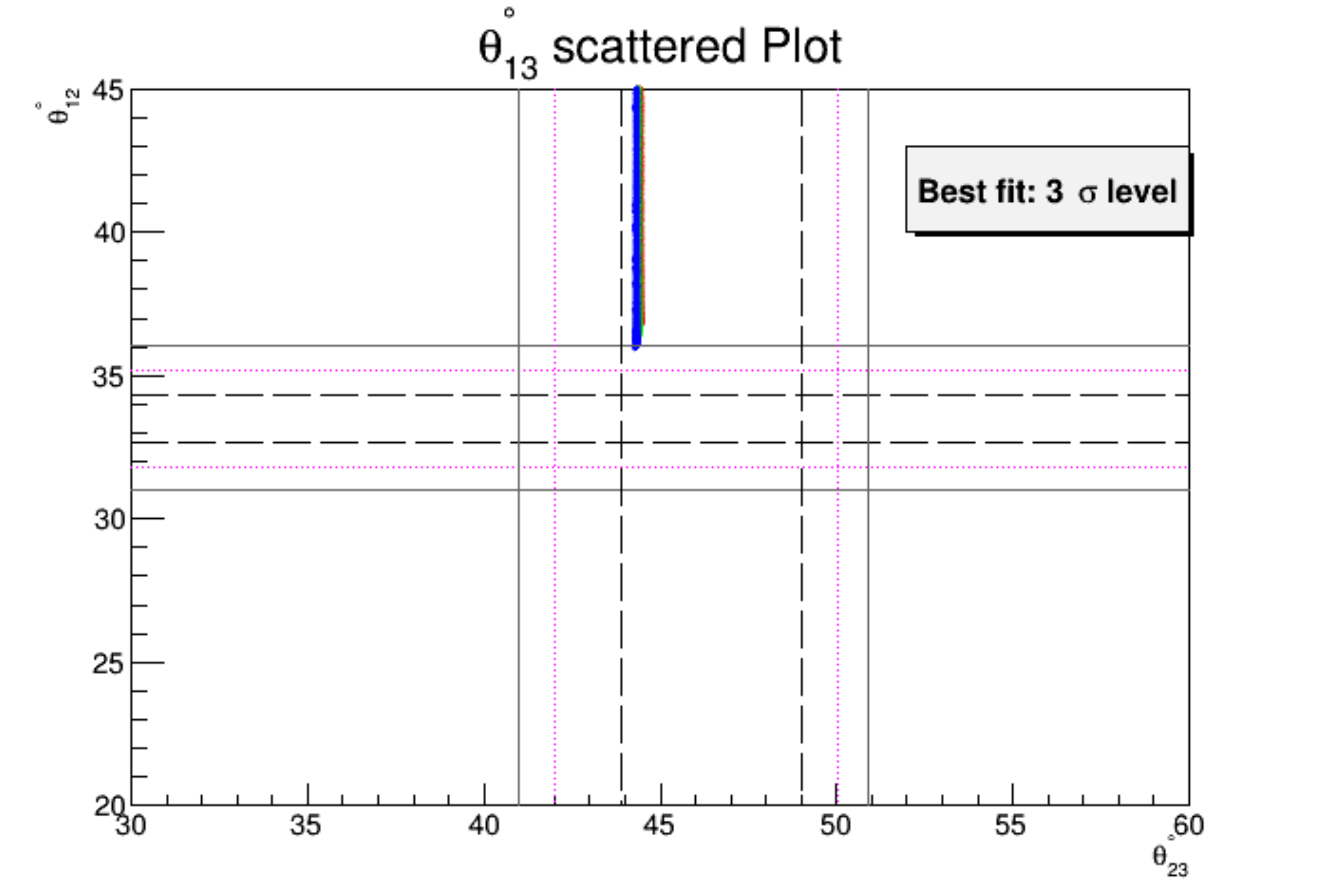}\\
\end{tabular}
\caption{\it{Scattered plot of $\chi^2$ (left fig.) over $\alpha-\sigma$ plane and $\theta_{13}$ (right fig.) 
over $\theta_{23}-\theta_{12}$ (in degrees) plane for $U^{BML}_{12}$ rotation scheme. The information about other color coding 
and various horizontal, vertical 
lines in right fig. is given in text.}}
\label{fig12L1}
\end{figure}

\begin{figure}[!t]\centering
\begin{tabular}{c c} 
\hspace{-5mm}
\includegraphics[angle=0,width=80mm]{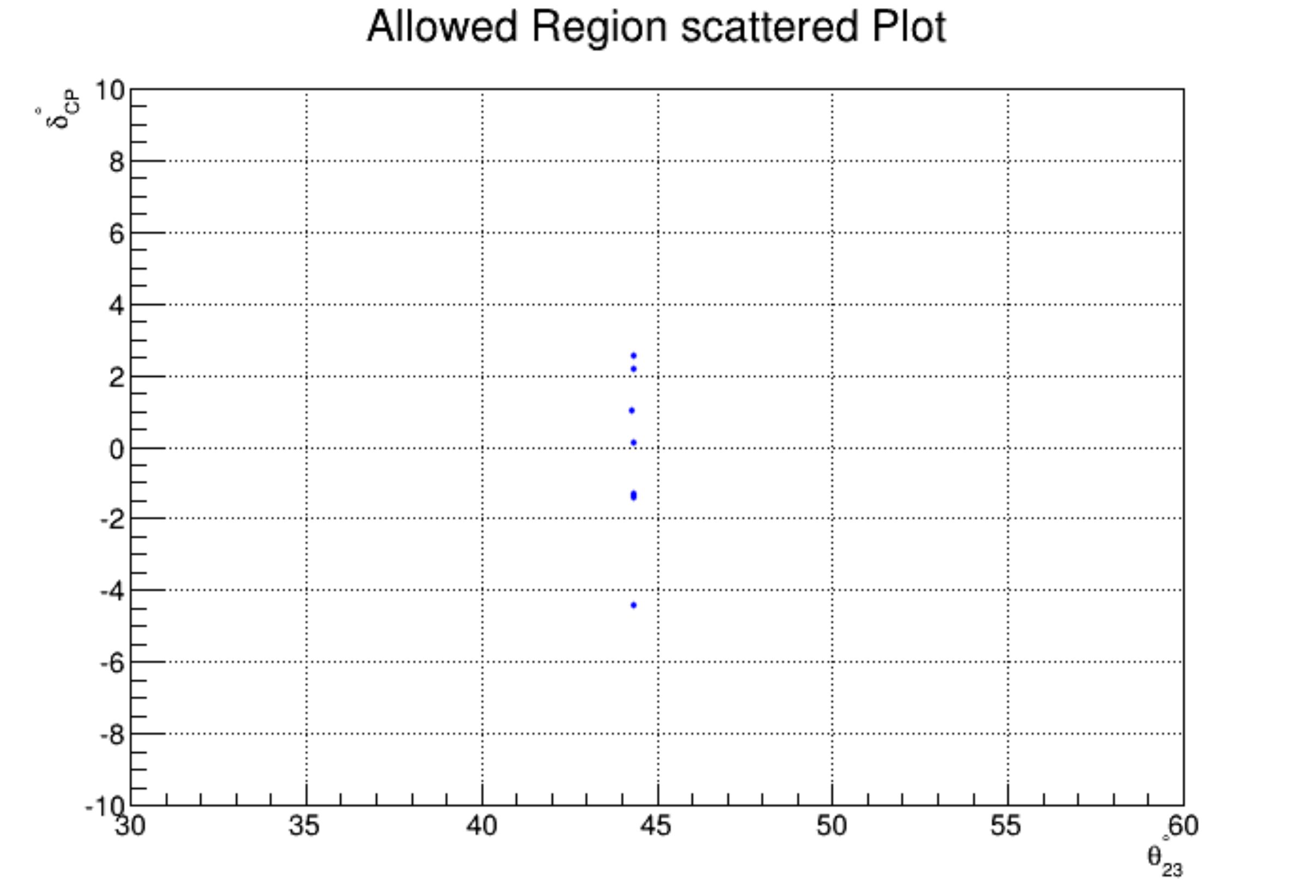} &
\includegraphics[angle=0,width=80mm]{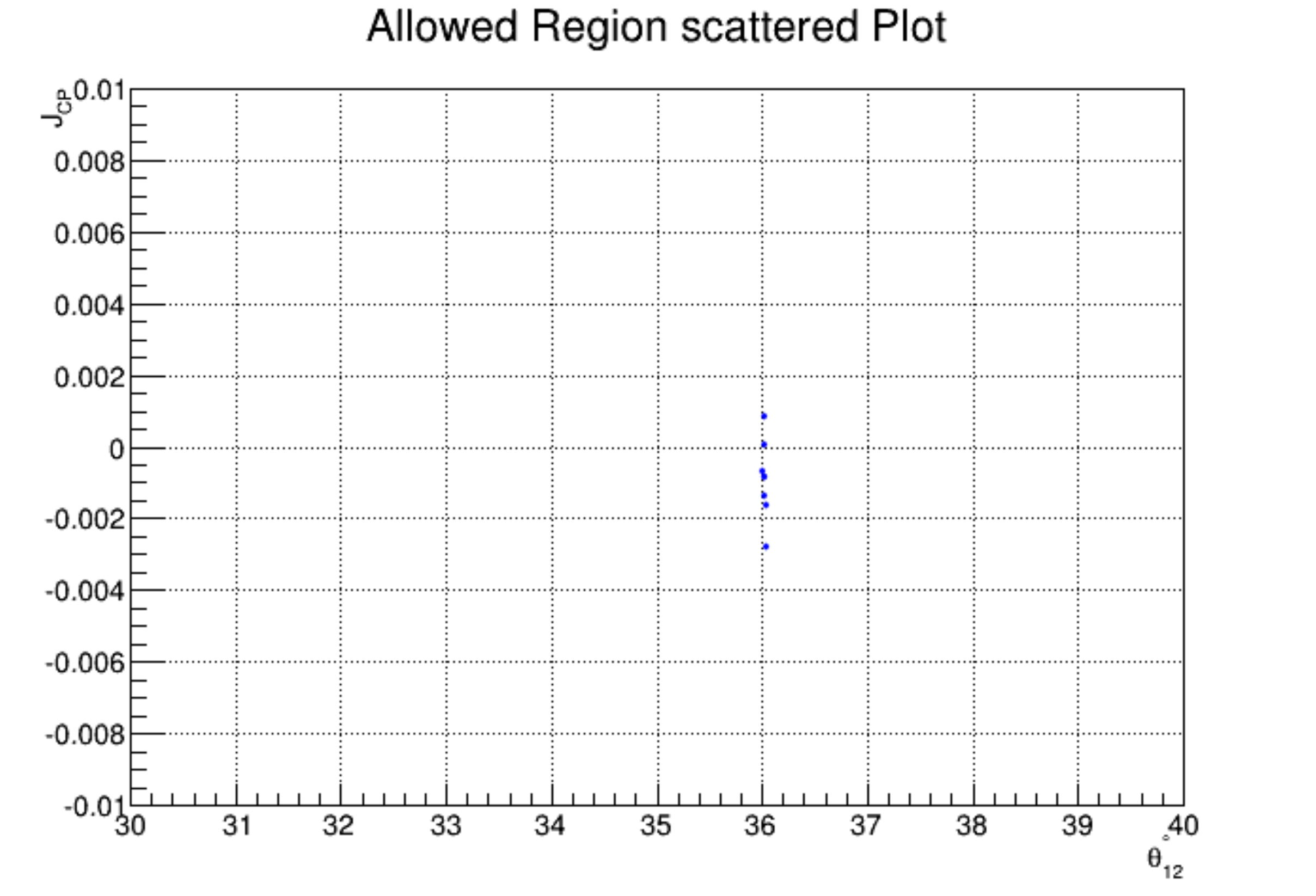}\\
\end{tabular}
\caption{\it{Scattered plot of $\delta_{CP}$ (left fig.) vs $\theta_{23}$ (in degrees) and scattered plot of $J_{CP}$ (right fig.) 
over $\theta_{12}$ (in degrees) plane for $U^{BML}_{12}$ rotation scheme.}}
\label{fig12L2}
\end{figure}

\begin{figure}[!t]\centering
\begin{tabular}{c c} 
\hspace{-5mm}
\includegraphics[angle=0,width=80mm]{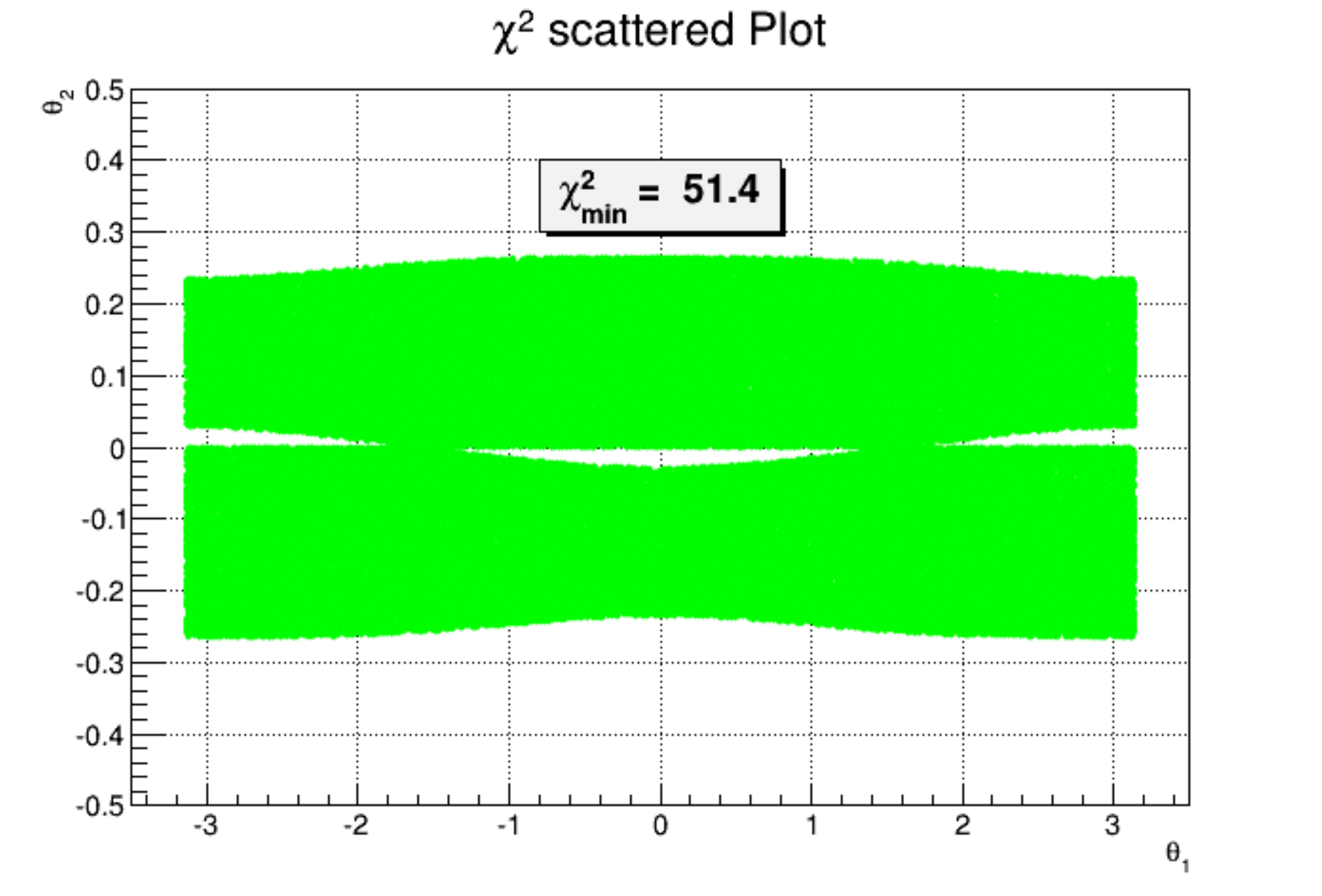} &
\includegraphics[angle=0,width=80mm]{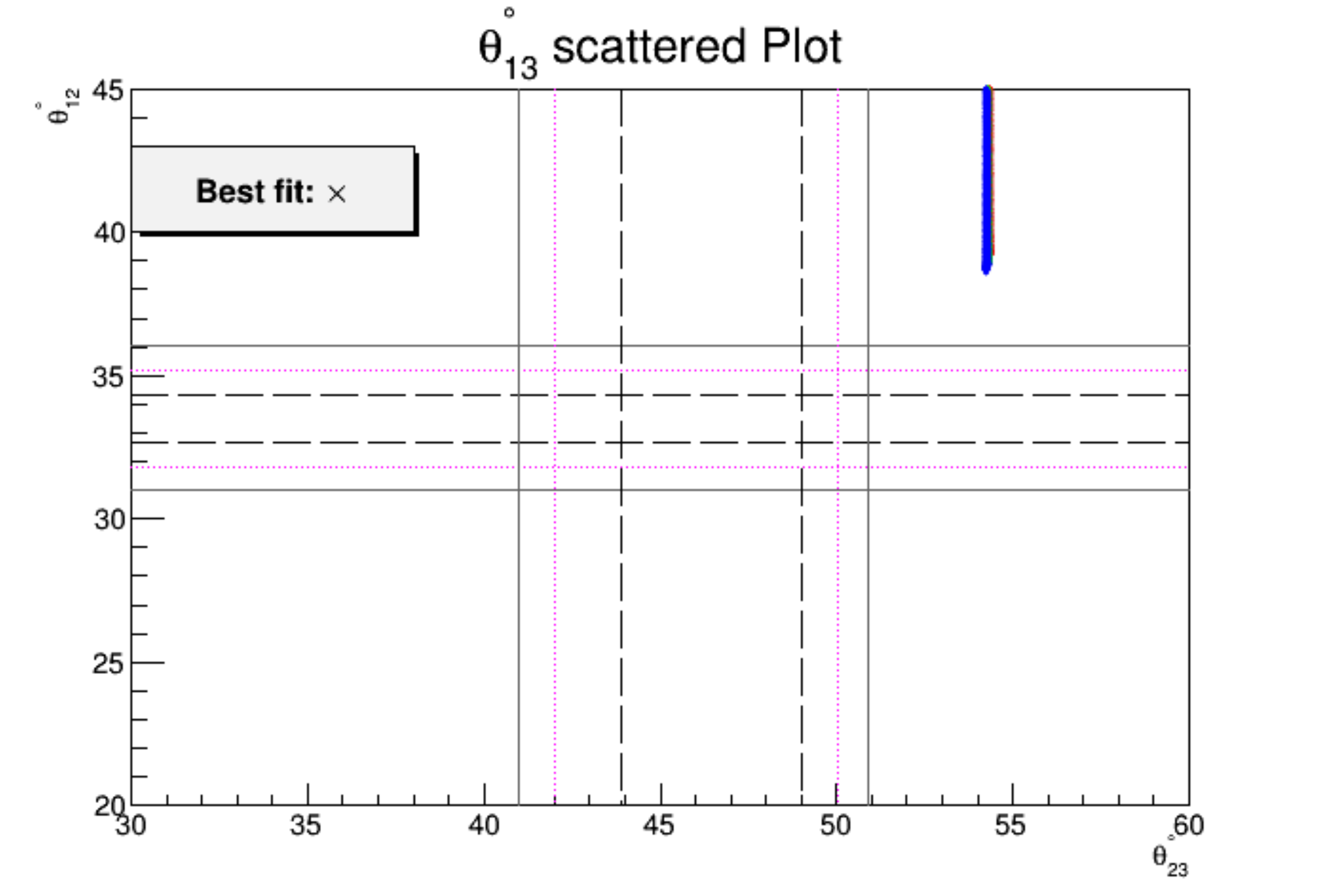}\\
\end{tabular}
\caption{\it{Scattered plot of $\chi^2$ (left fig.) over $\alpha-\sigma$ plane and  $\theta_{13}$ (right fig.) 
over $\theta_{23}-\theta_{12}$ (in degrees) plane for $U^{DCL}_{12}$ rotation scheme. The information about other color coding 
and various horizontal, vertical lines in right fig. is given in text.}}
\label{fig12L3}
\end{figure}

\begin{figure}[!t]\centering
\begin{tabular}{c c} 
\hspace{-5mm}
\includegraphics[angle=0,width=80mm]{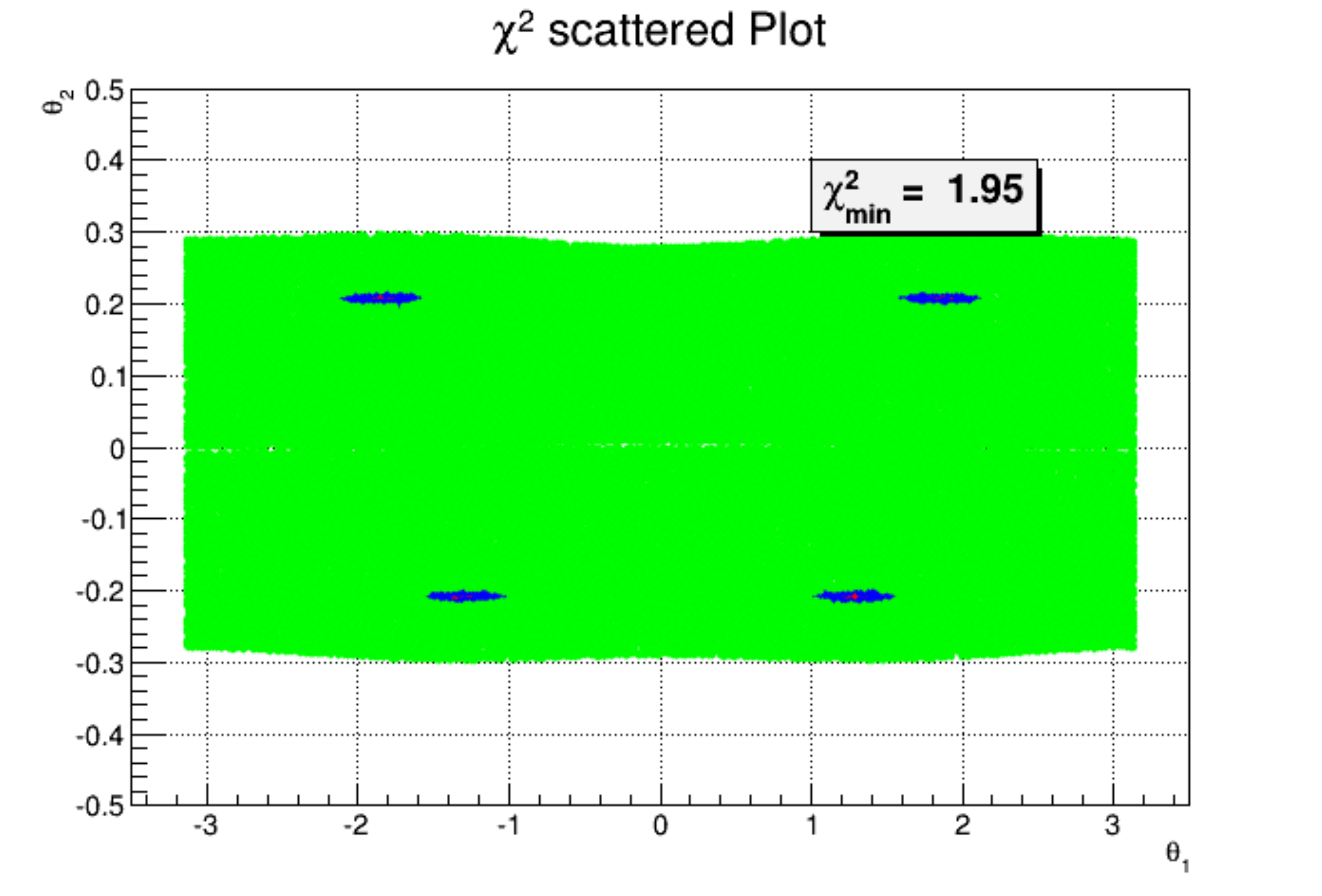} &
\includegraphics[angle=0,width=80mm]{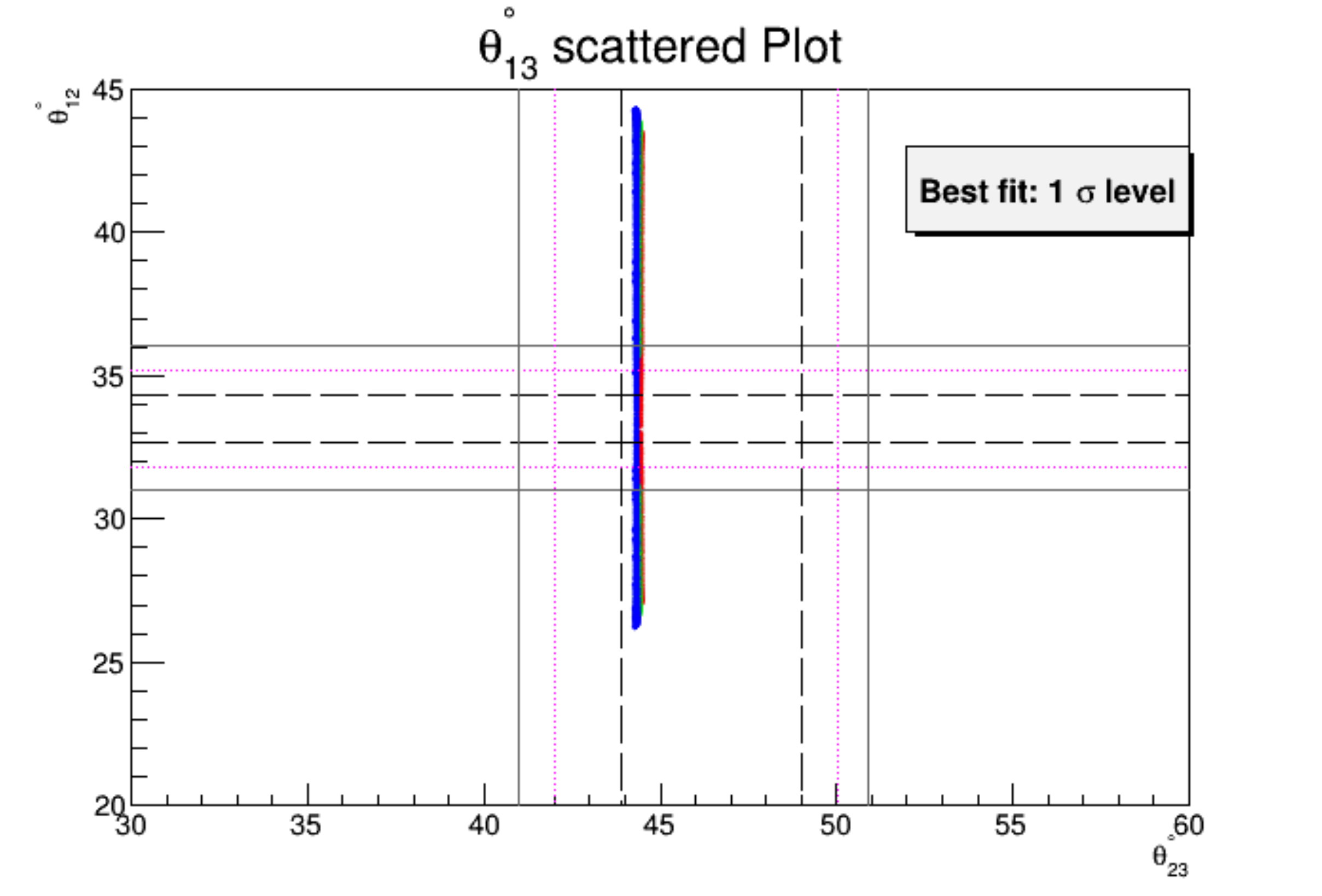}\\
\end{tabular}
\caption{\it{Scattered plot of $\chi^2$ (left fig.) over $\alpha-\sigma$ plane and  $\theta_{13}$ (right fig.) 
over $\theta_{23}-\theta_{12}$ (in degrees) plane for $U^{TBML}_{12}$ rotation scheme. }}
\label{fig12L6}
\end{figure}

\begin{figure}[!t]\centering
\begin{tabular}{c c} 
\hspace{-5mm}
\includegraphics[angle=0,width=80mm]{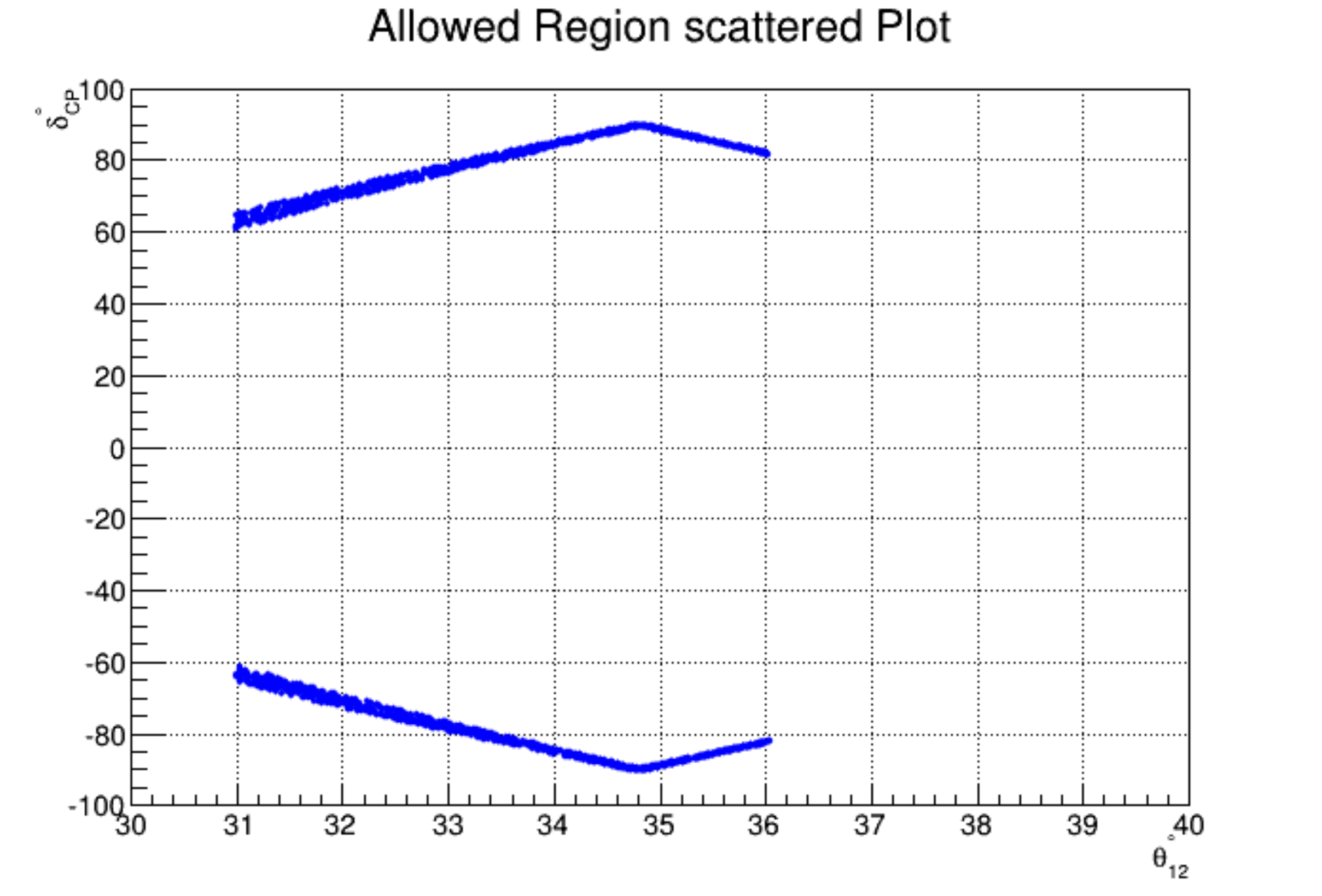} &
\includegraphics[angle=0,width=80mm]{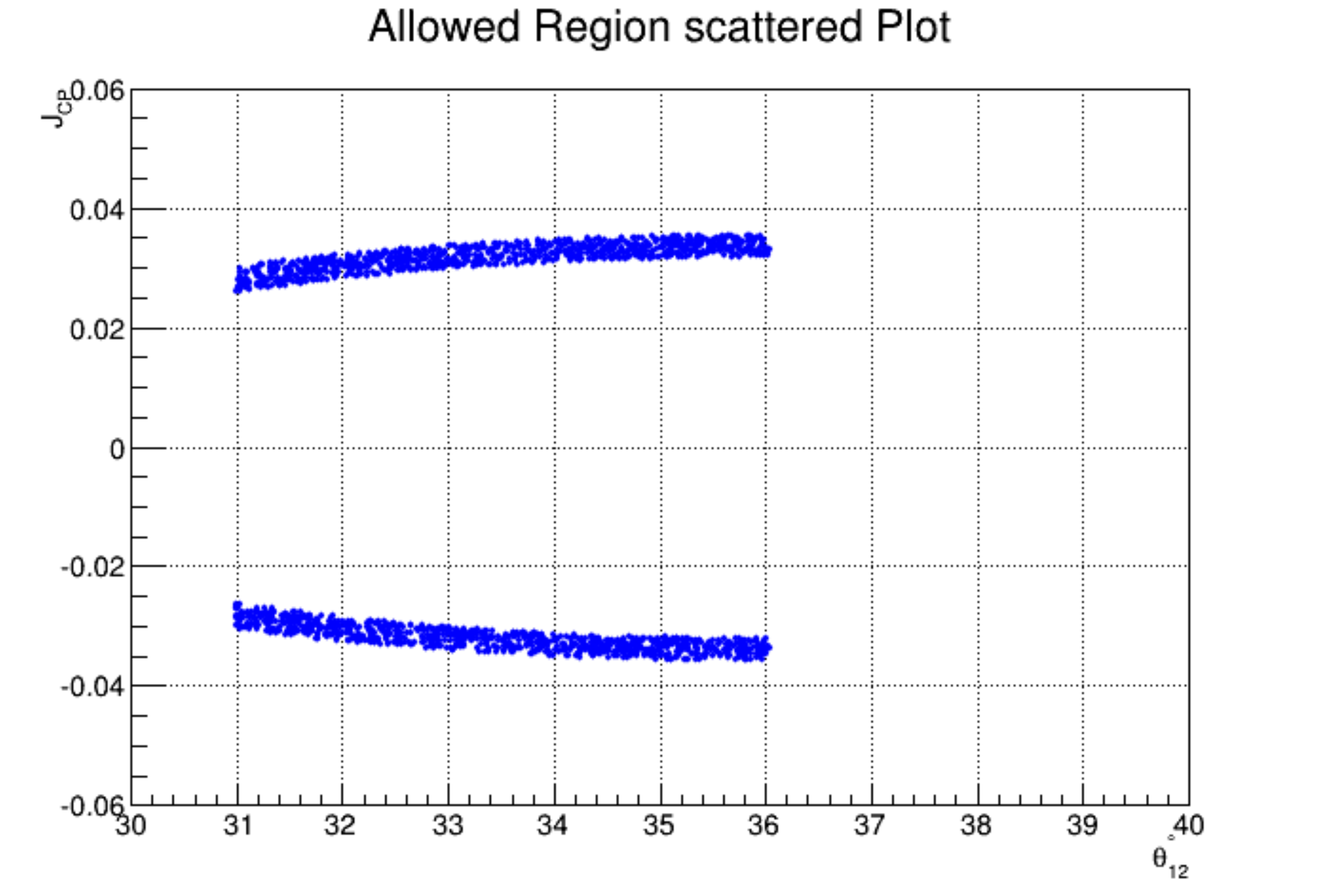}\\
\end{tabular}
\caption{\it{Scattered plot of $\delta_{CP}$ (left fig.) vs $\theta_{12}$ (in degrees) and scattered plot of $J_{CP}$ (right fig.) 
over $\theta_{12}$ (in degrees) plane for $U^{TBML}_{12}$ rotation scheme.}}
\label{fig12L7}
\end{figure}

Fig.~\ref{fig12L1}-\ref{fig12L7} show the numerical results corresponding to this
mixing case for normal hierarchy(NH). The notable features of this mixing are:\\
{\bf{(i)}} Here $\theta_{23}$ remain close to its unperturbed value since it receives corrections of $O(\theta^2)$
from parameter $\alpha$. Since for DC case unperturbed $\theta_{23}\sim 54.7$ so it is disfavored completely in this
mixing scheme. \\
{\bf{(ii)}} As fitting of $\theta_{13}$ and $\theta_{23}$ is only governed by $\alpha$
so its allowed range is much constrained in parameter space. e.g. for TBM case, the fitting of $\theta_{13}$ under its $3\sigma$ domain 
constraints the magnitude of correction parameter 
$|\alpha| \in [0.1962(0.1988), 0.2204(0.2223)]$ which in turn fixes $\theta_{23} \in [44.29^\circ(44.28^\circ), 44.44^\circ(44.43^\circ)]$ for 
corresponding $\alpha$ values in NH(IH) case. However $\theta_{12}$ possess much wider range of values 
since it receives corrections from $\alpha$ as well from phase parameter $\sigma$. \\
{\bf{(iii)}} The minimum value of $\chi^2 \sim 20.2(30.2),~51.4(70.4),~1.95(11.0)$ and $1.94(11.0)$ for BM, DC, TBM and HG respectively with 
NH(IH) case. Here BM barely manages to fit all mixing angles in $3\sigma$ level for a very minute region of parameter space while DC is not viable.  
However HG and TBM are much favorable as they are able to fit all mixing angles within $1\sigma$ level in NH case. Since $1\sigma$ range for 
$\theta_{23}$ in IH is quite constrained as compared to its NH counterpart so fitted value of $\theta_{23}$ in TBM and HG  goes outside its $1\sigma$ range. Thus 
this case is only allowed at $2\sigma$ level.\\
{\bf{(iv)}} Leptonic phase $\delta_{CP}$ lies in the range $-4.4(-4.7) \leq \delta_{CP} \leq 3.4(5.3)$ for BM while it is confined
in $ 39.0(40.4) \leq |\delta_{CP}|  \leq 78.7(79.2)$ for modified HG  and $61.0(60.9) \leq |\delta_{CP}| \leq 89.9(89.9)$  corrected TBM matrix. \\
{\bf{(v)}} The Jarkslog invariant($J_{CP}$) remains in range $-0.0027(-0.0029) \leq J_{CP} \leq 0.0021(0.0033)$ 
for BM while it is confined in $0.020(0.021) \leq |J_{CP}| \leq 0.032(0.032)$ for corrected HG while it remains 
$0.026 \leq |J_{CP}| \leq 0.035$  for TBM mixing matrix.

\begin{center}
\begin{tabular}{ |p{1.3cm}||p{1.9cm}|p{2.0cm}|p{2.0cm}|p{2.0cm}|p{2.0cm}|p{2.5cm}|  }
 \hline
 \multicolumn{7}{|c|}{Best fit with Mixing data} \\
 \hline
Rotation  & $\chi^2_{min}$ & $\theta_{12}^\circ$ & $\theta_{23}^\circ$ & $\theta_{13}^\circ$ & $|\delta_{CP}^\circ|$ & $|J_{CP}|$\\
 \hline
 BM   & $20.2(30.2)$    & $36.02(35.97)$ &   $44.3(44.2)$ & $8.86(8.92)$ &$0.498(1.888)$ & $0.0003(0.0011)$\\
  \hline
  DC   & $51.4(70.4)$    & ${\bf{38.9}}({\bf{38.8}})$ &   ${\bf{54.2}}({\bf{54.2}})$ & $8.54(8.62)$ &$0.152(1.879)$ & $0.00008(0.0011)$\\
  \hline
  TBM   & $1.95(11.0)$    & $33.37(33.39)$ &   $44.3(44.3)$ & $8.42(8.48)$ &$80.55(80.85)$ & $0.032(0.032)$\\
  \hline
  HG  & $1.94(11.0)$    & $33.51(33.38)$ &   $44.3(44.3)$ & $8.41(8.48)$ &$61.89(62.98)$ & $0.029(0.029)$\\
  \hline
\end{tabular}\captionof{table}{\it{Neutrino Mixing angles, $|\delta^\circ_{CP}|$ and $|J_{CP}|$ corresponding to $\chi^2_{min}$ numerical fit.
The mixing angle value that lies outside its best fit $3\sigma$ range is marked in boldface.}}
\label{chisqunperturb} 
\end{center}

\begin{figure}[!t]\centering
\begin{tabular}{c c} 
\hspace{-5mm}
\includegraphics[angle=0,width=80mm]{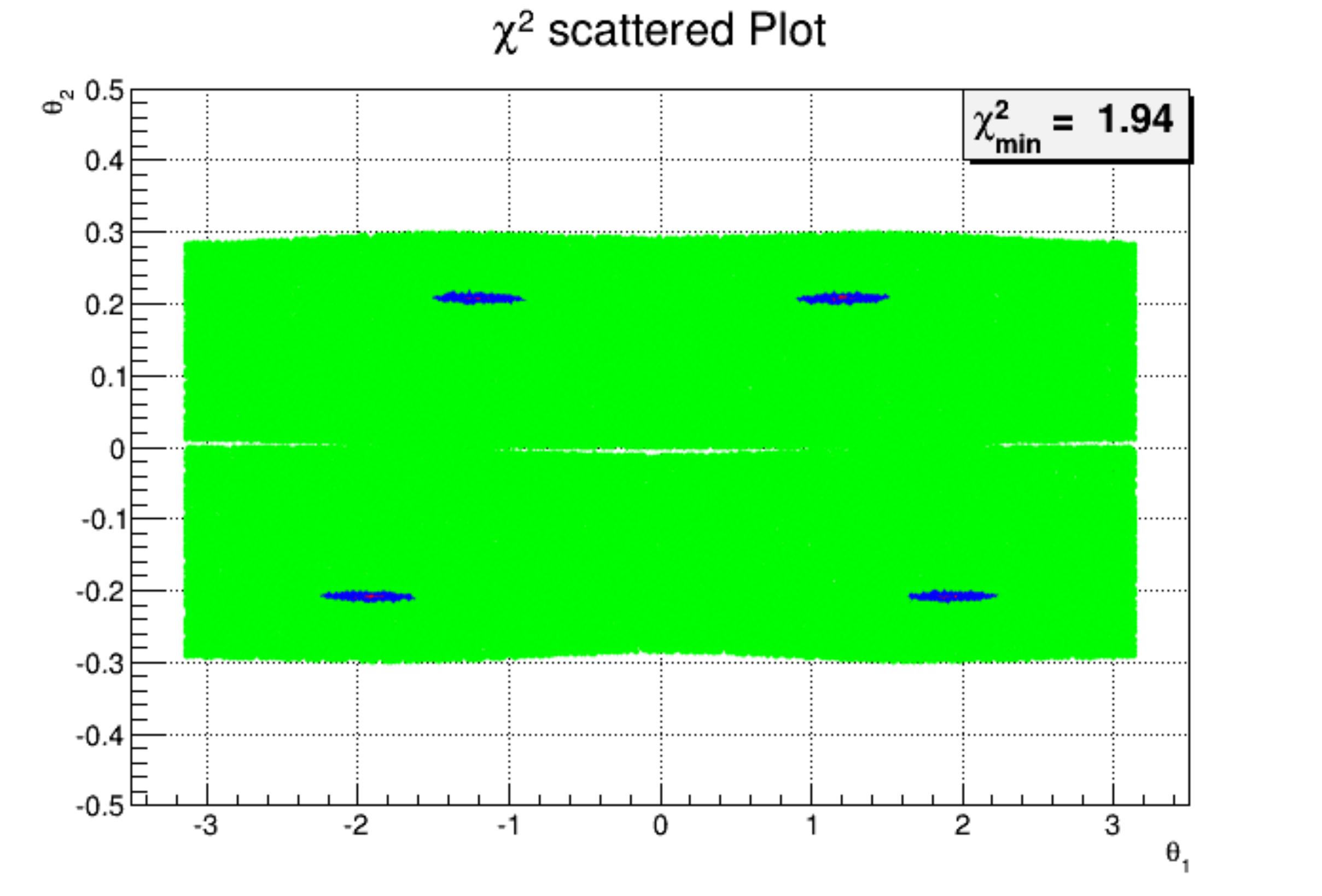} &
\includegraphics[angle=0,width=80mm]{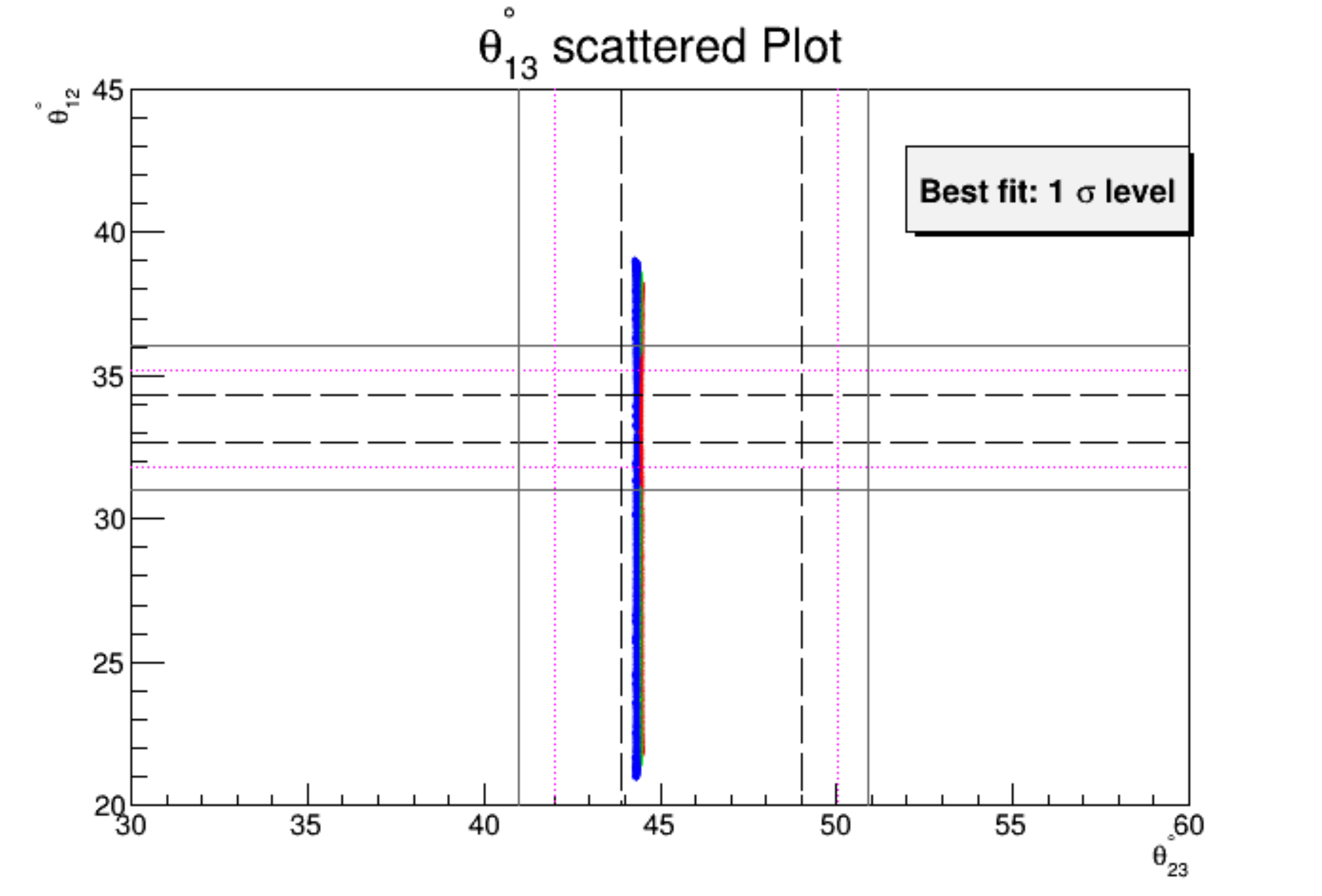}\\
\end{tabular}
\caption{\it{Scattered plot of $\chi^2$ (left fig.) over $\alpha-\sigma$ plane and $\theta_{13}$ (right fig.) 
over $\theta_{23}-\theta_{12}$ (in degrees) plane for $U^{HGL}_{12}$ rotation scheme. }}
\label{fig12L4}
\end{figure}

\begin{figure}[!t]\centering
\begin{tabular}{c c} 
\hspace{-5mm}
\includegraphics[angle=0,width=80mm]{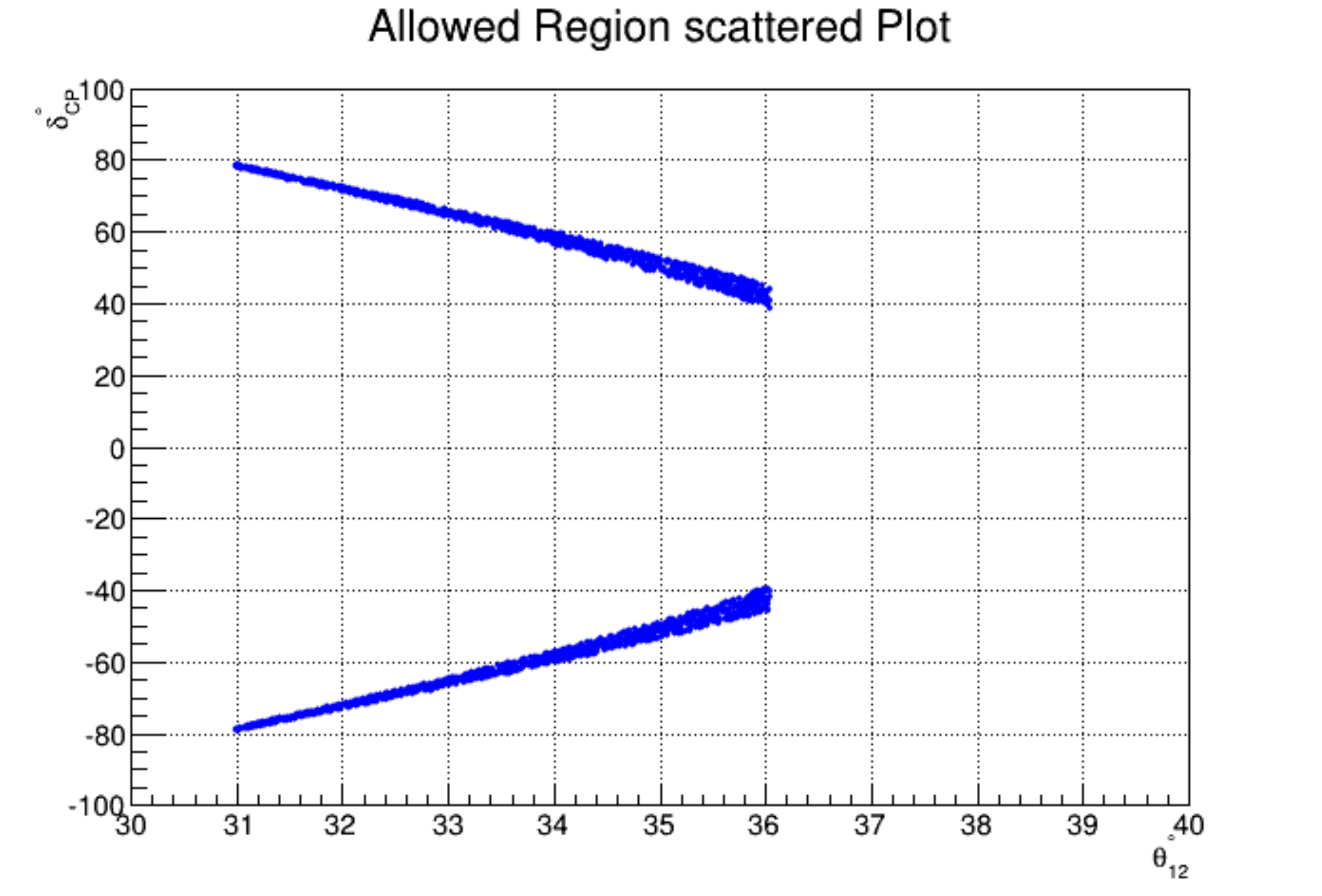} &
\includegraphics[angle=0,width=80mm]{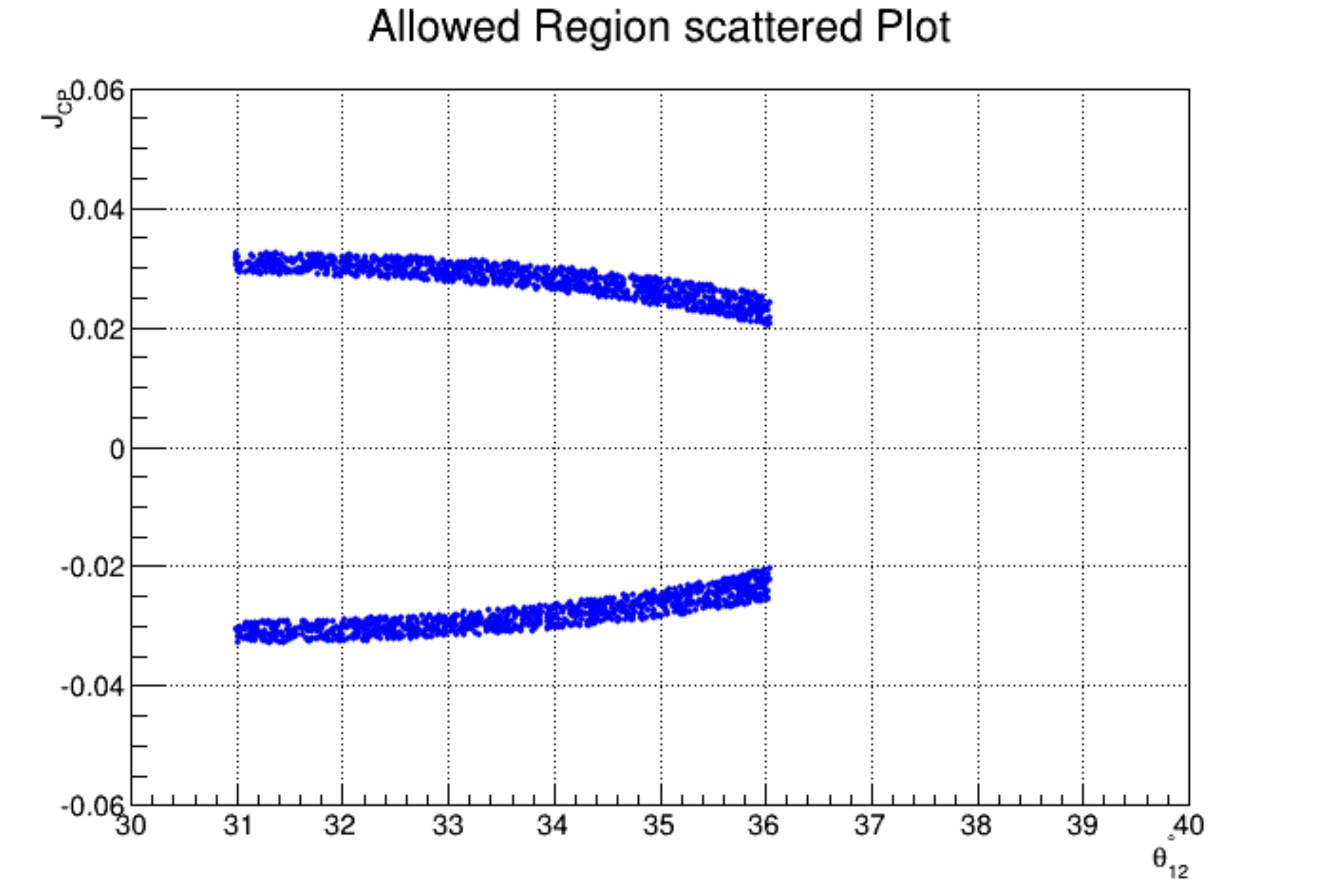}\\
\end{tabular}
\caption{\it{Scattered plot of $\delta_{CP}$ (left fig.) vs $\theta_{12}$ (in degrees) and scattered plot of $J_{CP}$ (right fig.) 
over $\theta_{12}$ (in degrees) plane for $U^{HGL}_{12}$ rotation scheme.}}
\label{fig12L5}
\end{figure}

\subsection{13 Rotation}

This case refers to rotation in 13 sector of  these special matrices that
bring modifications in Ist and 3rd row of unperturbed mixing matrix. 
The expressions of neutrino mixing angles for this case are given as


\beqa
 \sin^2\theta_{13} &=&  a_{33}^2 \sin^2\gamma,\\
 \sin^2\theta_{23} &=& \frac{a_{23}^2}{\cos^2\theta_{13}},\\
  \sin^2\theta_{12} &=& \frac{a_{12}^2\cos^2\gamma + a_{32}^2\sin^2\gamma + a_{12}a_{32}\sin 2\gamma \cos\sigma}{\cos^2\theta_{13}}
\eeqa

\beqa
\sin^2\delta_{CP} &=& C_{13L}^2 \left(\frac{p_{1\gamma}}{p_{2\gamma} p_{3\gamma\sigma} p_{4\gamma\sigma}}\right)\cos^2\gamma\sin^2\sigma,\\
J_{CP}&=& J_{13L} \sin2\gamma \sin\sigma 
\eeqa

where 
\beqa
 J_{13L} &=& \frac{1}{2} a_{23} a_{33}~C_{13L},\\
 C_{13L} &=& \frac{a_{21}a_{22}}{a_{23}a_{33}}(a_{11} a_{32}-a_{12} a_{31}),\\
 p_{1\gamma} &=& 1+a_{33}^4\sin^4\gamma-2 a_{33}^2\sin^2\gamma,\\
 p_{2\gamma} &=& 1-a_{23}^2 - a_{33}^2\sin^2\gamma,\\
 p_{3\gamma\sigma} &=& 1-a_{12}^2\cos^2\gamma - (a_{32}^2 +a_{33}^2)\sin^2\gamma - a_{12}a_{32}\cos\sigma \sin 2\gamma,\\
 p_{4\gamma\sigma} &=& a_{12}^2\cos^2\gamma + a_{32}^2\sin^2\gamma + a_{12}a_{32}\cos\sigma \sin 2\gamma 
\eeqa

\begin{figure}[!t]\centering
\begin{tabular}{c c} 
\hspace{-5mm}
\includegraphics[angle=0,width=80mm]{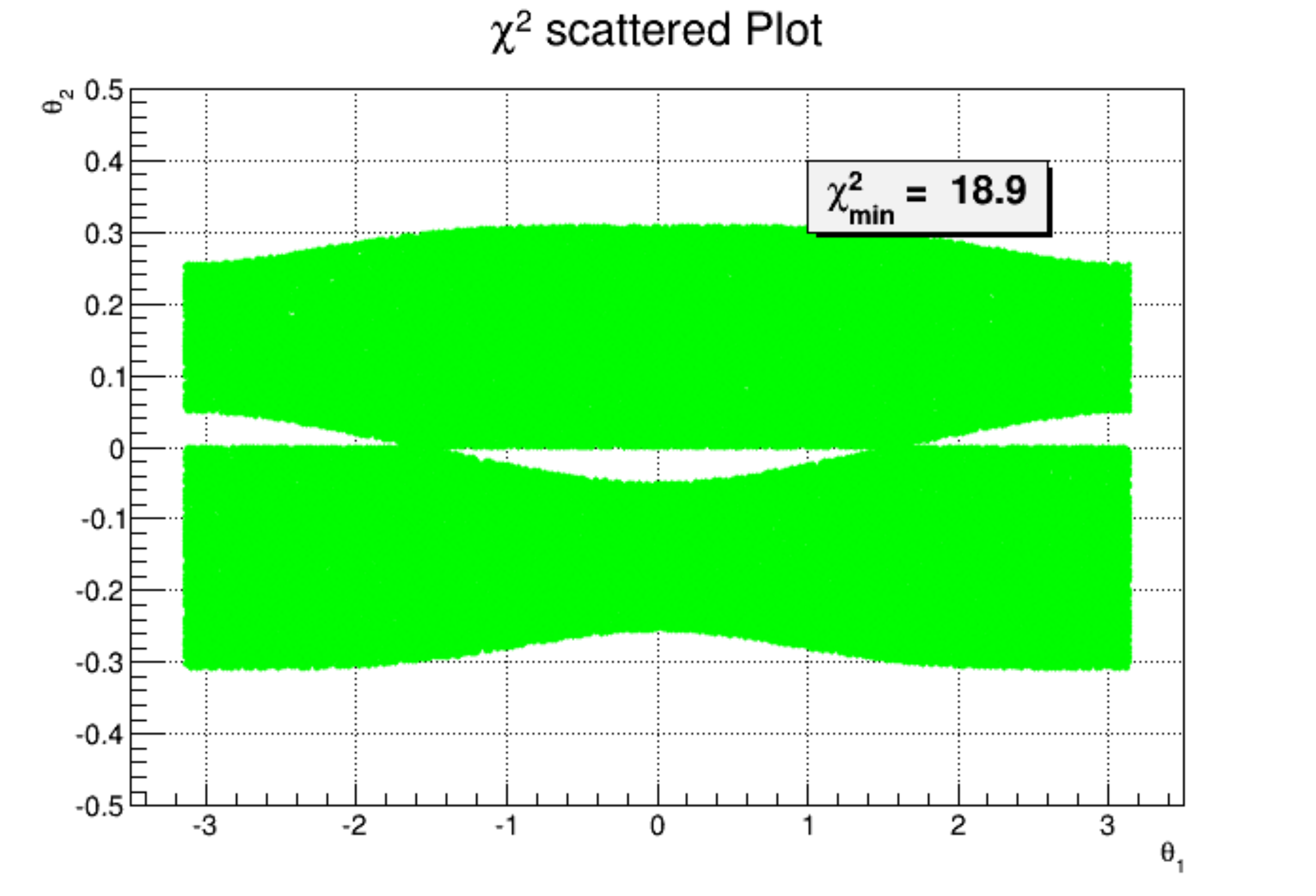} &
\includegraphics[angle=0,width=80mm]{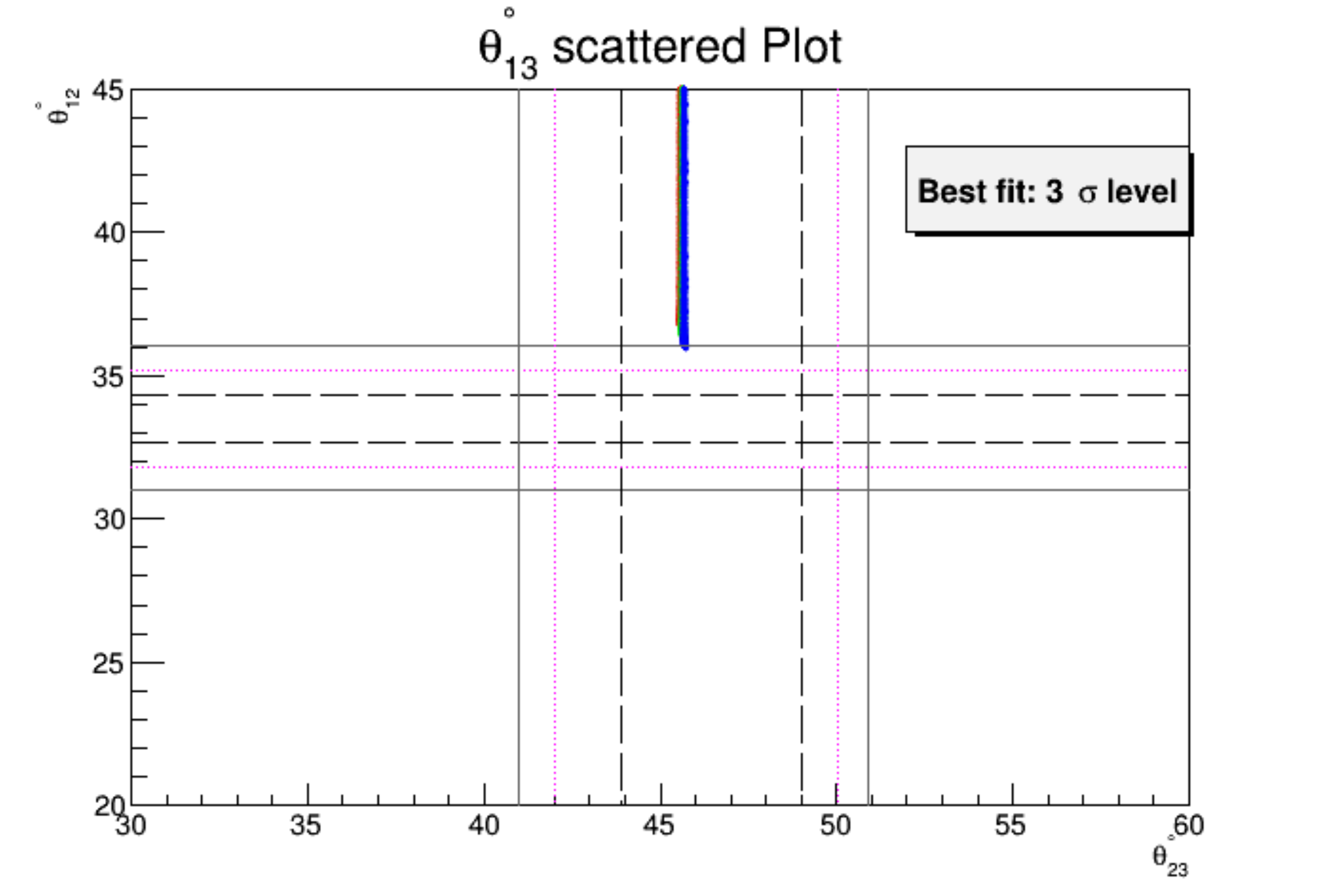}\\
\end{tabular}
\caption{\it{Scattered plot of $\chi^2$ (left fig.) over $\gamma-\sigma$ plane and $\theta_{13}$ (right fig.) 
over $\theta_{23}-\theta_{12}$ (in degrees) plane for $U^{BML}_{13}$ rotation scheme.}}
\label{fig13L1}
\end{figure}

\begin{figure}[!t]\centering
\begin{tabular}{c c} 
\hspace{-5mm}
\includegraphics[angle=0,width=80mm]{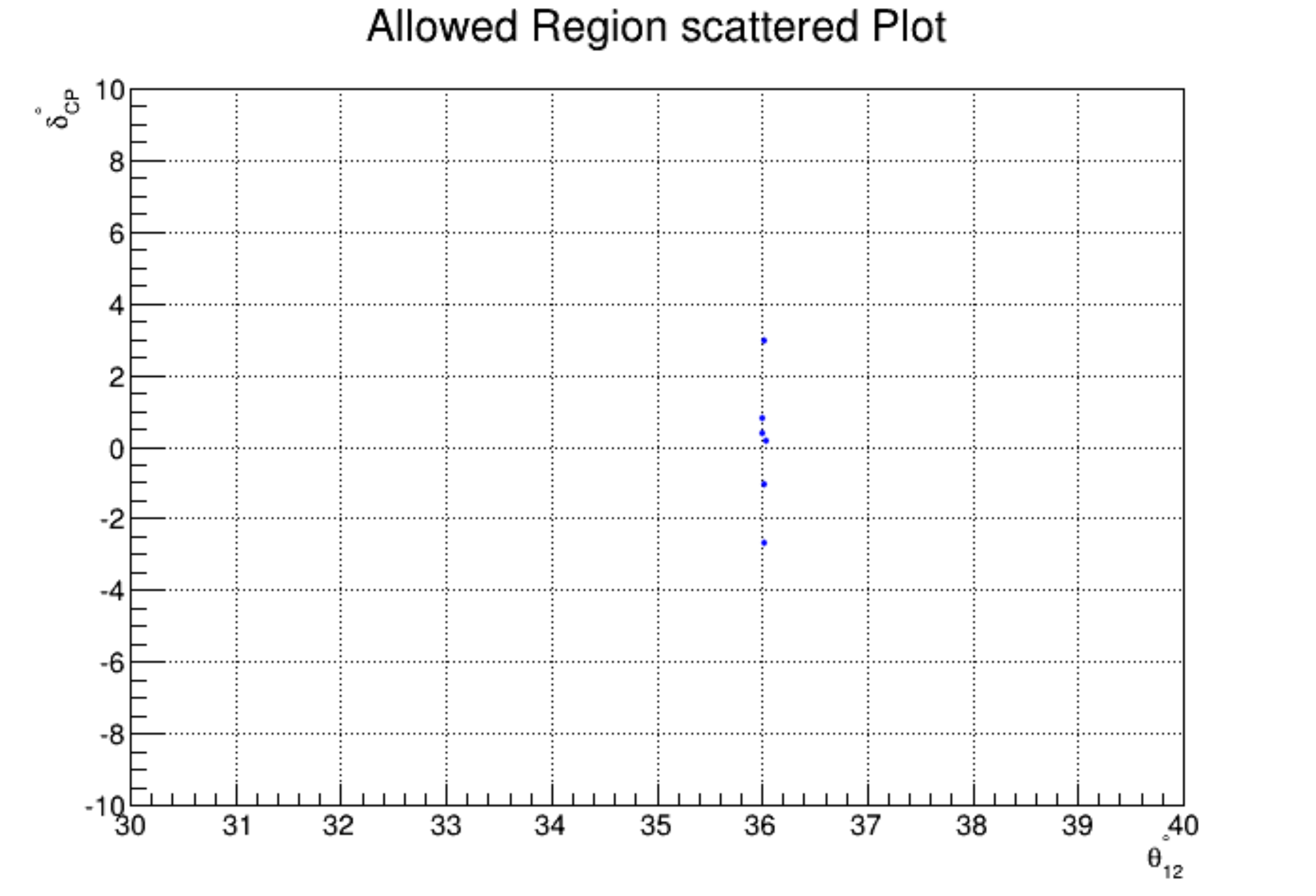} &
\includegraphics[angle=0,width=80mm]{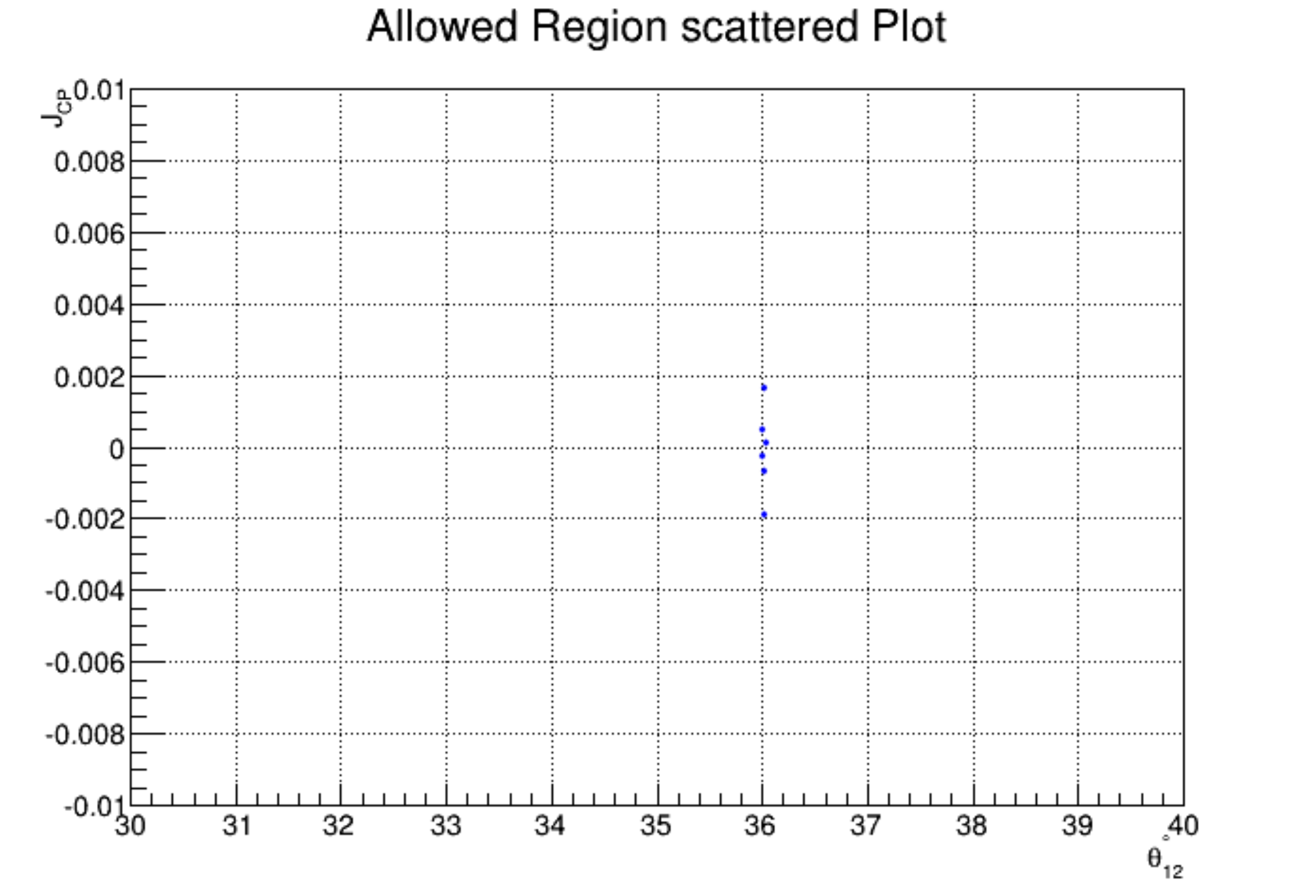}\\
\end{tabular}
\caption{\it{Scattered plot of $\delta_{CP}$ (left fig.) vs $\theta_{12}$ (in degrees) and scattered plot of $J_{CP}$ (right fig.) 
over $\theta_{12}$ (in degrees) plane for $U^{BML}_{13}$ rotation scheme.}}
\label{fig13L2}
\end{figure}

\begin{figure}[!t]\centering
\begin{tabular}{c c} 
\hspace{-5mm}
\includegraphics[angle=0,width=80mm]{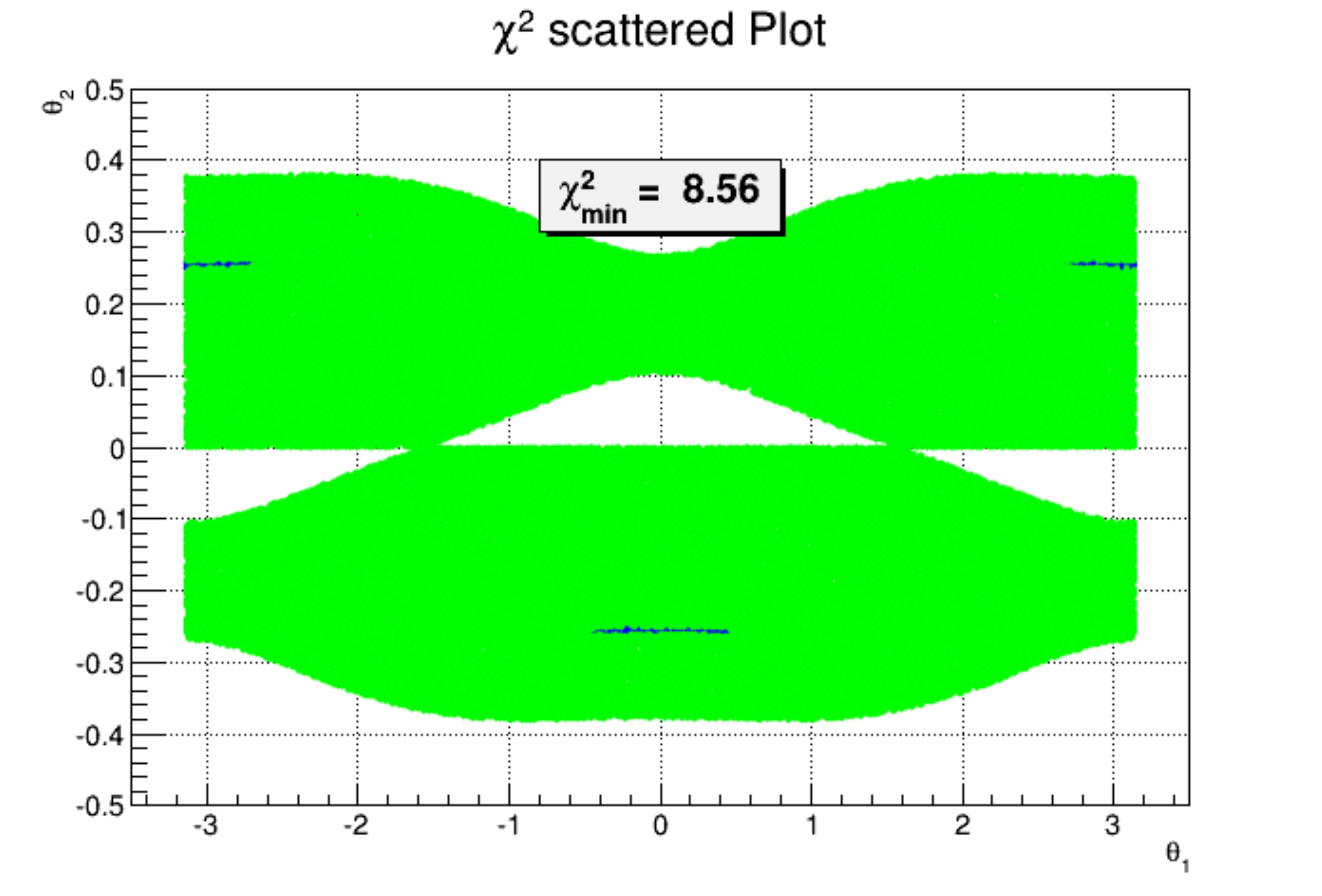} &
\includegraphics[angle=0,width=80mm]{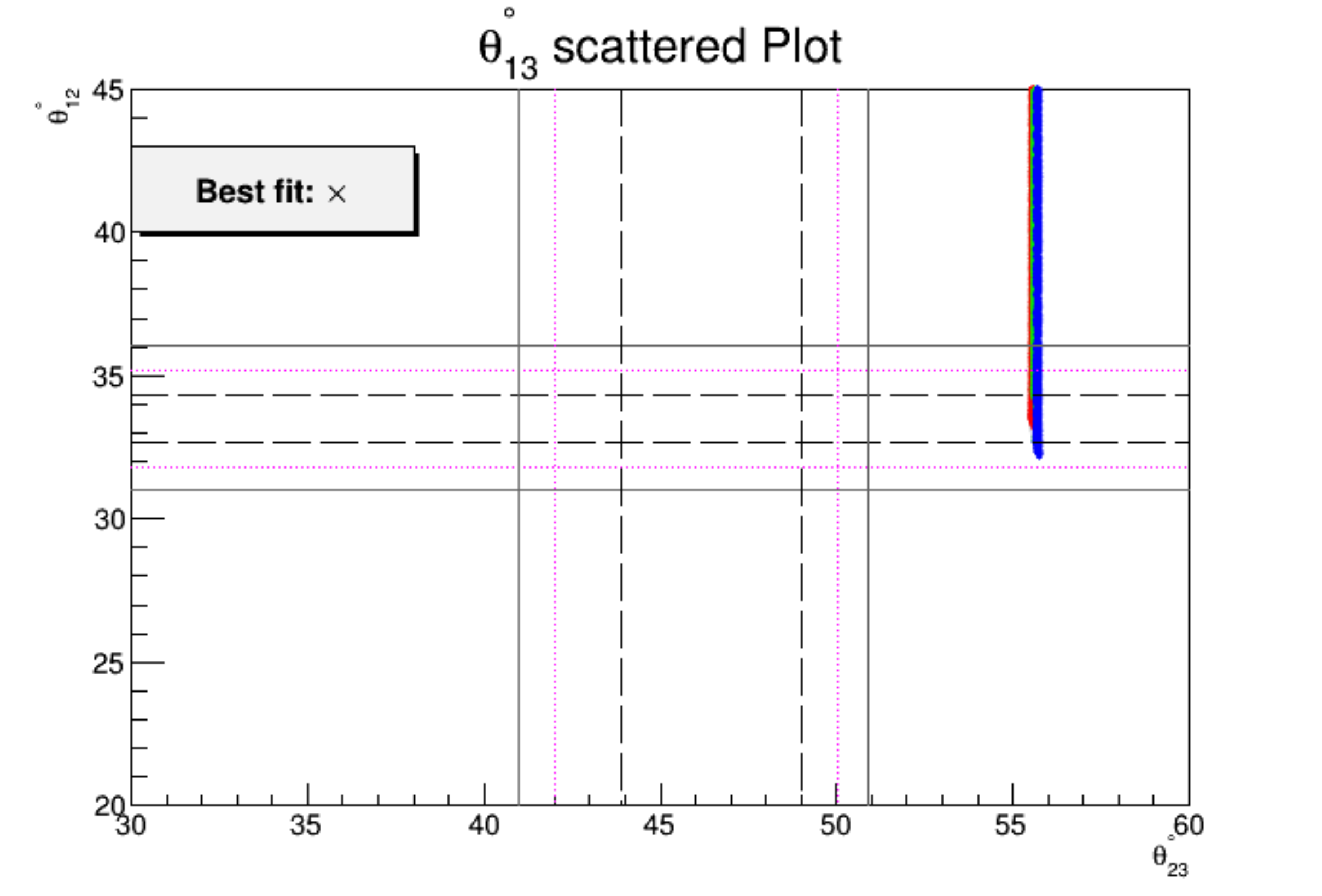}\\
\end{tabular}
\caption{\it{Scattered plot of $\chi^2$ (left fig.) over $\gamma-\sigma$ plane and $\theta_{13}$ (right fig.) 
over $\theta_{23}-\theta_{12}$ (in degrees) plane for $U^{DCL}_{13}$ rotation scheme.}}
\label{fig13L3}
\end{figure}

\begin{figure}[!t]\centering
\begin{tabular}{c c} 
\hspace{-5mm}
\includegraphics[angle=0,width=80mm]{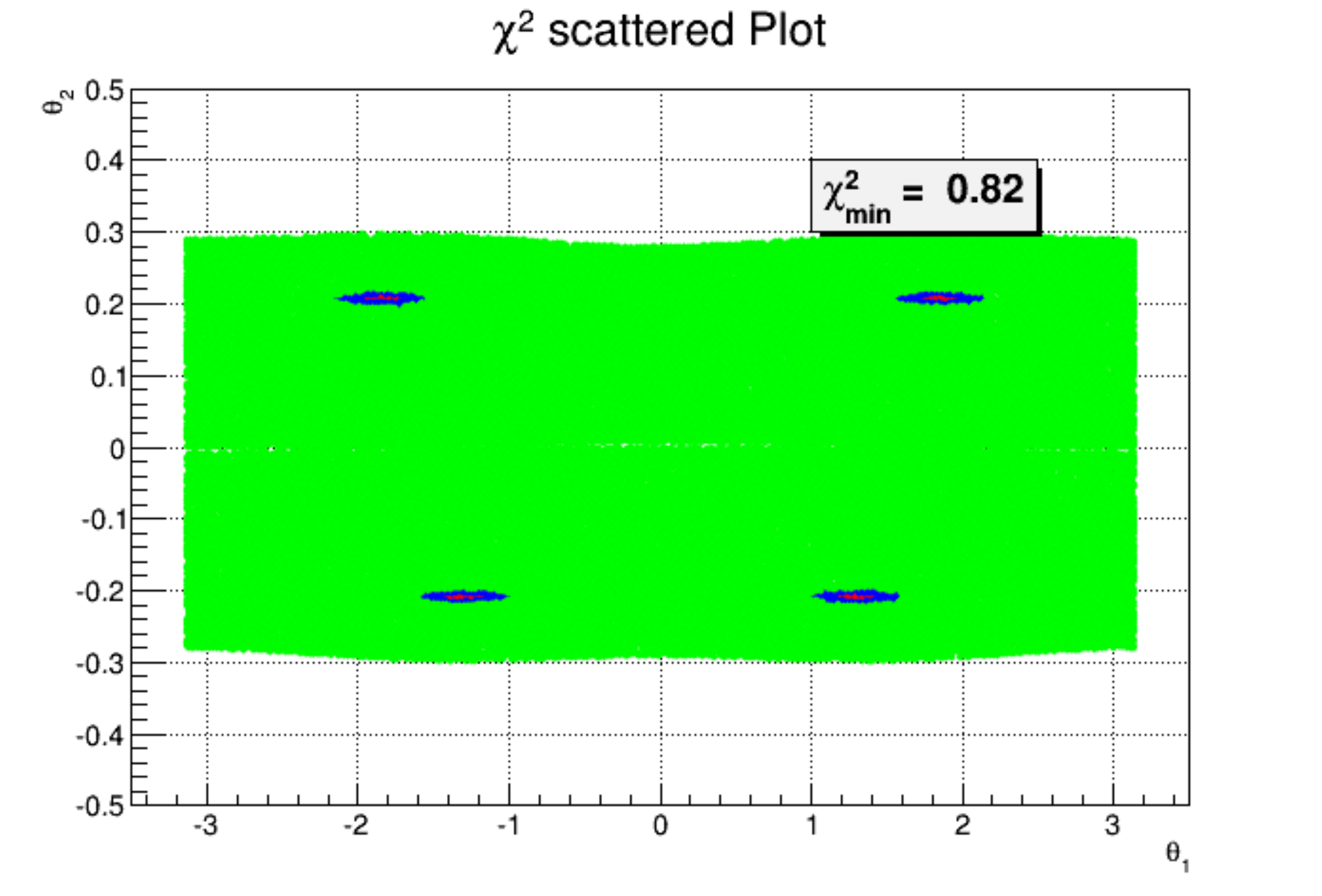} &
\includegraphics[angle=0,width=80mm]{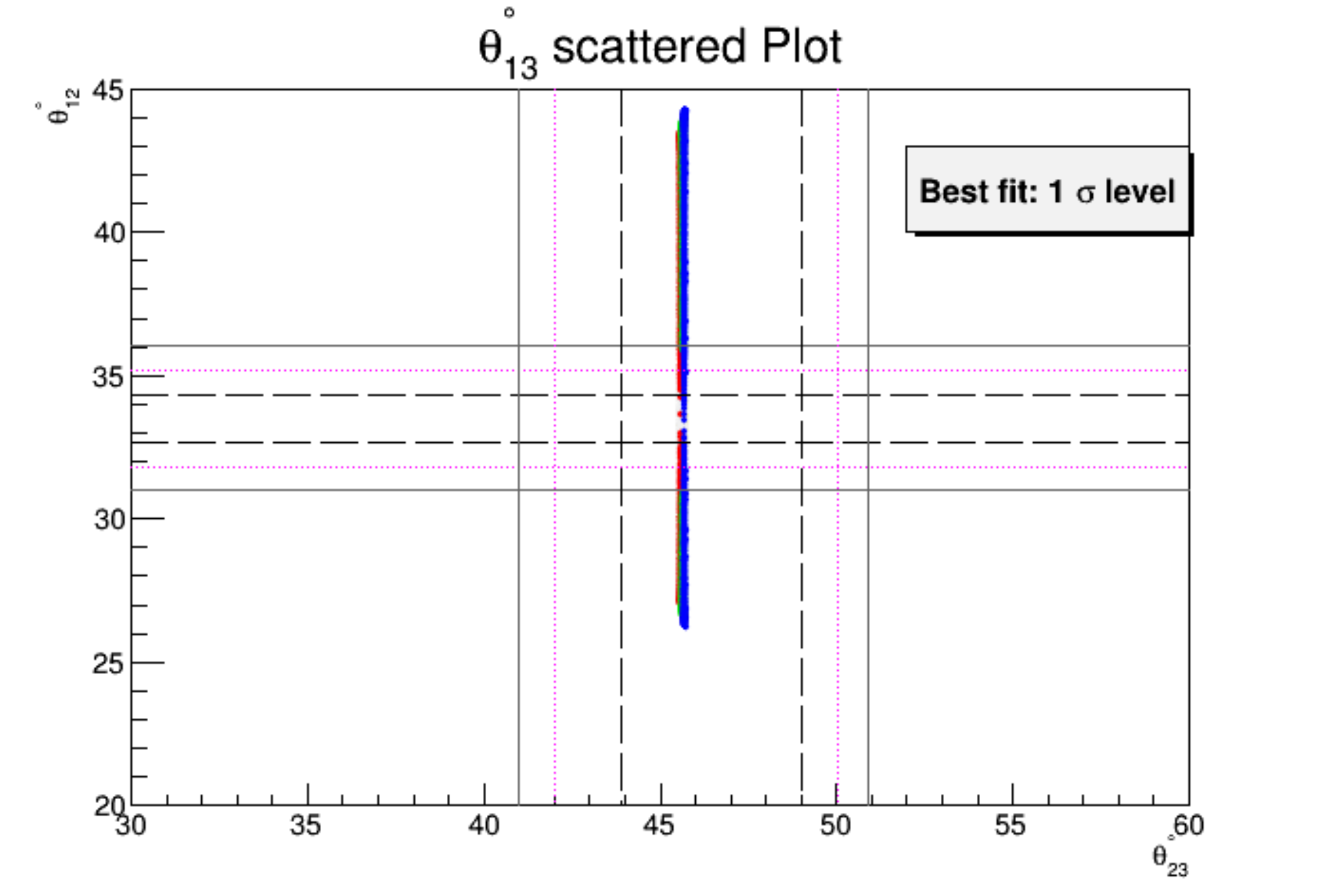}\\
\end{tabular}
\caption{\it{Scattered plot of $\chi^2$ (left fig.) over $\gamma-\sigma$ plane and $\theta_{13}$ (right fig.) 
over $\theta_{23}-\theta_{12}$ (in degrees) plane for $U^{TBML}_{13}$ rotation scheme. }}
\label{fig13L6}
\end{figure}

\begin{figure}[!t]\centering
\begin{tabular}{c c} 
\hspace{-5mm}
\includegraphics[angle=0,width=80mm]{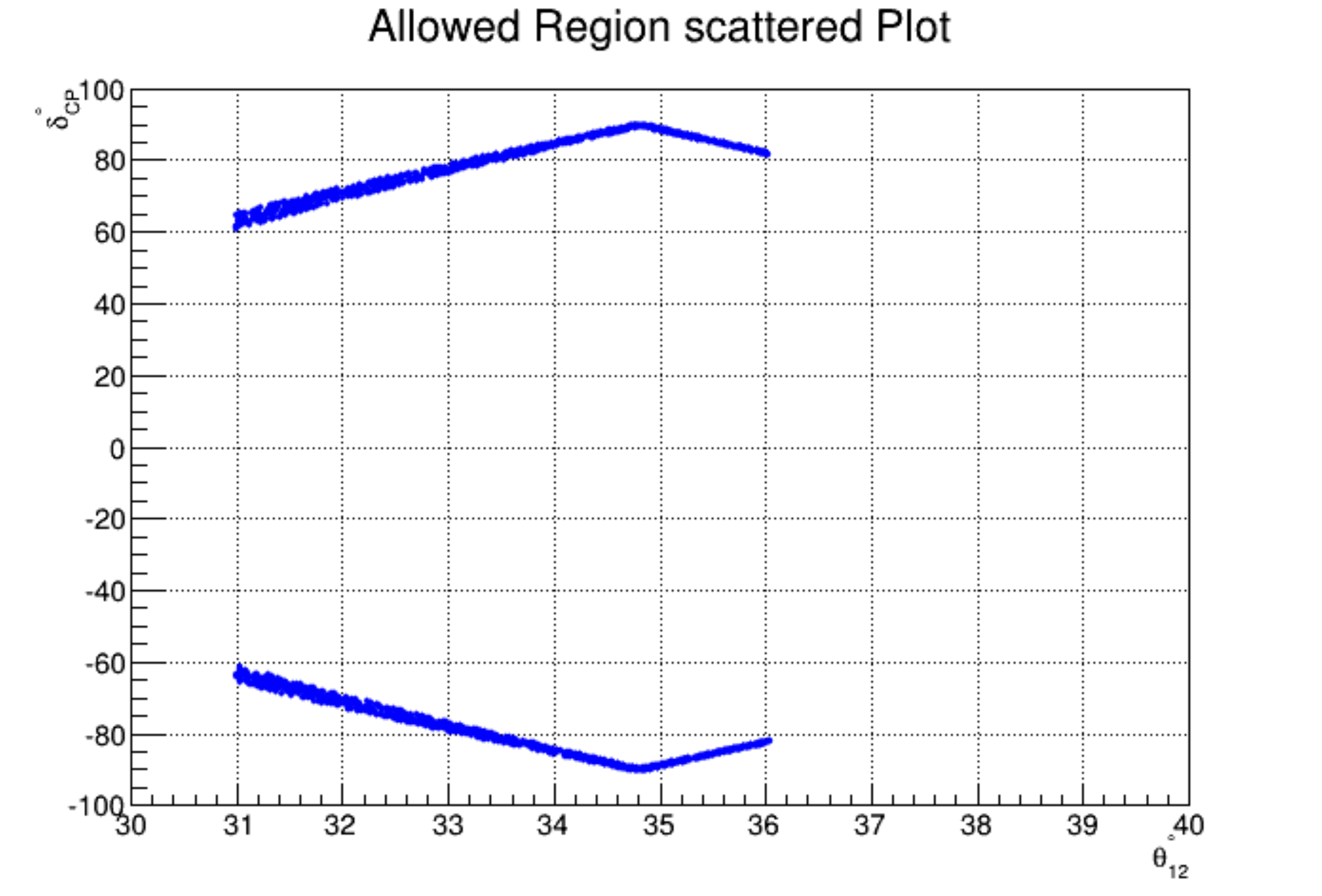} &
\includegraphics[angle=0,width=80mm]{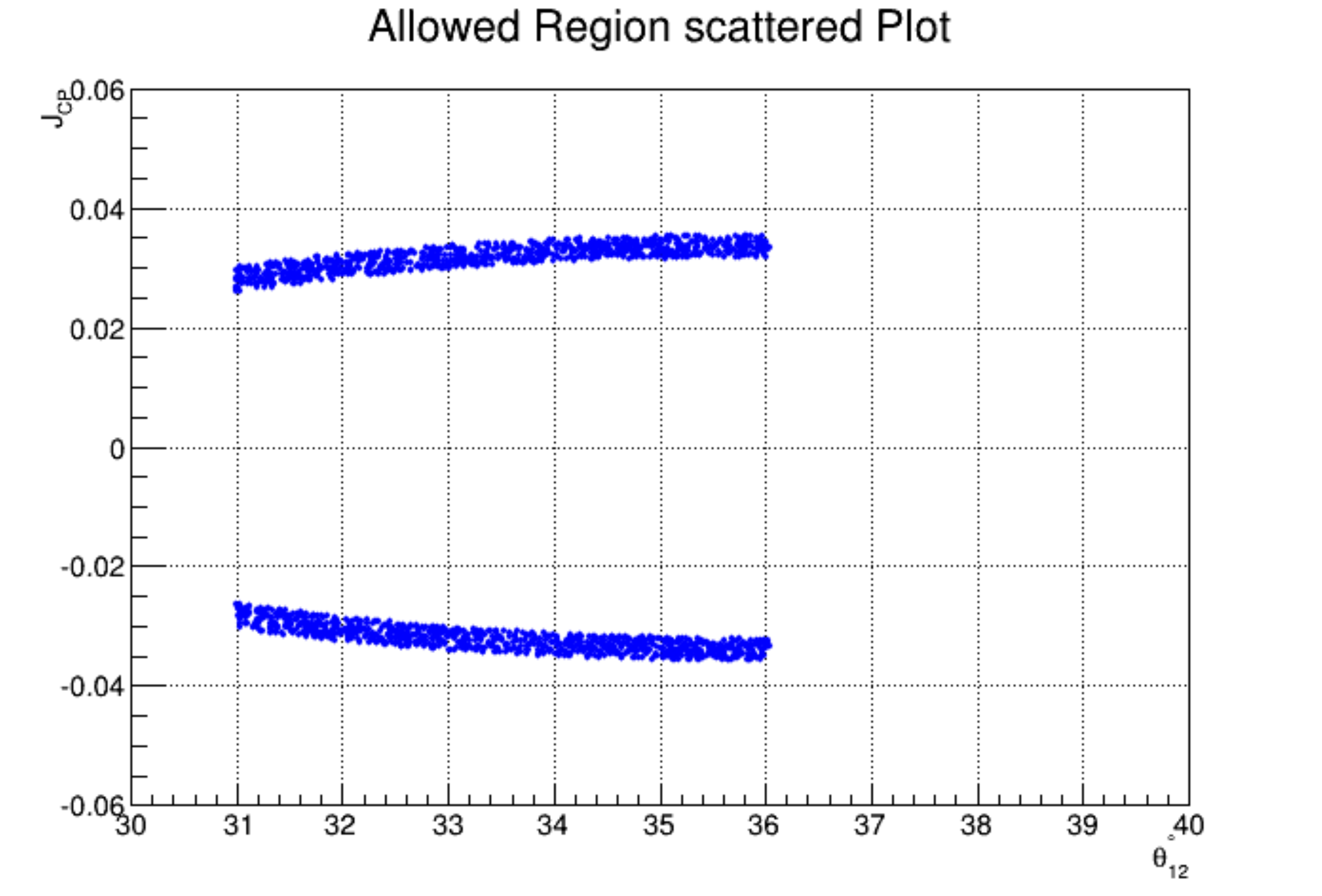}\\
\end{tabular}
\caption{\it{Scattered plot of $\delta_{CP}$ (left fig.) vs $\theta_{12}$ (in degrees) and scattered plot of $J_{CP}$ (right fig.) 
over $\theta_{12}$ (in degrees) plane for $U^{TBML}_{13}$ rotation scheme.}}
\label{fig13L7}
\end{figure}

Fig.~\ref{fig13L1}-\ref{fig13L7} show the numerical results corresponding to our considered HG mixing. The main features of 
this perturbative scheme are:\\
{\bf{(i)}} Like previous case, $\theta_{23}$ remains very close to its unperturbed value as corrections to this angle enters through $\theta_{13}$.
Thus DC case is disfavored completely. \\
{\bf{(ii)}} The numerical fitting of $\theta_{13}$ and $\theta_{23}$ is only governed by $\alpha$
so its allowed range is much restricted in parameter space. e.g. for TBM case, the fitting of $\theta_{13}$ under its $3\sigma$ domain 
constraints the magnitude of correction parameter 
$|\alpha| \in [0.1962(0.1988), 0.2204(0.2223)]$ which in turn fixes $\theta_{23} \in [45.55^\circ(45.56^\circ), 45.70^\circ(45.71^\circ)]$ for 
corresponding $\gamma$ values with NH(IH) case. 
However $\theta_{12}$ possess much wider range of values since it receives corrections from $\alpha$ as well from phase parameter $\sigma$. \\
{\bf{(iii)}} The minimum value of $\chi^2 \sim 18.9(23.4),~8.56(36.8),~0.82(5.0)$ and $0.81(5.0)$ for BM, DC, TBM and HG respectively with 
NH(IH) case. Here also BM barely manages to fit all mixing angles in $3\sigma$ level for a very minute region of parameter space while DC is not viable.  
However HG and TBM are much favorable as they are able to fit all mixing angles within $1\sigma$ level for NH. However for IH, $\theta_{23}$ stays
outside its $1\sigma$ range so it is only allowed at $2\sigma$ level.\\
{\bf{(iv)}} Leptonic phase $\delta_{CP}$ lies in the range $-3.6(-4.9) \leq \delta_{CP} \leq 4.2(5.5)$ for BM while it is confined
in $39.0(40.4) \leq |\delta_{CP}| \leq 78.7(79.2)$ for modified HG  and $61.0(60.9) \leq |\delta_{CP}| \leq 89.9(89.9)$  corrected TBM matrix. \\
{\bf{(v)}} The Jarkslog invariant($J_{CP}$) remains in range $-0.0025(-0.0034) \leq J_{CP} \leq 0.0026(0.0030)$ 
for BM while it is confined in $0.020(0.021) \leq |J_{CP}| \leq 0.032(0.032)$ for corrected HG and $0.026 \leq |J_{CP}| \leq 0.035$ TBM mixing matrix.

\begin{center}
\begin{tabular}{ |p{1.3cm}||p{2.0cm}|p{2.0cm}|p{2.0cm}|p{2.0cm}|p{2.0cm}|p{2.3cm}|  }
 \hline
 \multicolumn{7}{|c|}{Best Fit with Mixing data} \\
 \hline
Rotation  & $\chi^2_{min}$ & $\theta_{12}^\circ$ & $\theta_{23}^\circ$ & $\theta_{13}^\circ$ & $|\delta_{CP}^\circ|$ & $|J_{CP}|$\\
 \hline
 BM   & $18.9(23.4)$    & $36.03(35.96)$ &   $45.6(45.7)$ & $8.86(8.92)$         &$0.594(0.302)$ & $0.0003(0.0001)$\\
  \hline
  DC   & $8.56(36.8)$    & $33.42(33.35)$ &   ${\bf{55.6(55.6)}}$ & $8.41(8.46)$ &$17.51(17.62)$ & $0.009(0.009)$\\
  \hline
  TBM   & $0.82(5.0)$    & $33.37(33.41)$ &   $45.6(45.6)$ & $8.42(8.49)$        &$80.55(81.0)$ & $0.032(0.032)$\\
  \hline
  HG  & $0.81(5.0)$    & $33.51(33.39)$ &   $45.6(45.6)$ & $8.41(8.48)$          &$61.89(62.95)$ & $0.029(0.029)$\\
  \hline
\end{tabular}\captionof{table}{\it{Neutrino Mixing angles, $|\delta^\circ_{CP}|$ and $|J_{CP}|$ corresponding to $\chi^2_{min}$ numerical fit.
The mixing angle value that lies outside its best fit $3\sigma$ range is marked in boldface.}}
\label{chisqunperturb} 
\end{center}

\begin{figure}[!t]\centering
\begin{tabular}{c c} 
\hspace{-5mm}
\includegraphics[angle=0,width=80mm]{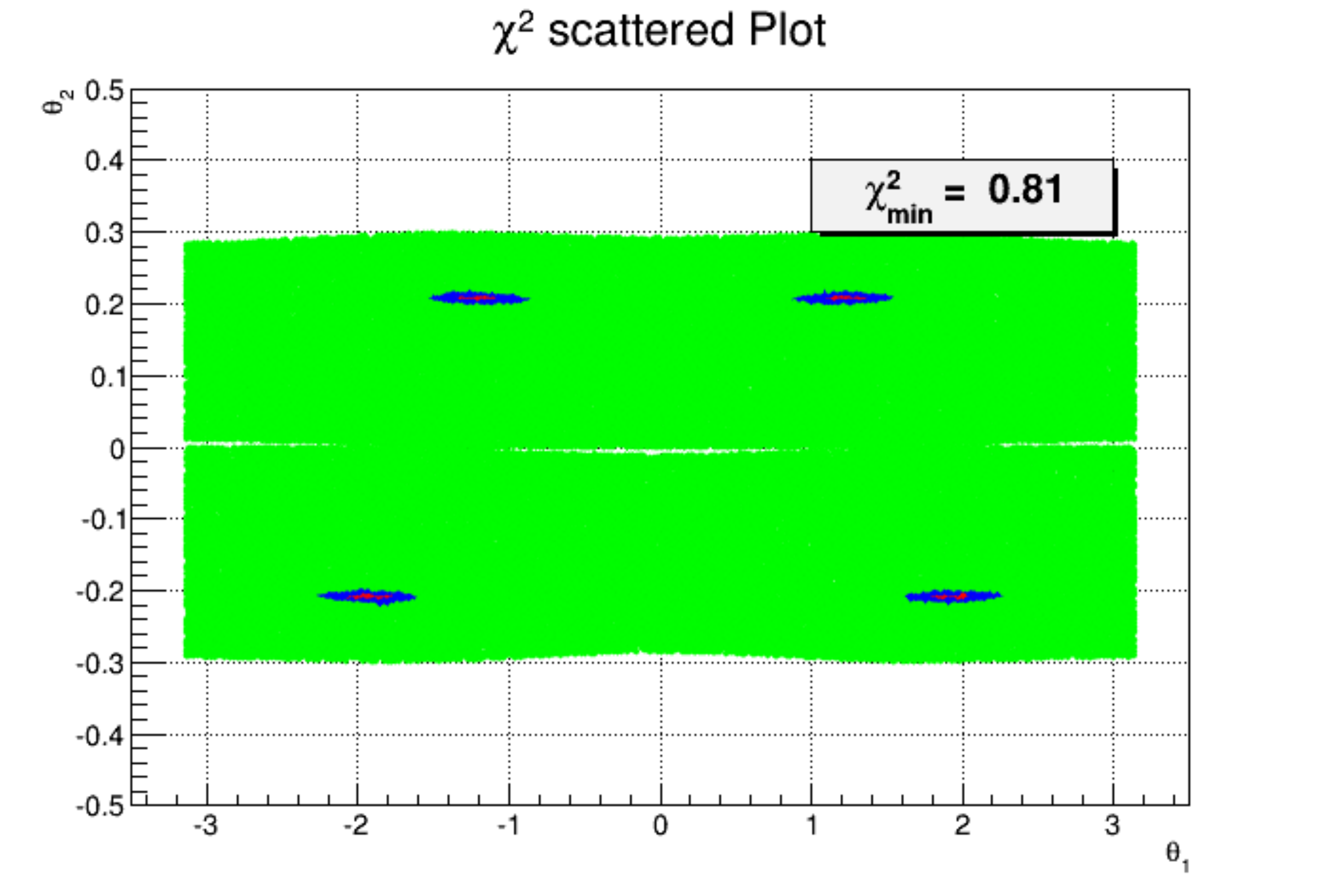} &
\includegraphics[angle=0,width=80mm]{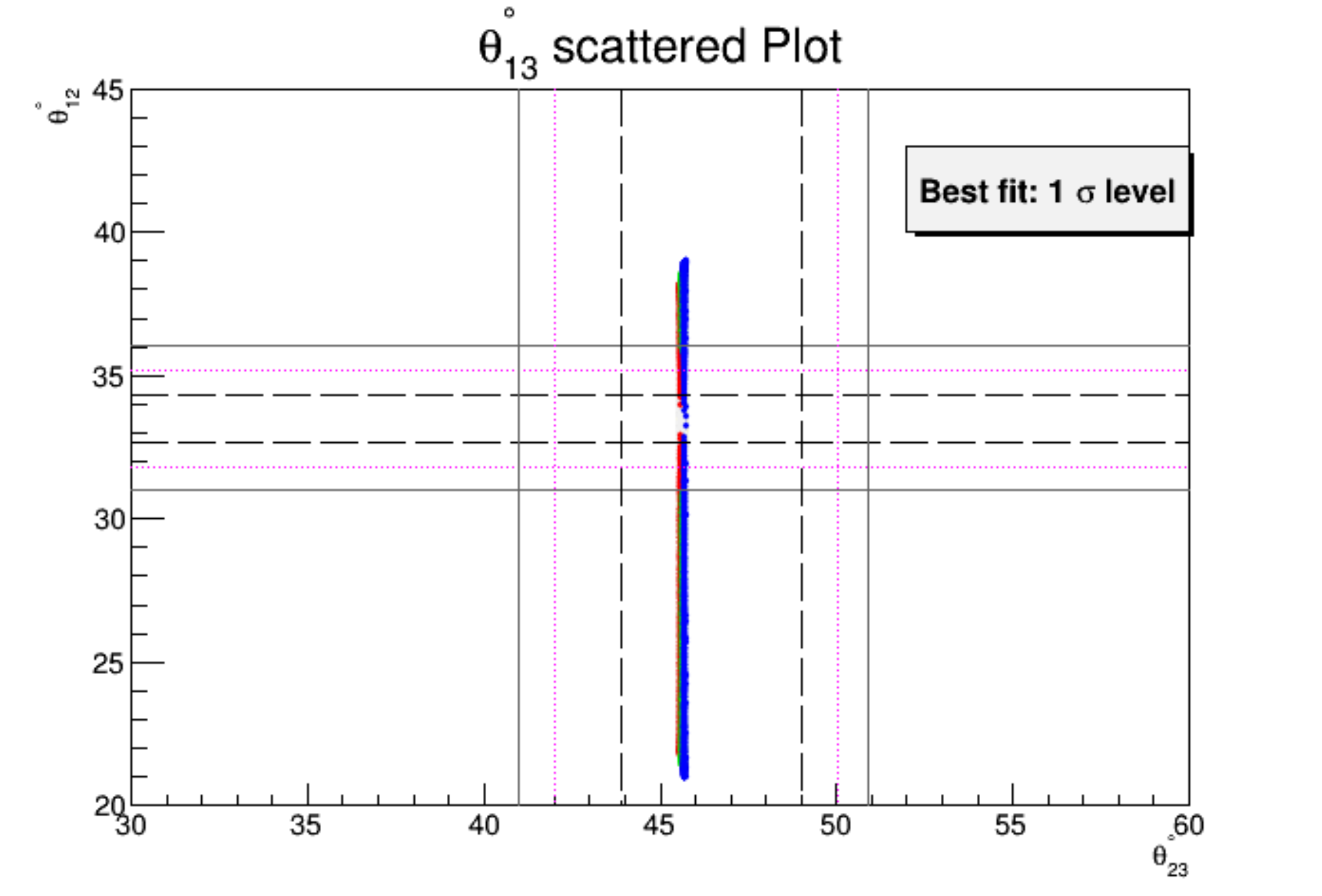}\\
\end{tabular}
\caption{\it{Scattered plot of $\chi^2$ (left fig.) over $\gamma-\sigma$ plane and $\theta_{13}$ (right fig.) 
over $\theta_{23}-\theta_{12}$ (in degrees) plane for $U^{HGL}_{13}$ rotation scheme. }}
\label{fig13L4}
\end{figure}

\begin{figure}[!t]\centering
\begin{tabular}{c c} 
\hspace{-5mm}
\includegraphics[angle=0,width=80mm]{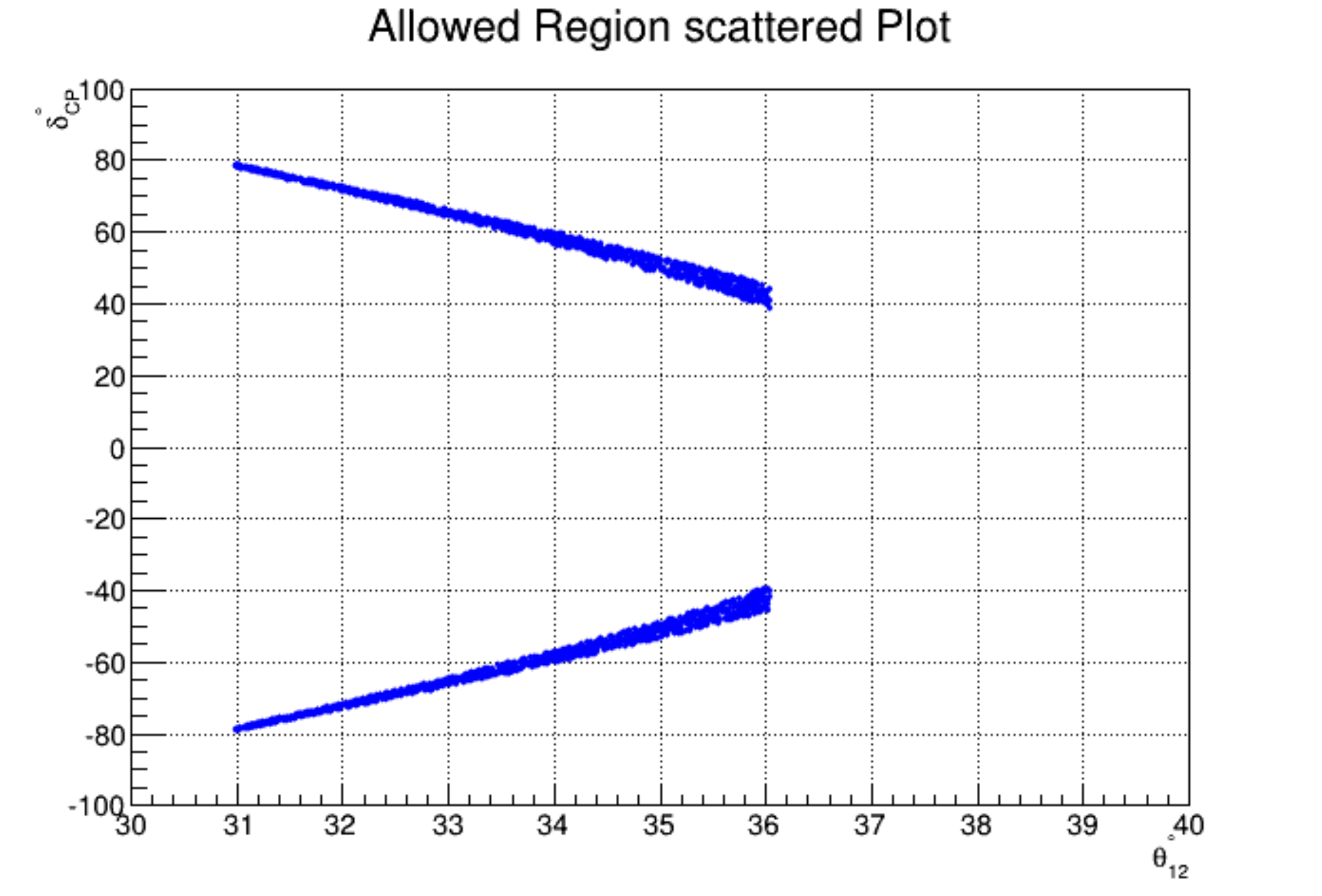} &
\includegraphics[angle=0,width=80mm]{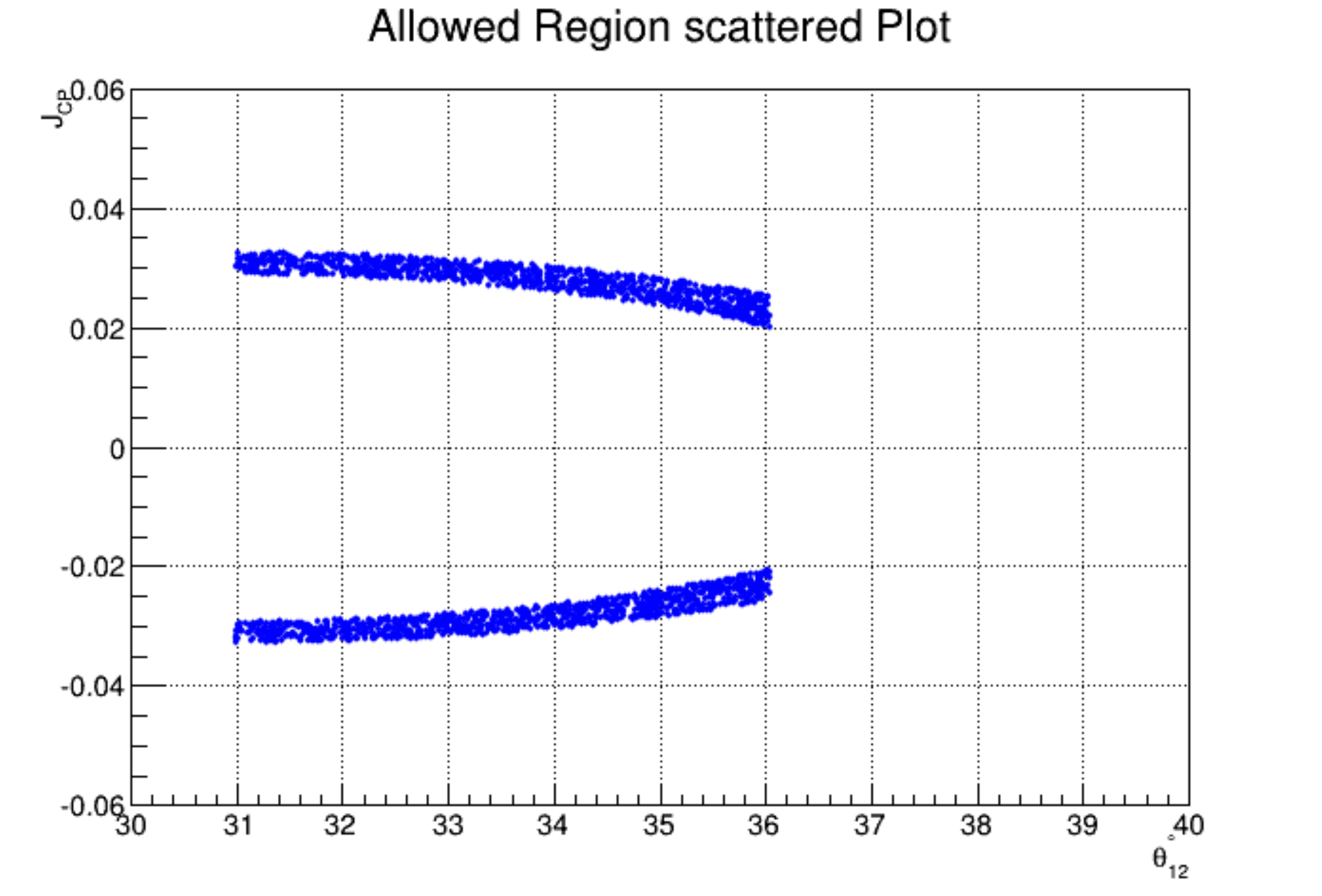}\\
\end{tabular}
\caption{\it{Scattered plot of $\delta_{CP}$ (left fig.) vs $\theta_{12}$ (in degrees) and scattered plot of $J_{CP}$ (right fig.) 
over $\theta_{12}$ (in degrees) plane for $U^{HGL}_{13}$ rotation scheme.}}
\label{fig13L5}
\end{figure}

\subsection{23 Rotation}

Here rotation matrix imparts corrections in last two rows of unperturbed matrix. Thus reactor mixing
angle, $\theta_{13}$ doesn't receive any corrections in this scheme. Thus we left this case without going for any
further discussion.

\section{Rotations-$V_{M}.U_{ij}^r$}

Here we take up the modifications for which PMNS matrix is given by $U_{PMNS} = V_{M}.U_{ij}^r$. This
scheme will introduce changes in $i^{\text{th}}$ and $j^{\text{th}}$ column of unperturbed mixing matrix. We will 
investigate the role of these perturbations in fitting the neutrino mixing angles and its prediction for Dirac
CP Phase($\delta_{CP}$). 

\subsection{12 Rotation}

In this case, rotation matrix imparts corrections in first two columns of unperturbed matrix. Thus reactor mixing
angle, $\theta_{13}$ doesn't get any modifications in this scheme. Hence this case is not of significance and we
left it for any further discussion.

\subsection{13 Rotation}

This case corresponds to rotation in 13 sector of these special matrices. 
The expressions for mixing angles in this case are given as


\beqa
 \sin^2\theta_{13} &=&  a_{11}^2 \sin^2\gamma,\\
 \sin^2\theta_{12} &=& \frac{a_{12}^2}{\cos^2\theta_{13}},\\
  \sin^2\theta_{23} &=& \frac{a_{23}^2\cos^2\gamma + a_{21}^2\sin^2\gamma + a_{21}a_{23}\sin 2\gamma \cos\sigma}{\cos^2\theta_{13}}
\eeqa

The Jarsklog invariant and CP Dirac Phase is given by expressions 

\beqa
\sin^2\delta_{CP} &=& C_{13R}^2 \left(\frac{p_{1\gamma}}{p_{2\gamma} p_{3\gamma\sigma} p_{4\gamma\sigma}}\right)\cos^2\gamma\sin^2\sigma,\\
J_{CP} &=& J_{13R} \sin2\gamma \sin\sigma
 \eeqa

where 

\beqa
C_{13R} &=& a_{22} a_{23},\\
J_{13R} &=& \frac{1}{2} a_{11}a_{12} C_{13R} ,\\
  p_{1\gamma} &=& 1+a_{11}^4\sin^4\gamma-2 a_{11}^2\sin^2\gamma,\\
 p_{2\gamma} &=& 1-a_{12}^2 - a_{11}^2\sin^2\gamma,\\
 p_{3\gamma\sigma} &=& 1-a_{23}^2\cos^2\gamma - (a_{11}^2 +a_{21}^2)\sin^2\gamma - a_{21}a_{23}\cos\sigma \sin 2\gamma,\\
 p_{4\gamma\sigma} &=& a_{23}^2\cos^2\gamma + a_{21}^2\sin^2\gamma + a_{21}a_{23}\cos\sigma \sin 2\gamma 
\eeqa

\begin{figure}[!t]\centering
\begin{tabular}{c c} 
\hspace{-5mm}
\includegraphics[angle=0,width=80mm]{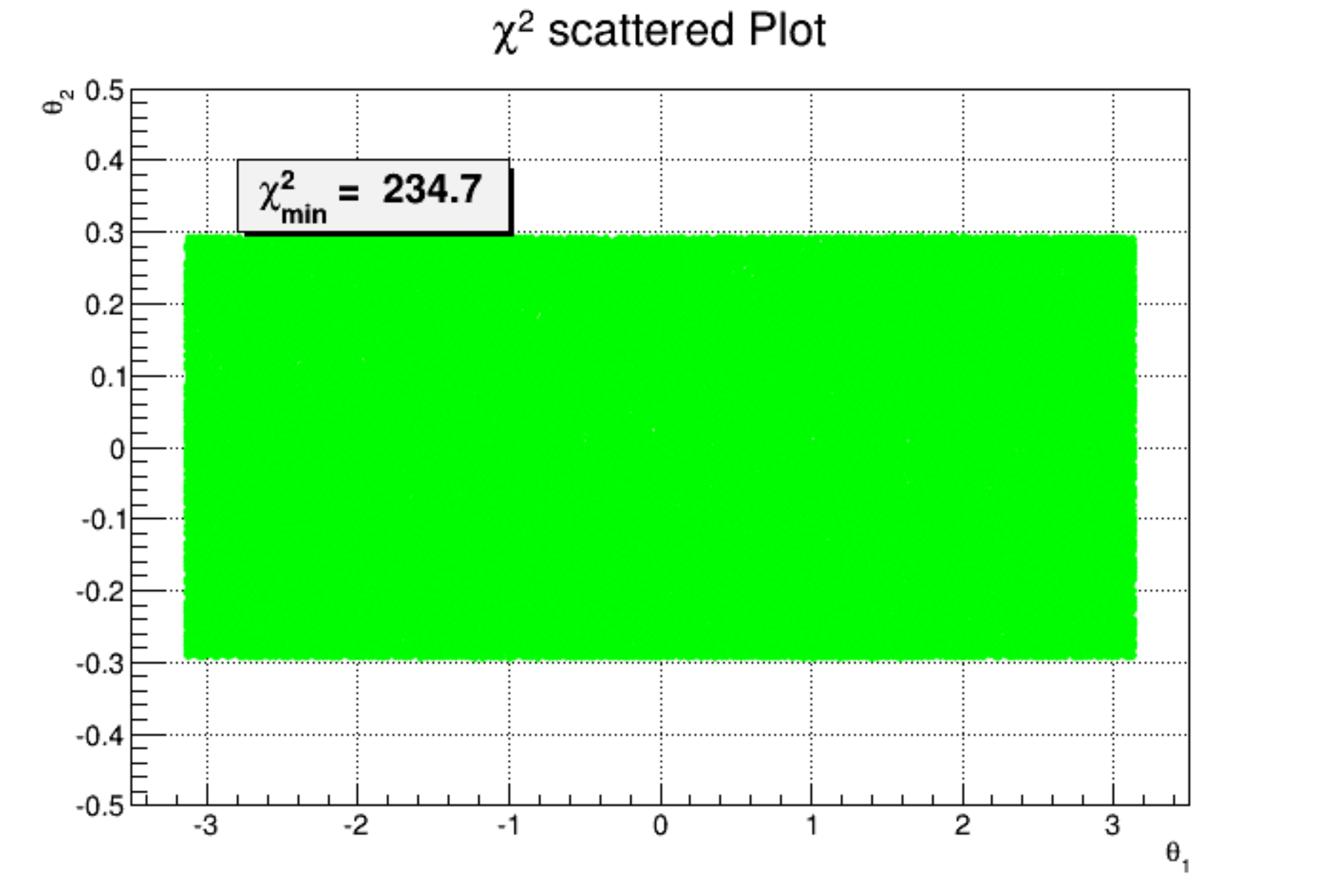} &
\includegraphics[angle=0,width=80mm]{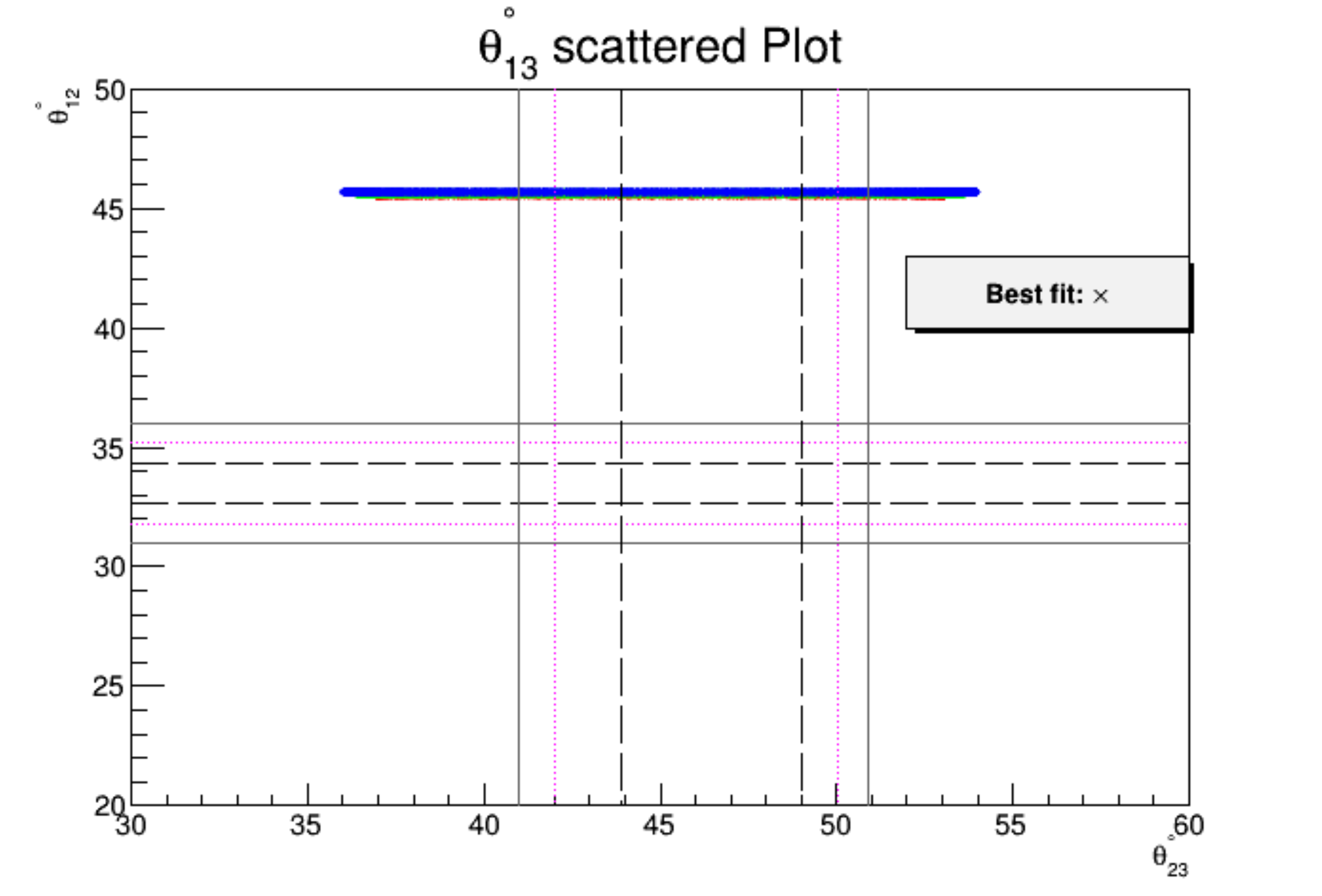}\\
\end{tabular}
\caption{\it{Scattered plot of $\chi^2$ (left fig.) over $\gamma-\sigma$ plane and $\theta_{13}$ (right fig.) 
over $\theta_{23}-\theta_{12}$ (in degrees) plane for $U^{BMR}_{13}$ rotation scheme. }}
\label{fig13R1}
\end{figure}

\begin{figure}[!t]\centering
\begin{tabular}{c c} 
\hspace{-5mm}
\includegraphics[angle=0,width=80mm]{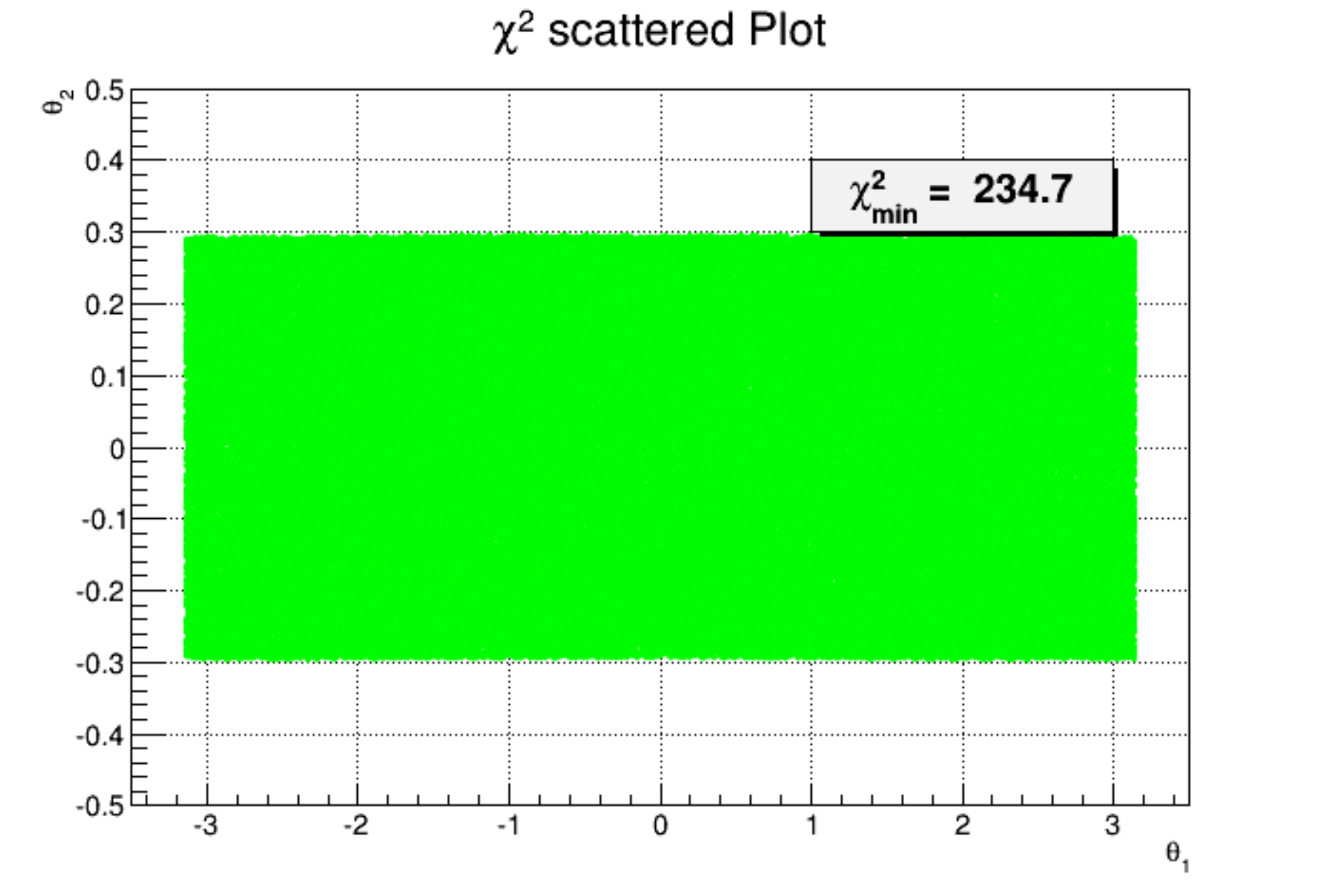} &
\includegraphics[angle=0,width=80mm]{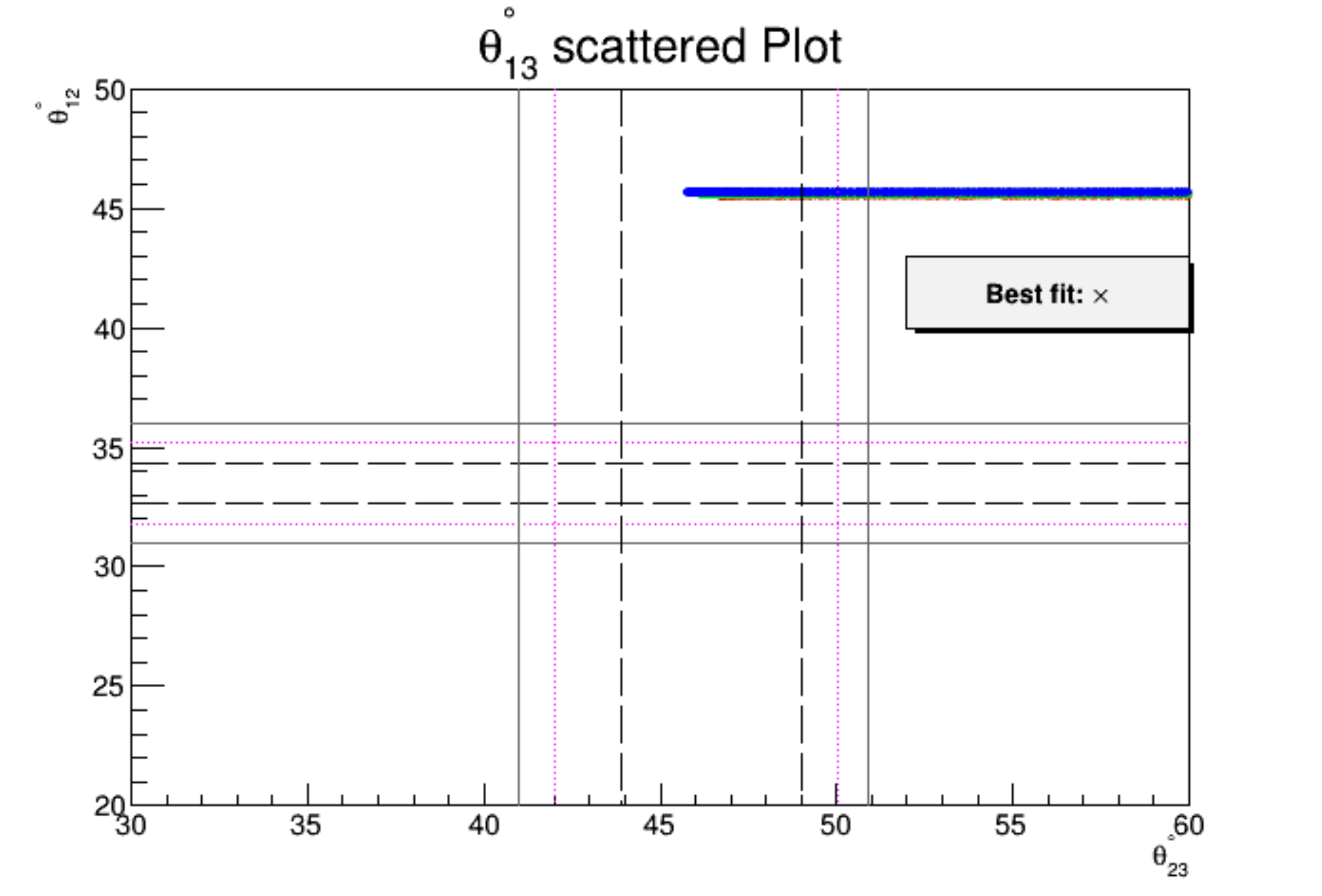}\\
\end{tabular}
\caption{\it{Scattered plot of $\chi^2$ (left fig.) over $\gamma-\sigma$ plane and $\theta_{13}$ (right fig.) 
over $\theta_{23}-\theta_{12}$ (in degrees) plane for $U^{DCR}_{13}$ rotation scheme. }}
\label{fig13R2}
\end{figure}

\begin{figure}[!t]\centering
\begin{tabular}{c c} 
\hspace{-5mm}
\includegraphics[angle=0,width=80mm]{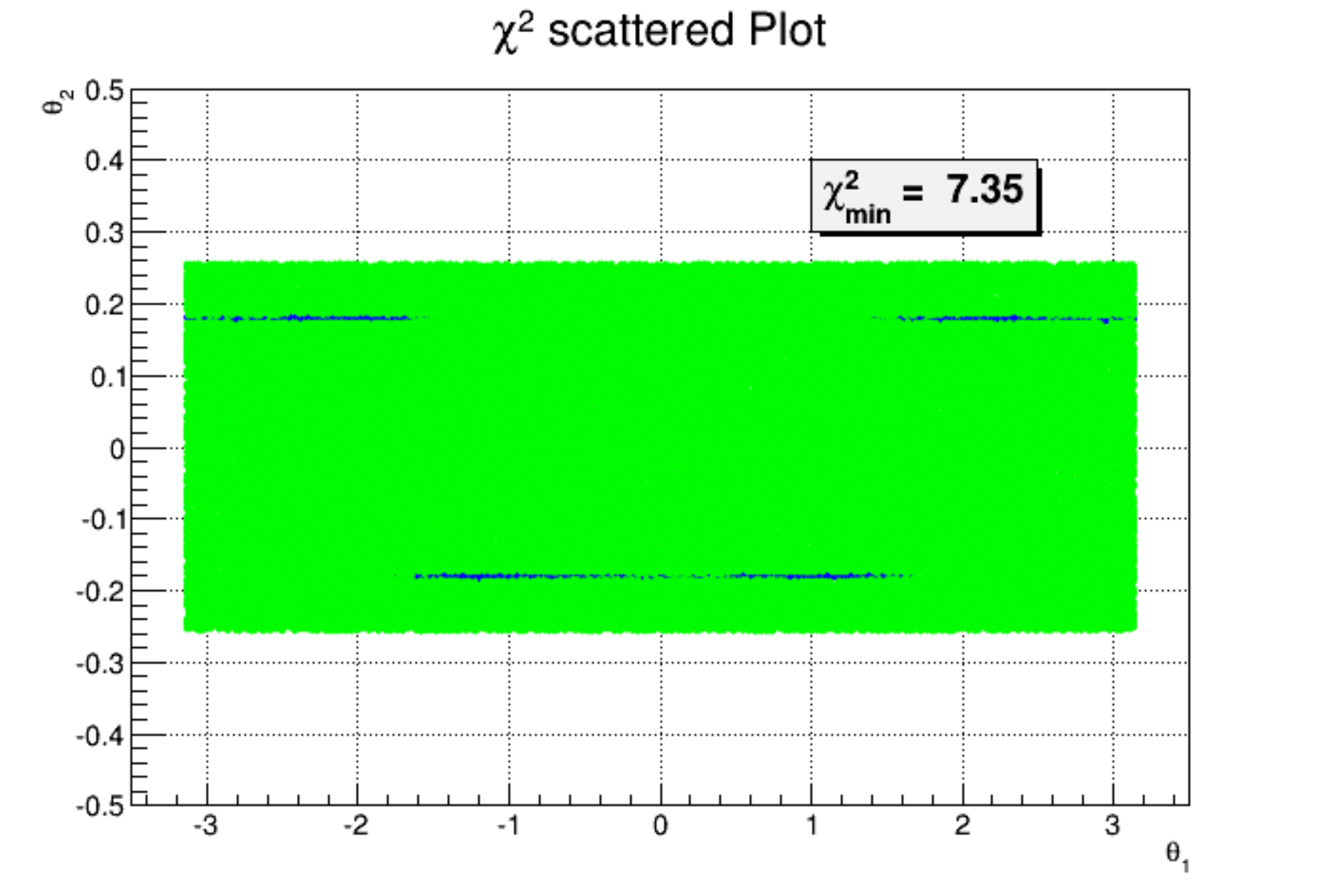} &
\includegraphics[angle=0,width=80mm]{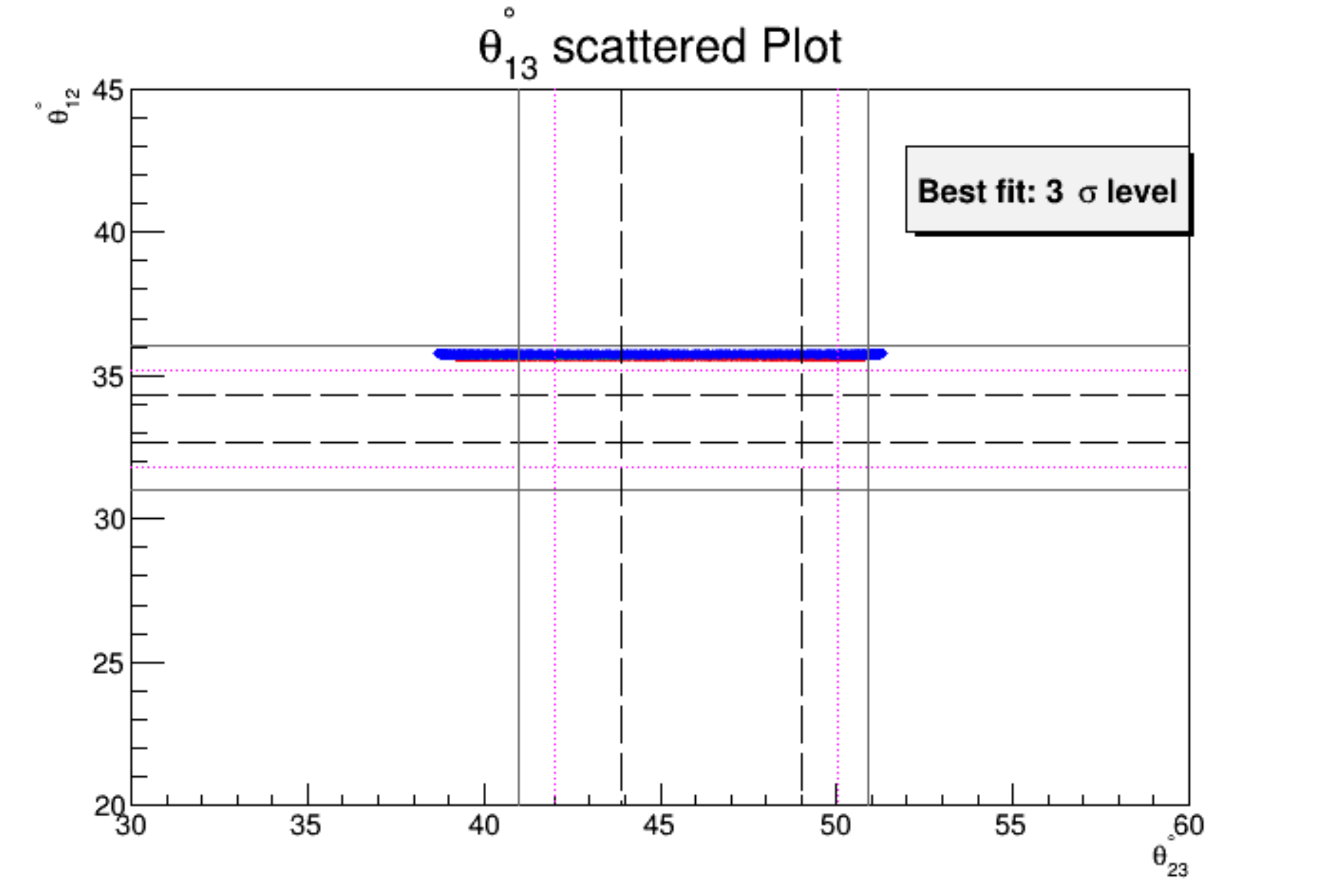}\\
\end{tabular}
\caption{\it{Scattered plot of $\chi^2$ (left fig.) over $\gamma-\sigma$ plane and $\theta_{13}$ (right fig.) 
over $\theta_{23}-\theta_{12}$ (in degrees) plane for $U^{TBMR}_{13}$ rotation scheme.}}
\label{fig13R4}
\end{figure}

\begin{figure}[!t]\centering
\begin{tabular}{c c} 
\hspace{-5mm}
\includegraphics[angle=0,width=80mm]{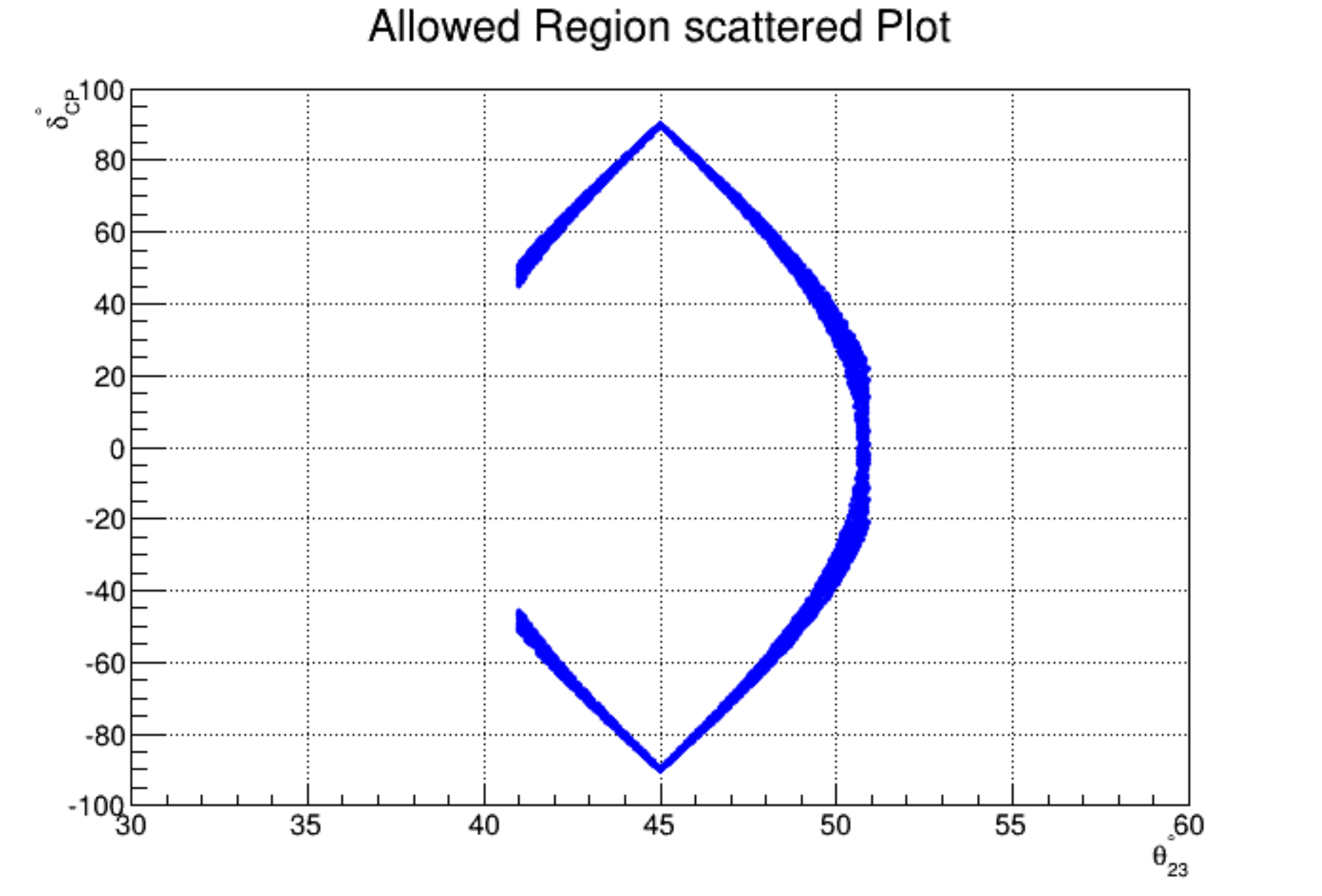} &
\includegraphics[angle=0,width=80mm]{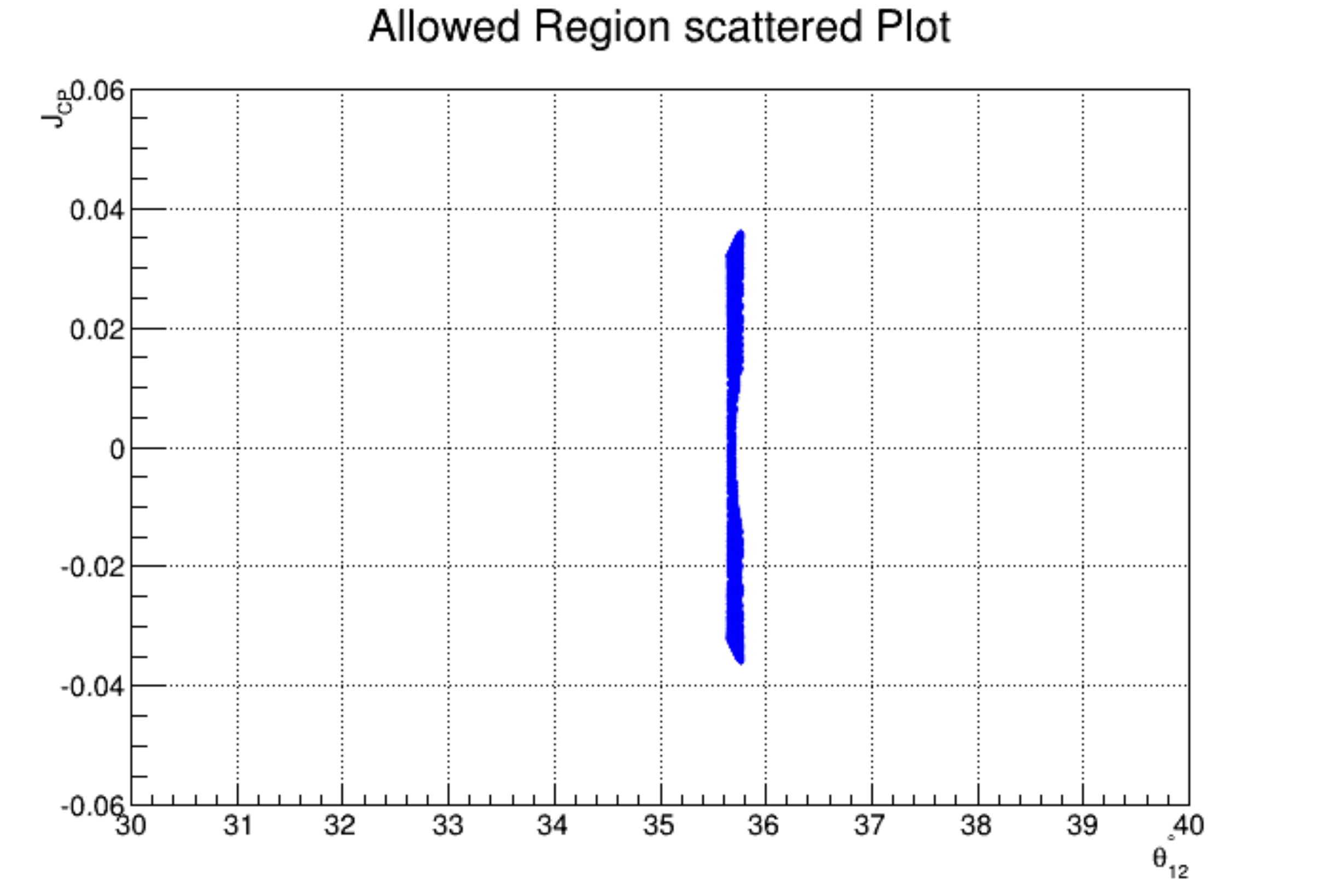}\\
\end{tabular}
\caption{\it{Scattered plot of $\delta_{CP}$ (left fig.) vs $\theta_{23}$ (in degrees) and scattered plot of $J_{CP}$ (right fig.) 
over $\theta_{12}$ (in degrees) plane for $U^{TBMR}_{13}$ rotation scheme.}}
\label{fig13R5}
\end{figure}

Fig.~\ref{fig13R1}-\ref{fig13R5} show the numerical results corresponding to perturbed HG case. The main features of these corrections are given as:\\
{\bf{(i)}} Here solar mixing angle($\theta_{12}$) receives very minor corrections through $\sin\theta_{13}$ and thus its value 
remain close to its original prediction. Thus BM and DC will be disfavored for this scheme.\\
{\bf{(ii)}} As fitting of $\theta_{13}$ and $\theta_{12}$ is only governed by $\gamma$
so its allowed range is much constrained in parameter space. e.g. for TBM case, the fitting of $\theta_{13}$ under its $3\sigma$ domain 
constraints the magnitude of correction parameter 
$|\gamma| \in [0.1696(0.1719), 0.1905(0.1921)]$ which in turn fixes $\theta_{12} \in [35.65^\circ(35.66^\circ), 35.75^\circ(35.76^\circ)]$ for corresponding $\gamma$ values. 
However $\theta_{12}$ possess much wider range of values 
since it receives corrections from $\alpha$ as well from phase parameter $\sigma$. \\
{\bf{(iii)}} The minimum value of $\chi^2 \sim 234.7(237.5),~234.7(237.5),~7.35(7.82)$ and $12.9(12.3)$ for BM, DC, TBM and HG respectively with 
NH(IH) case. Here only TBM manages to fit all mixing angles within $3\sigma$ level. The corrected HG predicts low value of $\theta_{12}$ which
is outside its $3\sigma$ range and hence it is not viable.\\
{\bf{(iv)}} Leptonic phase $\delta_{CP}$ lies in the range $0 \leq |\delta_{CP}| \leq 89.9$ while $J_{CP}$  in the range
$0(0) \leq |J_{CP}| \leq 0.0357(0.0360)$ for corrected TBM case.

\begin{center}
\begin{tabular}{ |p{1.3cm}||p{2.0cm}|p{2.0cm}|p{2.0cm}|p{2.0cm}|p{2.0cm}|p{2.0cm}|  }
 \hline
 \multicolumn{7}{|c|}{Best Fit with latest mixing data} \\
 \hline
Rotation  & $\chi^2_{min}$ & $\theta_{12}^\circ$ & $\theta_{23}^\circ$ & $\theta_{13}^\circ$ & $|\delta_{CP}^\circ|$ & $|J_{CP}|$\\
 \hline
 BM   & $234.7(237.5)$     & ${\bf{45.6}}({\bf{45.6}})$ &   $48.0(48.2)$ &   $8.32(8.42)$   &$69.6(68.2)$ & $0.033(0.033)$\\
  \hline
  DC   & $234.7(237.5)$   & ${\bf{45.6}}({\bf{45.6}})$ &   $48.0(48.2)$ &  $8.33(8.43)$     &$36.5(39.6)$ & $0.021(0.022)$\\
  \hline
  TBM   & $7.35(7.82)$    & $35.7(35.7)$ &                  $47.8(48.2)$ &  $8.39(8.49)$    &$61.6(58.0)$ & $0.029(0.028)$\\
  \hline
  HG  & $12.9(12.3)$      & ${\bf{30.3}}({\bf{30.3}})$ &   $47.9(48.2)$ &     $8.41(8.50)$  &$52.9(49.4)$ & $0.024(0.023)$\\
  \hline
\end{tabular}\captionof{table}{\it{Neutrino Mixing angles, $|\delta^\circ_{CP}|$ and $|J_{CP}|$ corresponding to $\chi^2_{min}$ numerical fit.
The mixing angle value that lies outside its best fit $3\sigma$ range is marked in boldface.}}
\label{chisqunperturb} 
\end{center}

\begin{figure}[!t]\centering
\begin{tabular}{c c} 
\hspace{-5mm}
\includegraphics[angle=0,width=80mm]{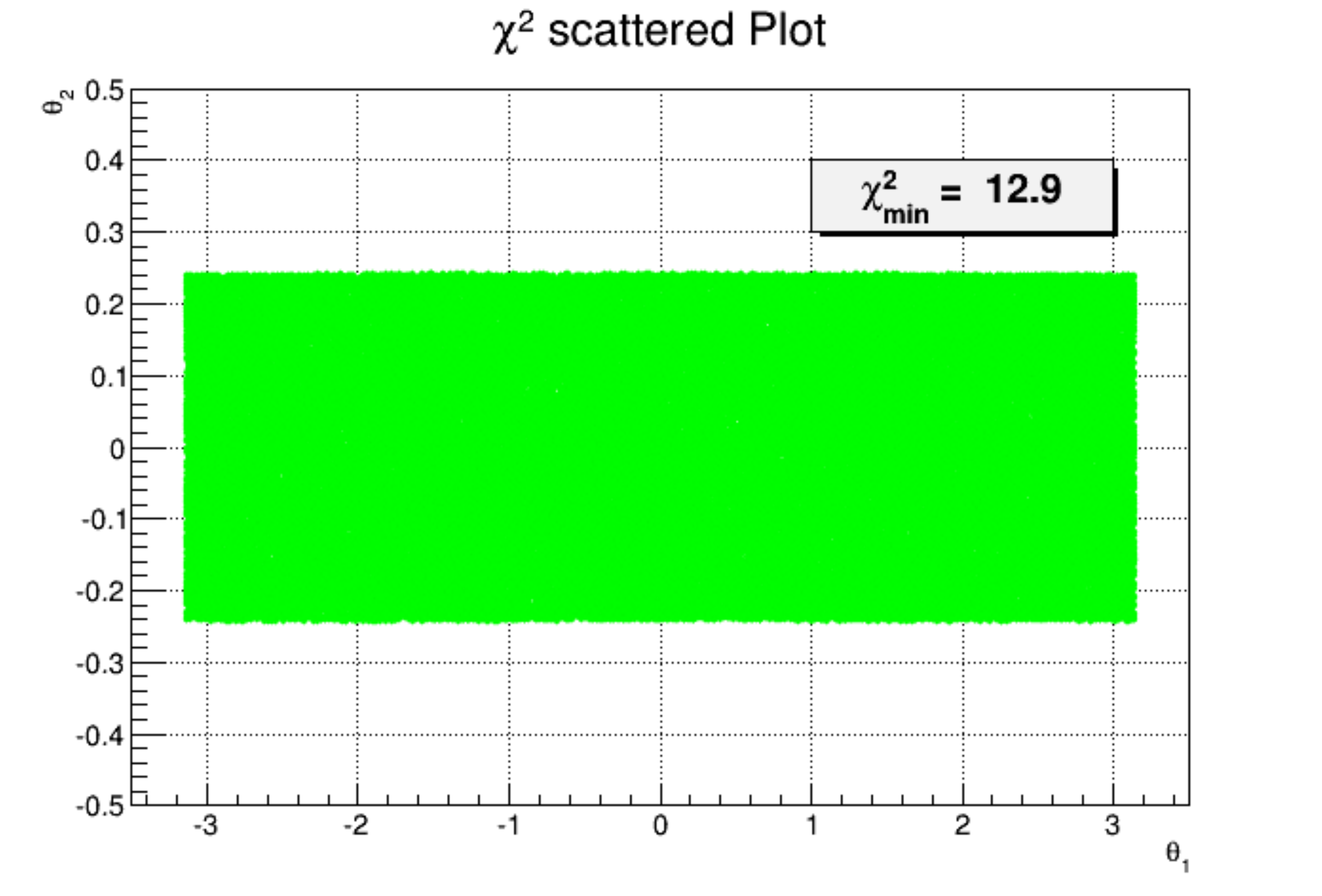} &
\includegraphics[angle=0,width=80mm]{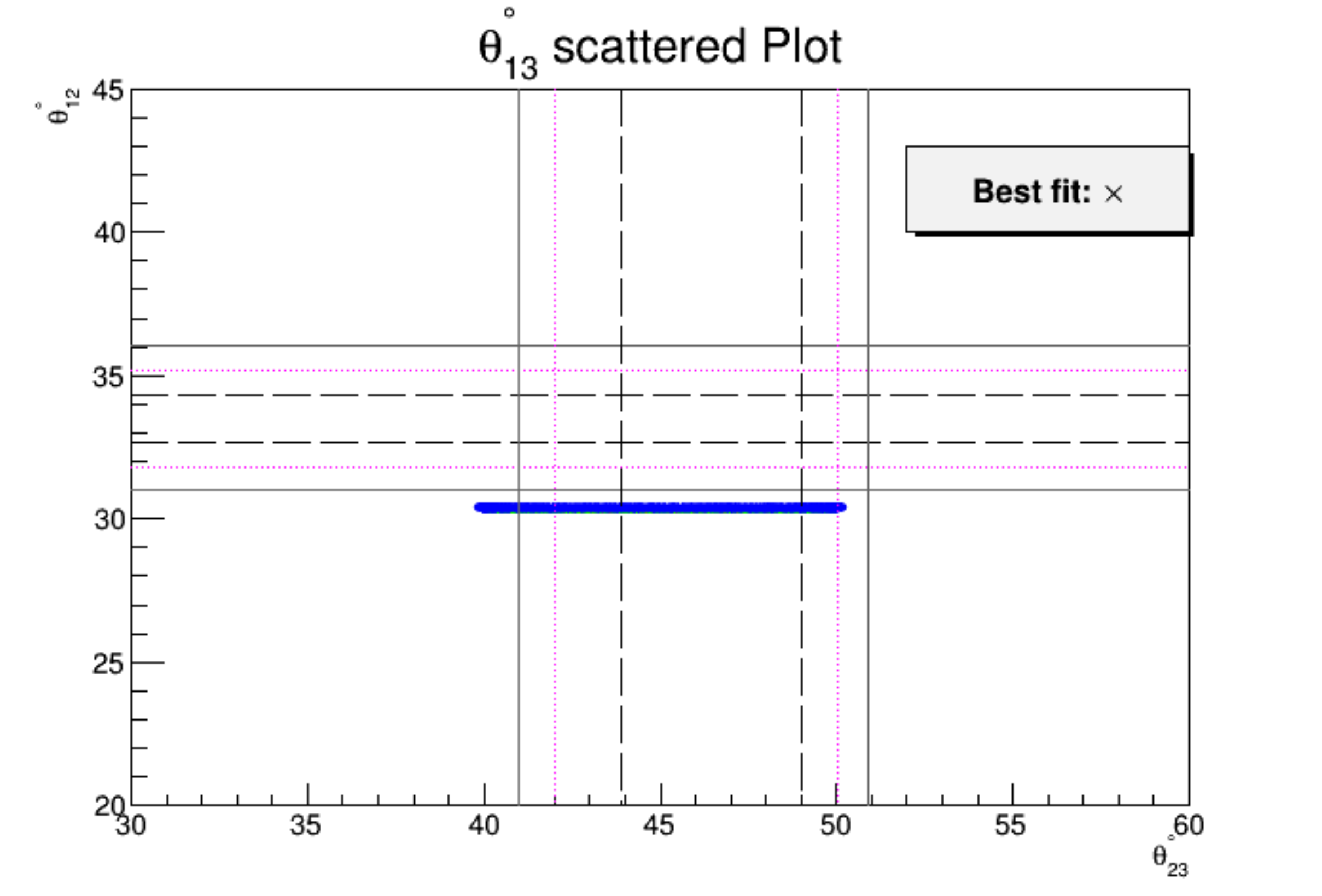}\\
\end{tabular}
\caption{\it{Scattered plot of $\chi^2$ (left fig.) over $\gamma-\sigma$ plane and $\theta_{13}$ (right fig.) 
over $\theta_{23}-\theta_{12}$ (in degrees) plane for $U^{HGR}_{13}$ rotation scheme. }}
\label{fig13R3}
\end{figure}

\subsection{23 Rotation}

This case pertains to rotation in 23 sector of  these special matrices. The expressions
for neutrino in this mixing scheme are given as


\beqa
 \sin^2\theta_{13} &=&  a_{12}^2 \sin^2\beta,\\
 \sin^2\theta_{12} &=& \frac{a_{12}^2\cos^2\beta}{\cos^2\theta_{13}},\\
  \sin^2\theta_{23} &=& \frac{a_{23}^2\cos^2\beta + a_{22}^2\sin^2\beta + a_{22}a_{23}\sin 2\beta \cos\sigma}{\cos^2\theta_{13}}
\eeqa

The Jarsklog invariant and CP Dirac Phase is given by expressions 

\beqa
J_{CP} &=& J_{23R} \sin2\beta \sin\sigma\\
 \sin^2\delta_{CP} &=& C_{23R}^2 \left(\frac{p_{1\beta}}{p_{2\beta\sigma} p_{3\beta\sigma}}\right)\sin^2\sigma
\eeqa

where

\beqa
 J_{23R} &=& \frac{1}{2}a_{12}^2 \sqrt{1-a_{12}^2}~C_{23R}  ,\\
 C_{23R} &=& -\frac{a_{11} a_{21} a_{23}}{a_{12} \sqrt{1-a_{12}^2}},\\
 p_{1\beta} &=& 1+a_{12}^4\sin^4\beta-2 a_{12}^2\sin^2\beta,\\
 p_{2\beta\sigma} &=& 1-a_{23}^2\cos^2\beta - (a_{12}^2 +a_{22}^2)\sin^2\beta - a_{22}a_{23}\cos\sigma \sin 2\beta,\\
 p_{3\beta\sigma} &=& a_{23}^2\cos^2\beta + a_{22}^2\sin^2\beta + a_{22}a_{23}\cos\sigma \sin 2\beta 
\eeqa

\begin{figure}[!t]\centering
\begin{tabular}{c c} 
\hspace{-5mm}
\includegraphics[angle=0,width=80mm]{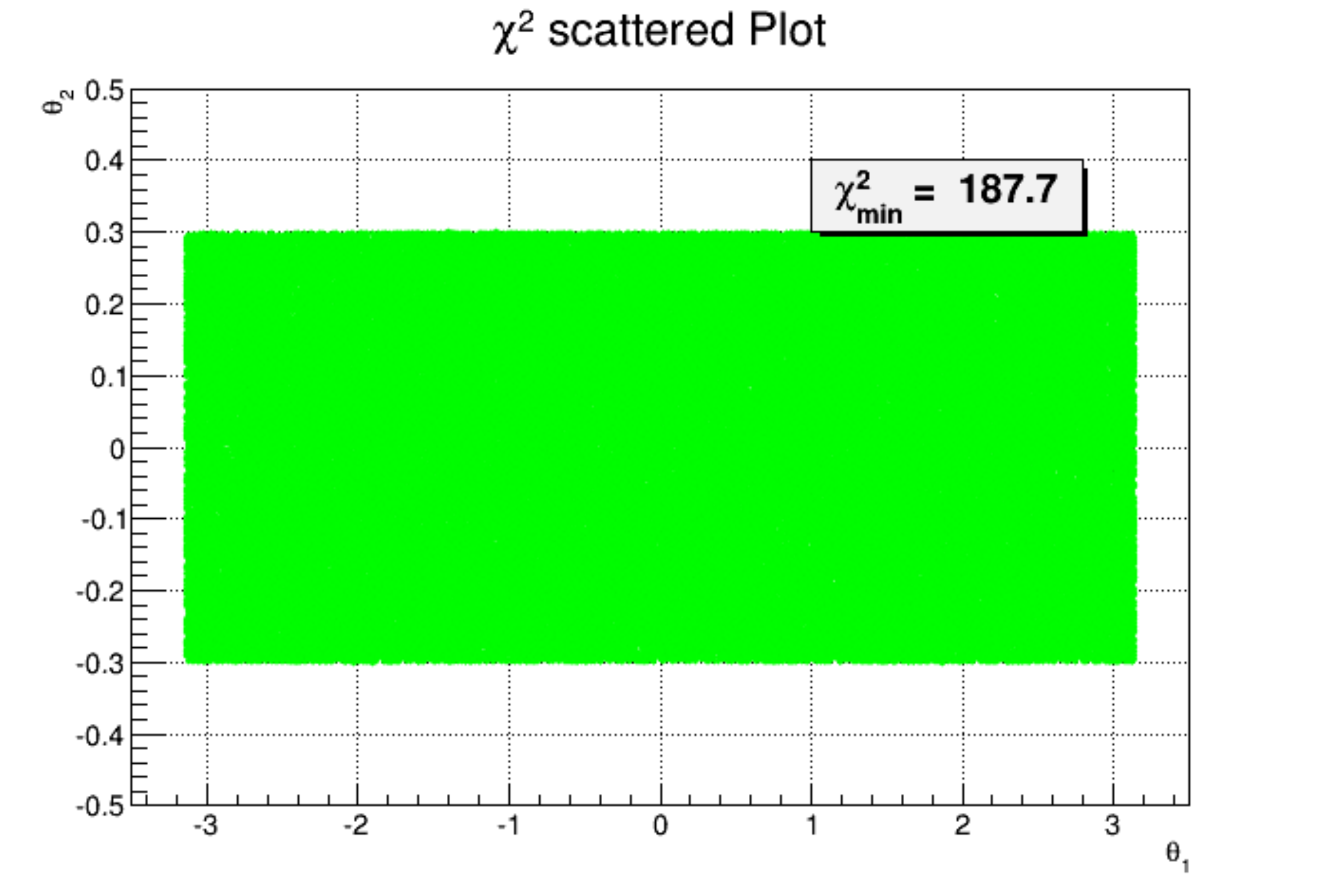} &
\includegraphics[angle=0,width=80mm]{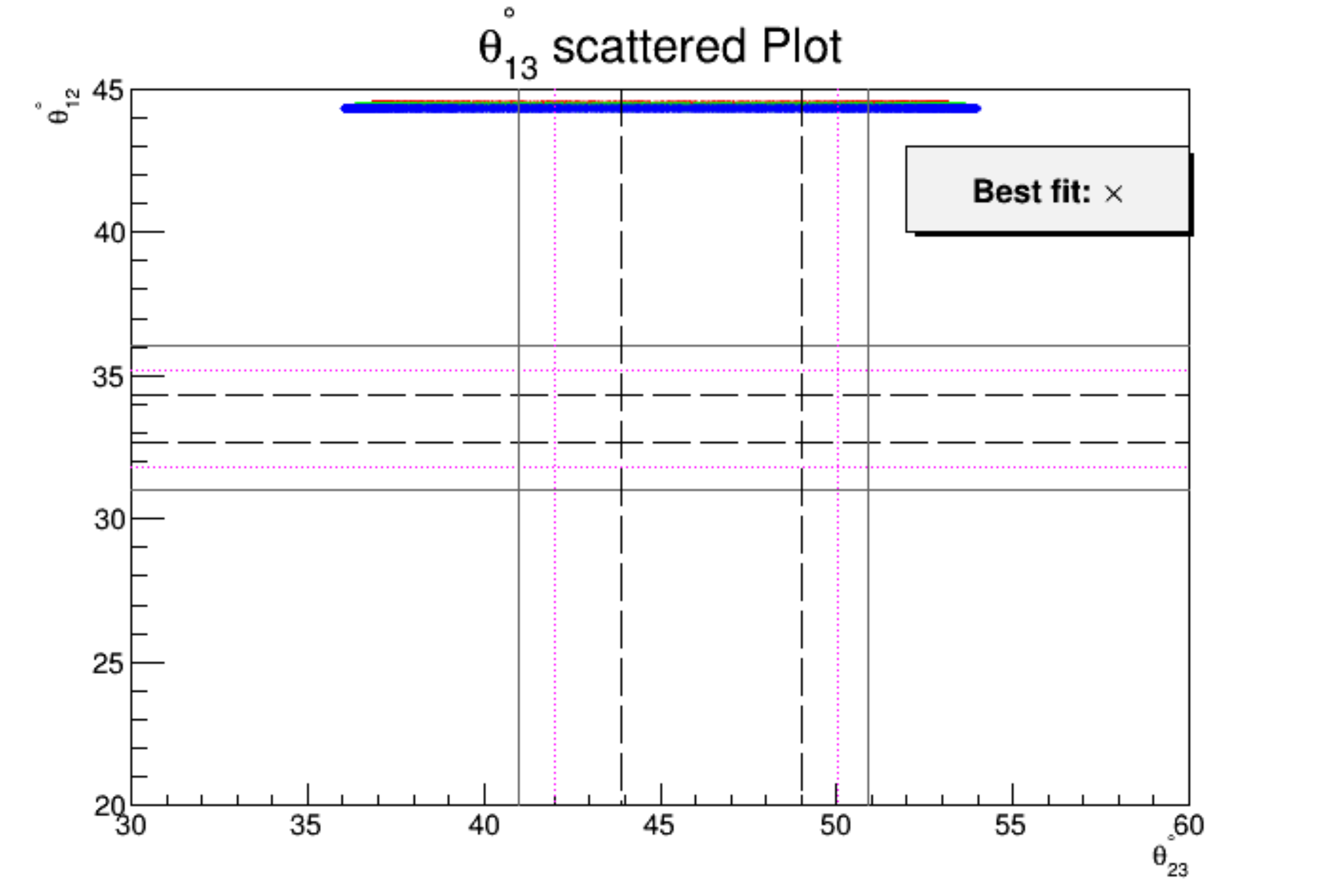}\\
\end{tabular}
\caption{\it{Scattered plot of $\chi^2$ (left fig.) over $\beta-\sigma$ plane and $\theta_{13}$ (right fig.) 
over $\theta_{23}-\theta_{12}$ (in degrees) plane for $U^{BMR}_{23}$ rotation scheme. }}
\label{fig23R1}
\end{figure}

\begin{figure}[!t]\centering
\begin{tabular}{c c} 
\hspace{-5mm}
\includegraphics[angle=0,width=80mm]{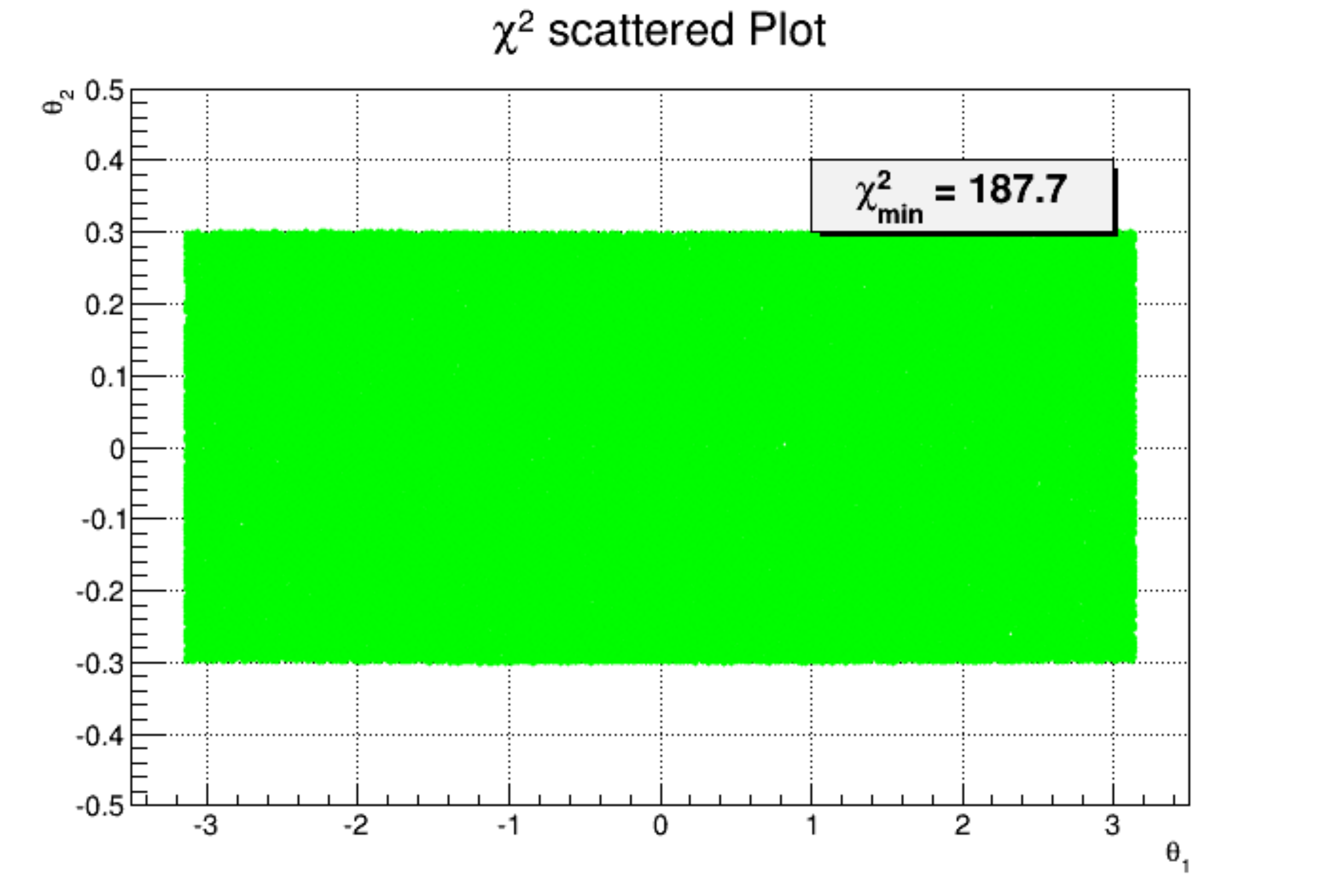} &
\includegraphics[angle=0,width=80mm]{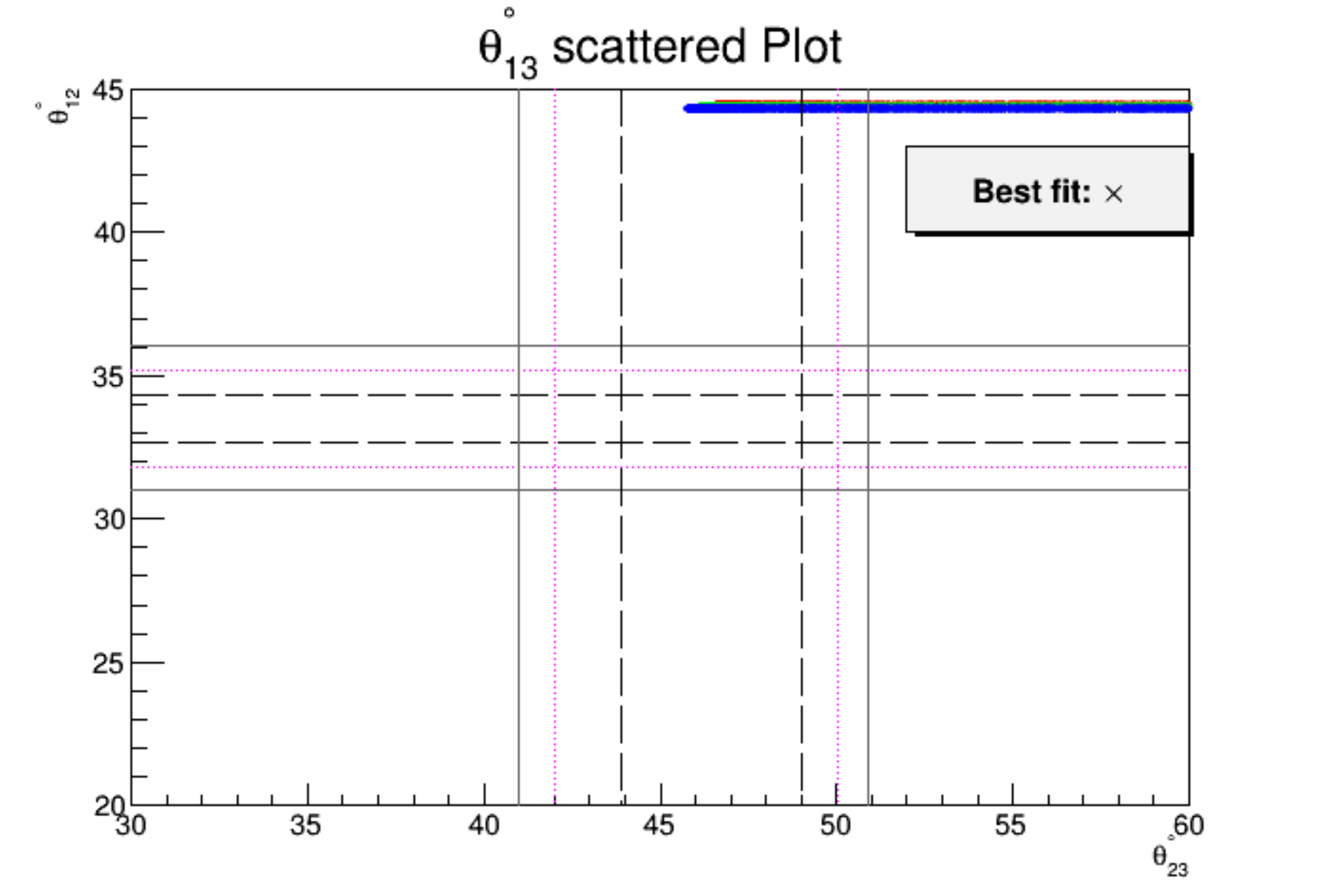}\\
\end{tabular}
\caption{\it{Scattered plot of $\chi^2$ (left fig.) over $\beta-\sigma$ plane and $\theta_{13}$ (right fig.) 
over $\theta_{23}-\theta_{12}$ (in degrees) plane for $U^{DCR}_{23}$ rotation scheme. }}
\label{fig23R2}
\end{figure}

\begin{figure}[!t]\centering
\begin{tabular}{c c} 
\hspace{-5mm}
\includegraphics[angle=0,width=80mm]{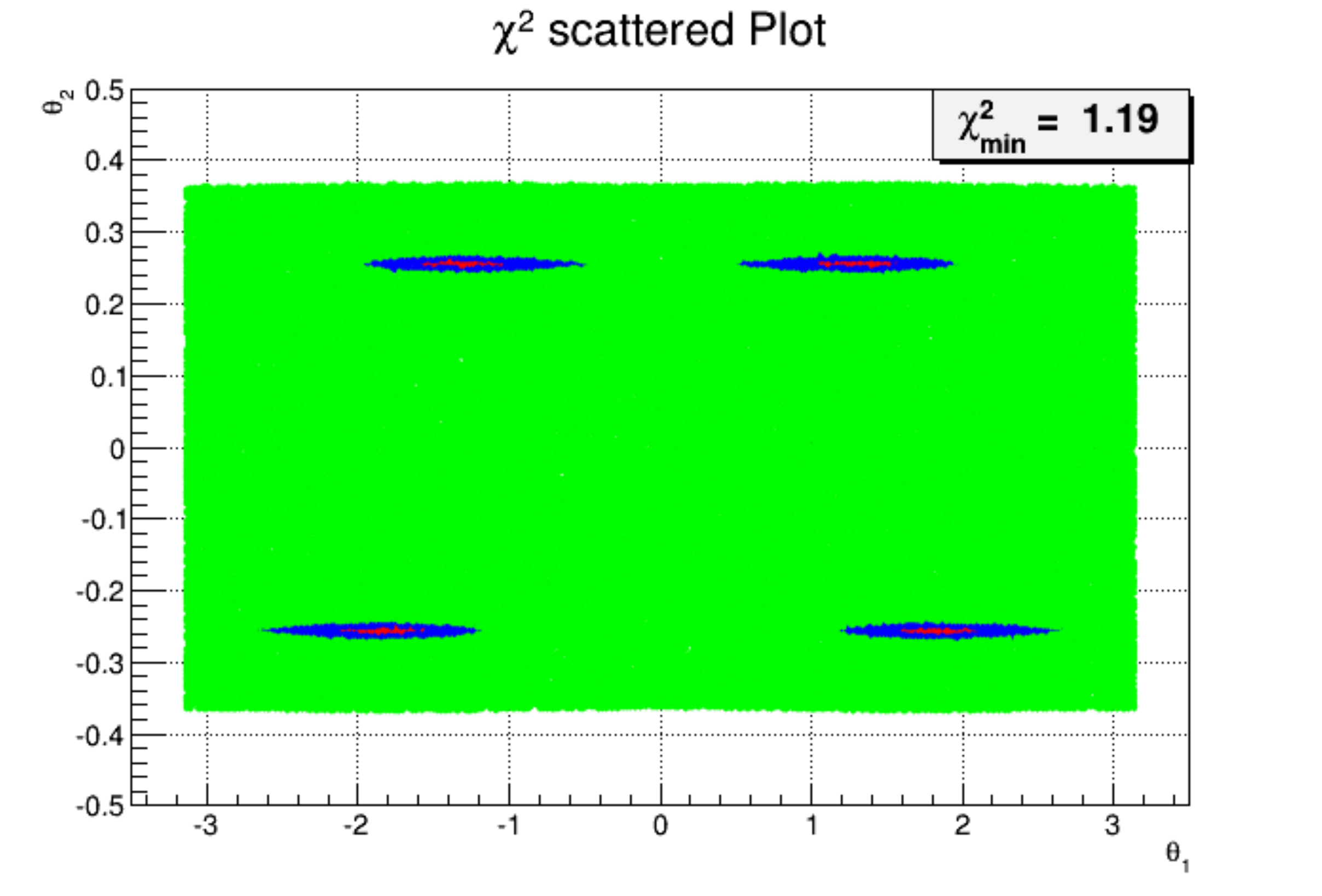} &
\includegraphics[angle=0,width=80mm]{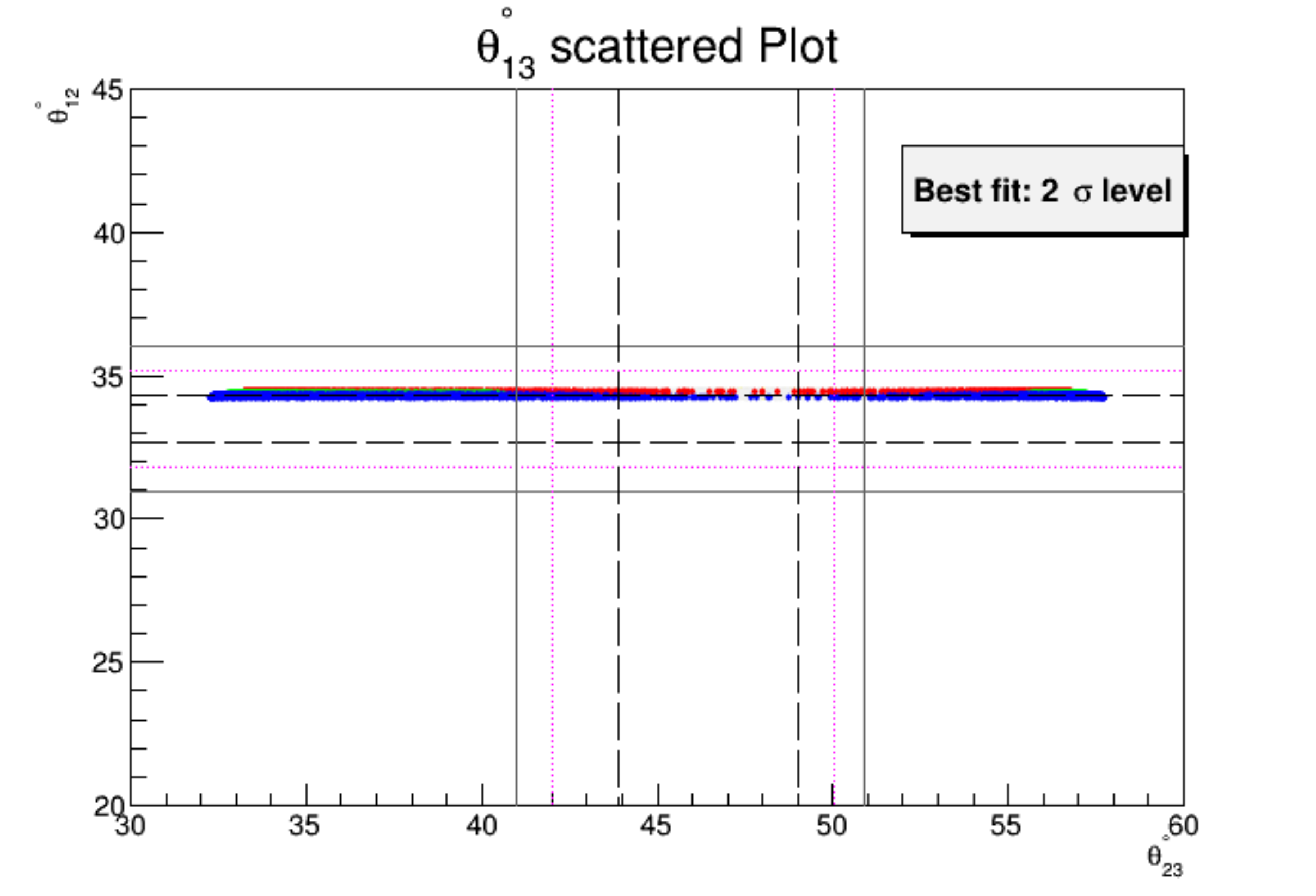}\\
\end{tabular}
\caption{\it{Scattered plot of $\chi^2$ (left fig.) over $\beta-\sigma$ plane and $\theta_{13}$ (right fig.) 
over $\theta_{23}-\theta_{12}$ (in degrees) plane for $U^{TBMR}_{23}$ rotation scheme.}}
\label{fig23R4}
\end{figure}

\begin{figure}[!t]\centering
\begin{tabular}{c c} 
\hspace{-5mm}
\includegraphics[angle=0,width=80mm]{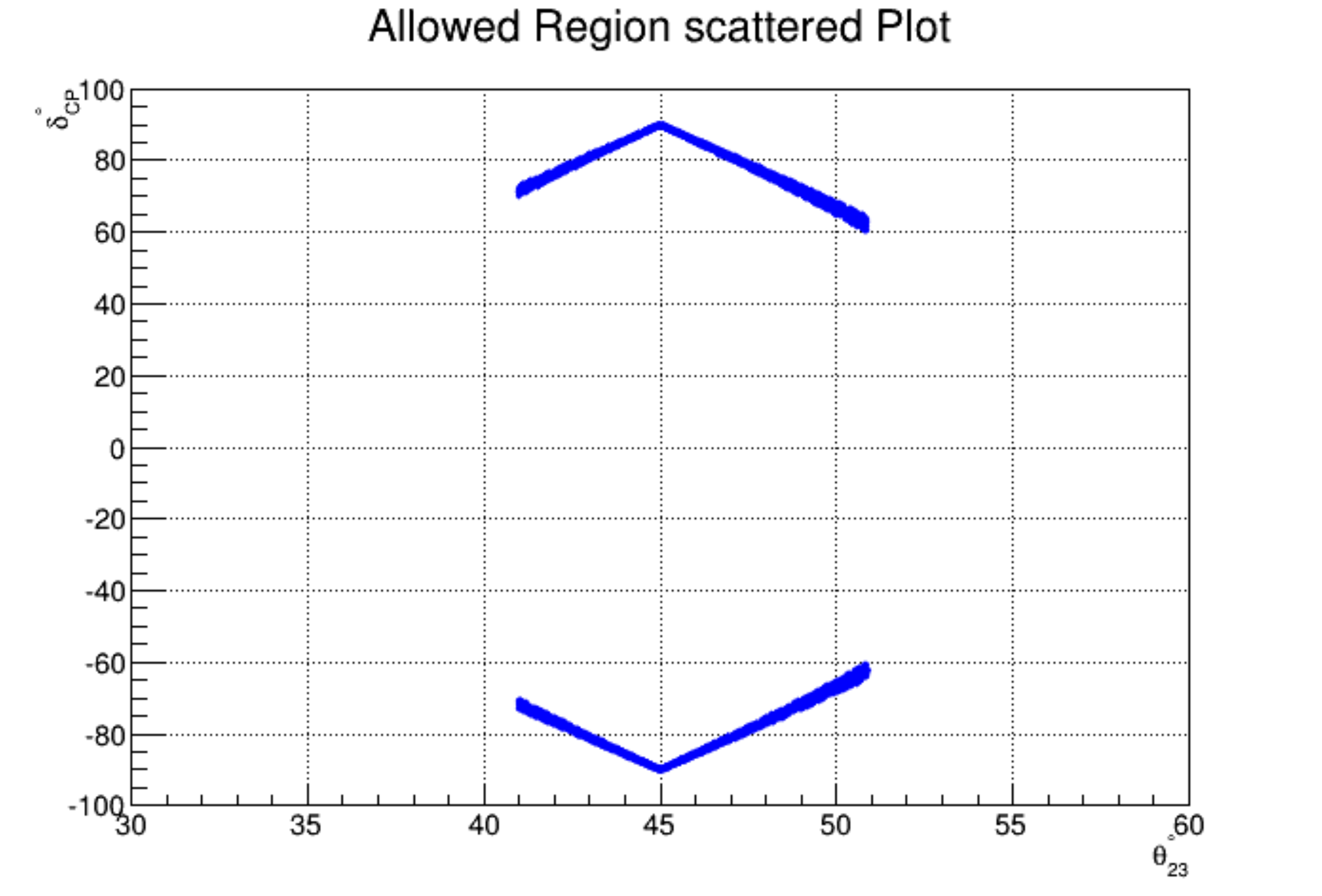} &
\includegraphics[angle=0,width=80mm]{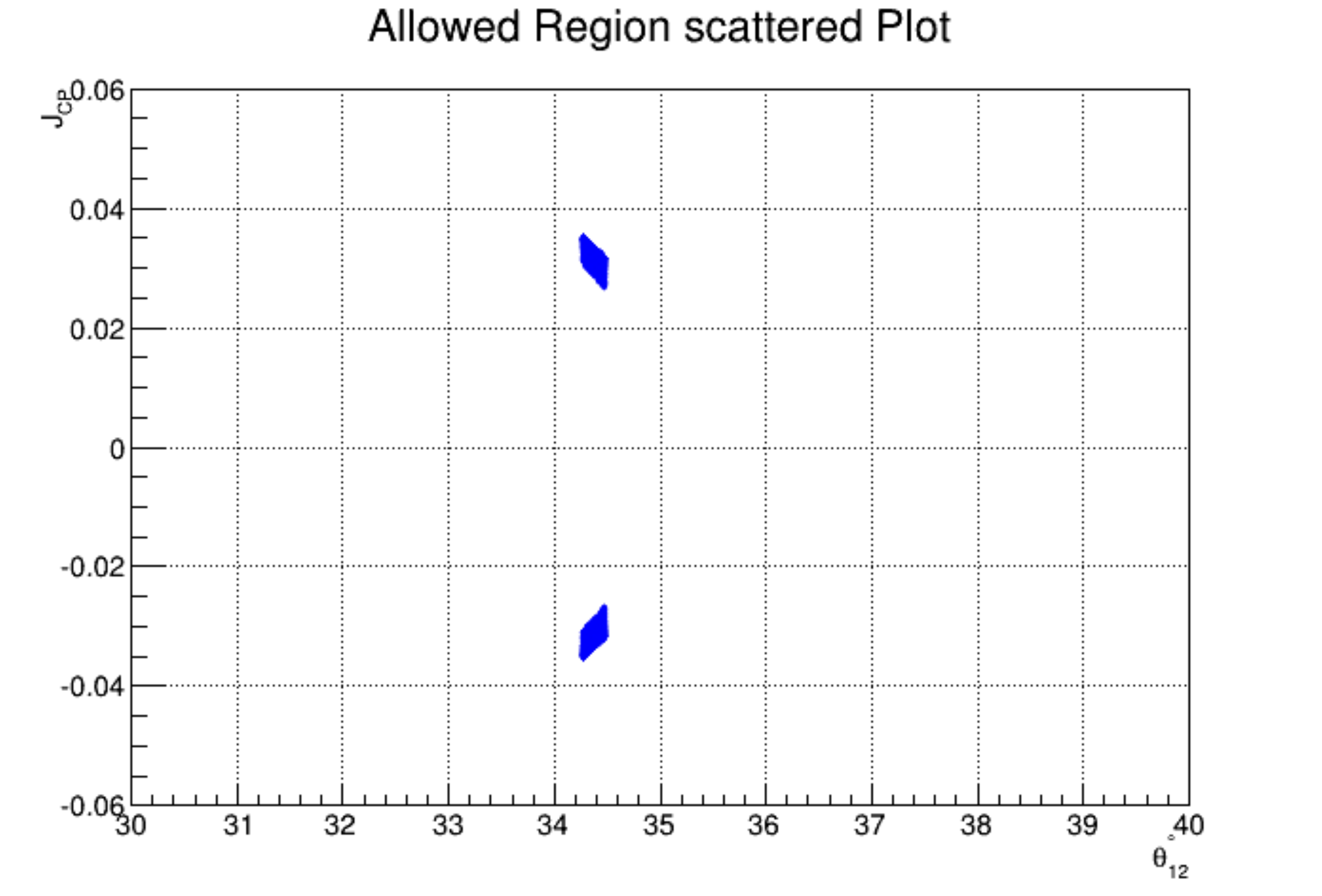}\\
\end{tabular}
\caption{\it{Scattered plot of $\delta_{CP}$ (left fig.) vs $\theta_{23}$ (in degrees) and scattered plot of $J_{CP}$ (right fig.) 
over $\theta_{12}$ (in degrees) plane for $U^{TBMR}_{23}$ rotation scheme.}}
\label{fig23R5}
\end{figure}

Fig.~\ref{fig23R1}-\ref{fig23R5} show the numerical results corresponding to this mixing scheme. The salient features in this perturbative scheme are:\\
{\bf{(i)}} Like previous case, $\theta_{12}$ receives corrections only of the $O(\theta^2)$ so its value
remains close to its original prediction. Thus BM and DC will not be preferred in this mixing scheme.\\
{\bf{(ii)}} As fitting of $\theta_{13}$ and $\theta_{12}$ is only governed by $\beta$
so its allowed range is much constrained in parameter space. e.g. for TBM case, the fitting of $\theta_{13}$ under its $3\sigma$ domain 
constraints the magnitude of correction parameter 
$|\alpha| \in [0.2412(0.2443), 0.2711(0.2734)]$ which in turn fixes $\theta_{12} \in [34.26^\circ(34.24^\circ), 34.47^\circ(34.45^\circ)]$ 
for corresponding $\beta$ values with NH(IH). 
However $\theta_{23}$ possess much wider range of values since it receives corrections from $\beta$ as well from phase parameter $\sigma$. \\
{\bf{(iii)}} The minimum value of $\chi^2 \sim 187.7(189.3),~187.7(189.3),~1.19(1.31)$ and $40.8(86.7)$ for BM, DC, TBM and HG respectively with 
NH(IH) case. Here BM, DC and HG are unable to bring $\theta_{12}$ in its allowed range so these cases are not consistent. However TBM is much
favored as it can fit all mixing angles within $1\sigma$ range for NH and IH.\\
{\bf{(iv)}} Leptonic phase $\delta_{CP}$ lies in the range $60.1(60.3) \leq |\delta_{CP}| \leq 89.9(89.9)$ and $J_{CP}$ confined 
in range $0.026(0.027) \leq |J_{CP}| \leq 0.035(0.035)$ for TBM matrix.

\begin{center}
\begin{tabular}{ |p{1.3cm}||p{2.0cm}|p{2.0cm}|p{2.0cm}|p{2.0cm}|p{2.0cm}|p{2.0cm}|  }
 \hline
 \multicolumn{7}{|c|}{Best Fit with Mixing data} \\
 \hline
Rotation  & $\chi^2_{min}$ & $\theta_{12}^\circ$ & $\theta_{23}^\circ$ & $\theta_{13}^\circ$ & $|\delta_{CP}^\circ|$ & $|J_{CP}|$\\
 \hline
 BM   & $187.7(189.3)$    & ${\bf{44.3}}({\bf{44.3}})$ &   $47.9(48.3)$ &                $8.48(8.54)$ &$70.2(67.9)$ & $0.033(0.033)$\\
  \hline
  DC   & $187.7(189.3)$    & ${\bf{44.3}}({\bf{44.3}})$ &   $47.7(48.1)$ &               $8.47(8.54)$ &$34.7(39.4)$ & $0.020(0.022)$\\
  \hline
  TBM   & $1.19(1.31)$    & $34.3(34.3)$ &                  $47.8(48.2)$ &               $8.41(8.50)$ &$77.0(75.2)$ & $0.032(0.032)$\\
  \hline
  HG  & $40.8(86.7)$    & ${\bf{28.9}}({\bf{28.9}})$ &      ${\bf{57.7}}({\bf{57.6}})$ & $8.33(8.23)$ &$32.0(31.9)$ & $0.014(0.014)$\\
  \hline
\end{tabular}\captionof{table}{\it{Neutrino Mixing angles, $|\delta^\circ_{CP}|$ and $|J_{CP}|$ corresponding to $\chi^2_{min}$ numerical fit.
The mixing angle value that lies outside its best fit $3\sigma$ range is marked in boldface.}}
\label{chisqunperturb} 
\end{center}

\begin{figure}[!t]\centering
\begin{tabular}{c c} 
\hspace{-5mm}
\includegraphics[angle=0,width=80mm]{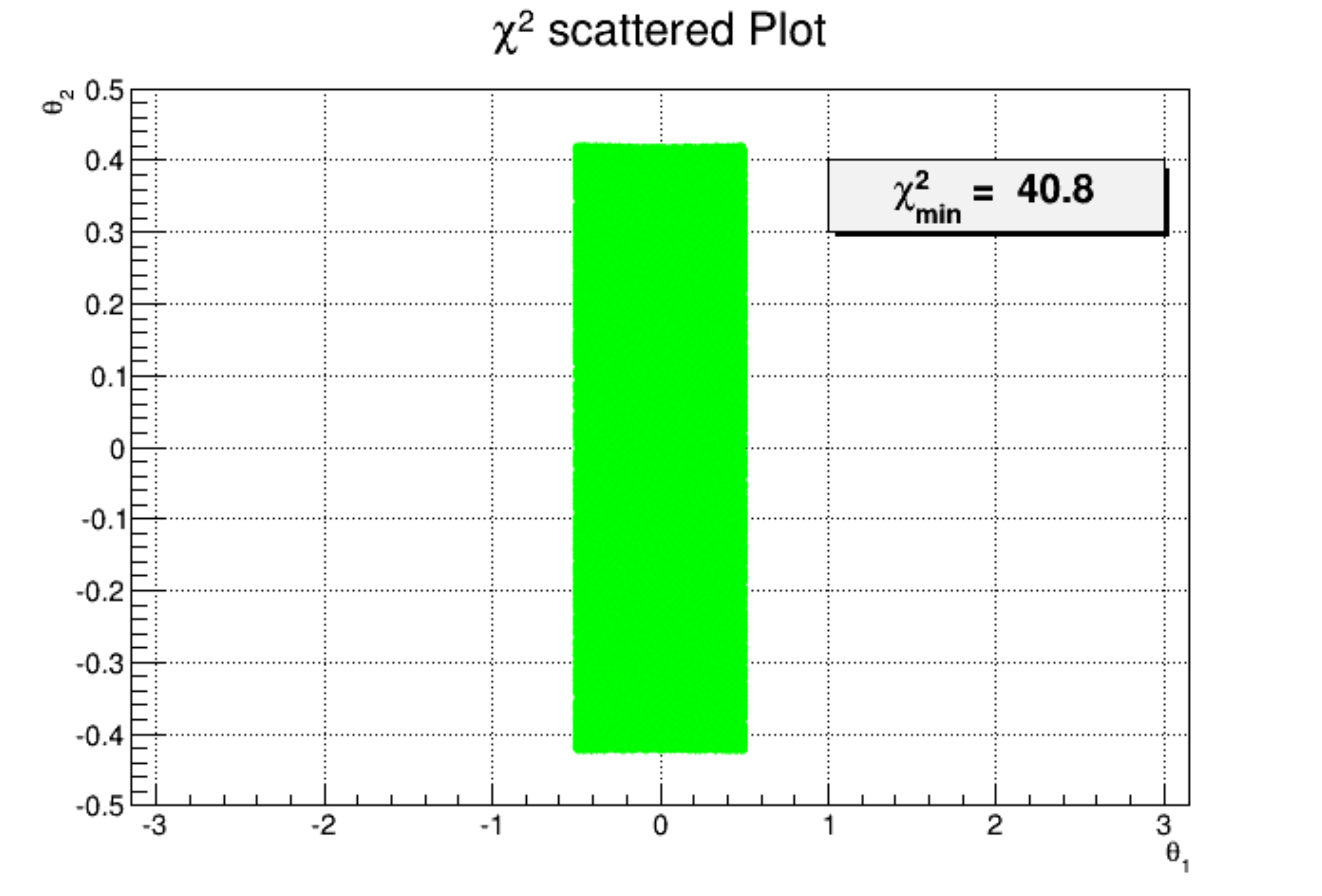} &
\includegraphics[angle=0,width=80mm]{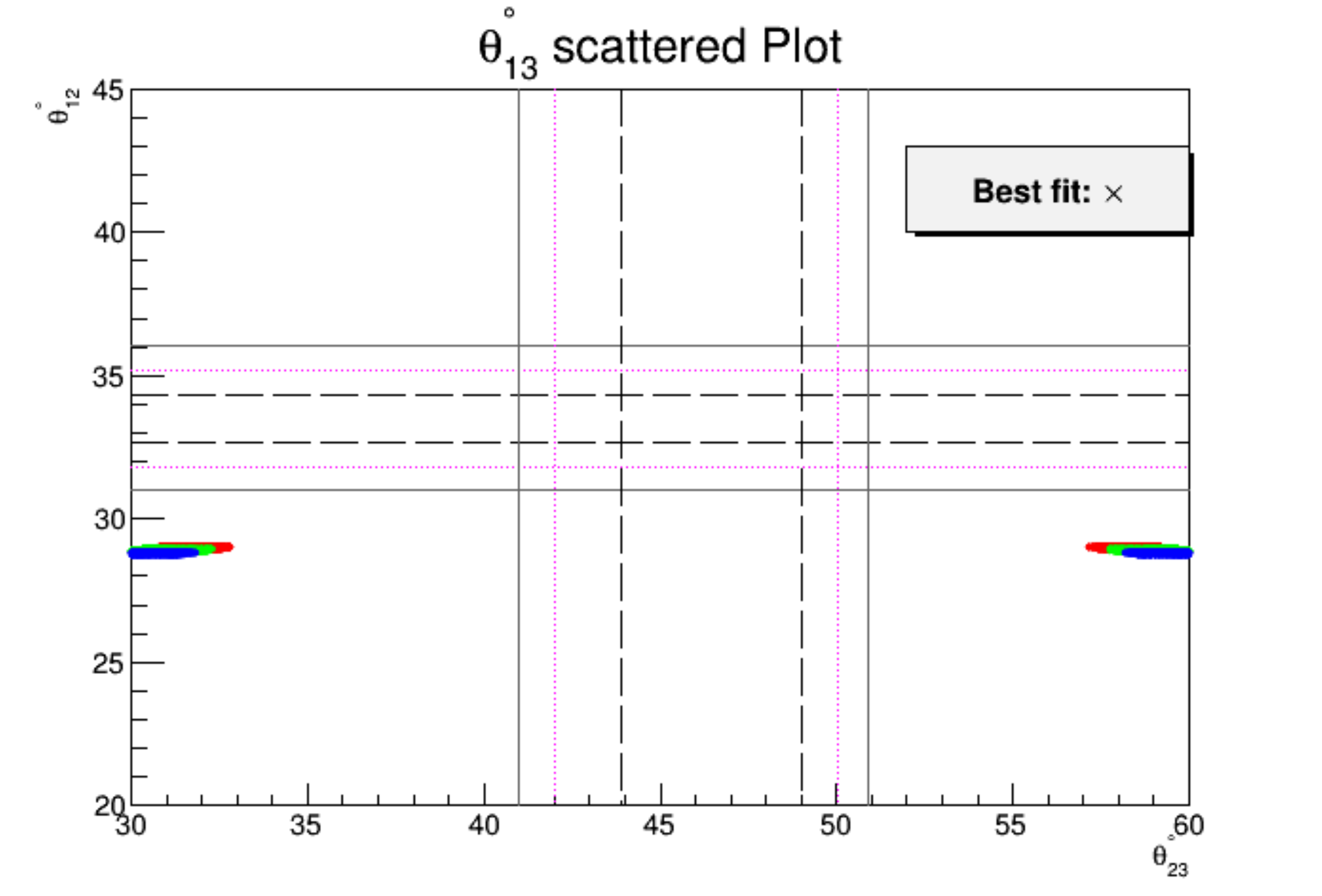}\\
\end{tabular}
\caption{\it{Scattered plot of $\chi^2$ (left fig.) over $\beta-\sigma$ plane and $\theta_{13}$ (right fig.) 
over $\theta_{23}-\theta_{12}$ (in degrees) plane for $U^{HGR}_{23}$ rotation scheme.}}
\label{fig23R3}
\end{figure}



\section{Summary and Conclusions}
Tribimaximal(TBM), Bimaximal(BM), Democratic(DC) and Hexagonal(HG) mixing attracted much attention in literature for explaining the 
neutrino mixing data. All these scenarios comes with a common prediction of vanishing reactor mixing angle. The atmospheric mixing 
angle($\theta_{23}$) is maximal in TBM, BM  and HG mixing while it takes a larger value of $54.7^\circ$ for DC case. The solar mixing 
angle ($\theta_{12}$) is maximal in BM and DC scenarios while its  value is $35.3^\circ$  and $30.0^\circ$ for TBM and HG case respectively. 
However experimental observation of non zero reactor mixing angle ($\theta_{13}\approx 8^\circ$) and departure of other two mixing angles  
from maximality is asking for corrections in these mixing schemes. 

In this study, we performed a detailed analysis of corrections around these mixing scenarios. These modifications are expressed in terms of 
three Unitary rotation matrices U$_{12}$, U$_{13}$ and U$_{23}$ which acts on 12, 13 and 23 sector of unperturbed PMNS matrix respectively. 
We investigated all possible cases that are governed by one rotation matrix with corresponding modified PMNS matrices of the forms
\big($U_{ij}\cdot V_{M},~V_{M}\cdot U_{ij}$\big) where $V_M$ is any one of these special matrices. Here $U_{ij}$ is a complex rotation matrix 
that will act on $ij$ sector of unperturbed mixing matrix and is described by a rotation angle and a phase parameter. As the form of PMNS matrix is given by $U_{PMNS} = U_l^{\dagger} U_\nu$ so 
these corrections can originate from charged lepton and neutrino sector respectively. For our numerical analysis, we invoked $\chi^2$ function 
which is a combined measure of deviation of values of mixing angles in parameter space to that coming from experimental best fit values. 
The resulting value of $\delta_{CP}$ and $J_{CP}$ in allowed parameter space is treated as the prediction of that particular mixing scheme. The numerical findings are presented in terms of $\chi^2$ vs perturbation 
parameters and as correlations among different neutrino mixing angles. The scattered plots for $\delta_{CP}$ and $J_{CP}$ vs mixing angles
are also presented for determining the allowed ranges of these quantities. In Table~\ref{Table2} and Table~\ref{Table3} we presented the formulae
for mixing angles, $\delta_{CP}$ and $J_{CP}$ for all considered cases. Table~\ref{Table4} contains final results of our investigation 
in terms of $(\chi^2,~\text{Best Fit})$. The obtained ranges of $\delta_{CP}$ and $J_{CP}$ are given in Table~\ref{Table4} and 
Table~\ref{Table5} respectively.

The mixing $U_{12}\cdot V_{M}$ imparts $O(\theta^2)$ corrections to $\theta_{23}$ and thus it prefers to stay close to 
its unperturbed prediction in parameter space. However $\theta_{12}$ can possess wide range of values since it gets correction
from rotation as well as phase parameter. Since for DC case $\theta_{23}\sim 54.7^\circ$ so it is disfavored completely. The modified
BM mixing can fit all mixing angles within $3\sigma$ range with $\chi^2_{min}\sim 
20.2(30.2)$ whereas TBM and HG can fit angles in $1\sigma(2\sigma)$ range with $\chi^2_{min}\sim 1.95(11.0)$ and $\chi^2_{min}\sim 1.94(11.0)$ 
respectively for Normal Hierarchy(Inverted Hierarchy). Thus this mixing scheme shows preference towards NH in parameter space. As far as 
leptonic CP phase is concerned, it prefers a smaller value of $[-4.4^\circ(-4.7^\circ) \leq \delta_{CP} \leq 3.4^\circ(5.3^\circ)]$ with corresponding 
$[-0.0027(-0.0029) \leq J_{CP} \leq 0.0021(0.0033)]$ in allowed parameter space for modified BM case. However allowed parameter space prefers
larger value of CP phase in range $[61.0^\circ(60.9^\circ) \leq |\delta_{CP}| \leq 89.9^\circ(89.9^\circ)]$ and  
$[39.0^\circ(40.4^\circ) \leq |\delta_{CP}| \leq 78.7^\circ(79.2^\circ)]$ with corresponding $J_{CP}$ in the range  
$[0.026 \leq |J_{CP}| \leq 0.035]$  and $[0.020(0.021) \leq |J_{CP}| \leq 0.032(0.032)]$ for corrected TBM and BM respectively.

The $U_{13}\cdot V_{M}$ mixing scheme is quite similar to previous case. Here $\theta_{23}$ receives very minor corrections which enters through $\theta_{13}$.
Thus modified value of $\theta_{23}$ remains quite close to its original prediction. However $\theta_{12}$ can have wide range in parameter space since
it receives correction from rotation as well as phase parameter. Here also DC case is not viable. The modified 
BM can fit all mixing angles within $3\sigma$ range with $\chi^2_{min}\sim 18.9(23.4)$ while TBM and HG can fit all angles 
in $1\sigma(2\sigma)$ range with $\chi^2_{min}\sim 0.82(5.0)$ and $\chi^2_{min}\sim 0.81(5.0)$ respectively. Thus this case also prefers 
NH and overall picture of fitting is better then previous scheme due to lower value of $\chi^2_{min}$.
The leptonic phase $\delta_{CP}$ prefers a smaller value of $[-3.6^\circ(-4.9^\circ)\leq \delta_{CP} \leq 4.2^\circ(5.5^\circ)]$ with corresponding 
$[-0.0025(-0.0034) \leq J_{CP} \leq 0.0026(0.0030)]$ in allowed parameter space for modified BM case. However allowed parameter have 
same ranges of $\delta_{CP}$ and $J_{CP}$ as that obtained in previous case for $U_{12}\cdot V_{M}$ mixing.

For $U_{23}\cdot V_{M}$  and  $V_{M}\cdot U_{12}$ mixing scheme,  13 element of modified matrix is still zero and thus 
$\theta_{13}$ remains stick to its unperturbed value i.e. $\theta_{13}=0$. Thus these two cases are not suitable for any further
investigation.

For $V_{M}\cdot U_{13}$ mixing scheme,
$\theta_{12}$ receives very minor corrections only through $\theta_{13}$ and thus it remains quite close to its unperturbed value. 
However $\theta_{23}$ can possess wide range of values in parameter space since it receives correction from rotation as well as
phase parameter. Thus BM, DC and HG are not favorable as their unperturbed values are outside $3\sigma$ range. However TBM is still consistent as
its unperturbed value is $\theta_{12}\sim 35.3^\circ$ which lies within global fit $3\sigma$ range.  It can fit all mixing angles with
$\chi^2_{min}\sim 7.35(7.82)$ for NH(IH). This mixing scheme allows much larger range of 
 $[-89.9^\circ \leq \delta_{CP} \leq 89.9^\circ]$  and  $[-0.035(-0.036) \leq J_{CP} \leq 0.035(0.036)]$ in allowed parameter space for NH and IH. 
 Thus overall this mixing scheme is not much favorable. 

In this case,  $\theta_{12}$ receives $O(\theta^2)$  corrections and 
thus its modified value remains close to its unperturbed value. However $\theta_{23}$ can have wide range of values in parameter space 
as it gets corrections from rotation and phase parameter. The BM, DC and HG are not viable but TBM is preferable as it can fit 
all mixing angles within $2\sigma$ level with $\chi^2_{min}\sim 1.19(1.31)$. The predicted value of Leptonic CP phase lies in range
 $[60.1^\circ(60.3^\circ) \leq |\delta_{CP}|\leq 89.9^\circ(89.9^\circ)]$ for NH(IH). However $J_{CP}$ remains in range $[0.026(0.027)\leq |J_{CP}| \leq 
 0.035(0.035)]$  for NH as well as IH.

This completes our discussion on checking the consistency of these schemes with mixing data and corresponding prediction for CP Violating Phase 
and Jarkslog invariant for various cases. This model independent study might turn out to be useful in restricting vast number of possible models 
which offers different corrections to this mixing scheme in neutrino model building physics. It thus can be a guideline for neutrino model building.
Moreover the predictions of these mixing scenarios can be checked from current and future neutrino experiments.

\section*{Acknowledgments}
The author is thankful to Prof. B. Ananthnarayan for hospitality during visit to CHEP, IISC Bengaluru where version v2 of 
this study was completed.


\appendix
\section{Results: Summary} \label{App:AppendixA}
In this appendix, we collect all our formulae and results for considered mixing schemes.  
In Table~\ref{Table2} and Table~\ref{Table3}, we supplied the expressions for mixing angles, Dirac CP Phase($\delta_{CP}$) and 
Jarkslog invariant($J_{CP}$) in terms of correction parameters. In Table~\ref{Table4},
we presented $(\chi^2_{min}, {\text{Best Fit}})$ for various studied cases while in Table~\ref{Table5} and Table~\ref{Table6} we 
gave allowed ranges for $\delta_{CP}$ and $J_{CP}$ respectively.\\

\begin{landscape}
\begin{center}
\renewcommand{\arraystretch}{1.8}
\begin{tabular}{ |p{1.4cm}||p{5.2cm}|p{5.2cm}|p{5.2cm}|p{5.8cm}|  }
 \hline
 \multicolumn{5}{|c|}{Relevant formuale for $U_{12}^l\cdot V_M$ and  $U_{13}^l\cdot V_M$ Mixing Scheme } \\
 \hline
 $U_{12}^l\cdot V_M$  & BM &DC&TBM & HG\\
 \hline
 $\sin^2\theta_{13}$   
                       & $\frac{1}{2}\sin^2\alpha$   
                       &  $\frac{2}{3}\sin^2\alpha$ 
                       &  $\frac{1}{2}\sin^2\alpha$ 
                       & $\frac{1}{2}\sin^2\alpha$\\
  \hline
  $\sin^2\theta_{23}$   
                       & $\frac{\cos^2\alpha}{1+\cos^2\alpha}$   
                       & $\frac{2\cos^2\alpha}{1+2\cos^2\alpha}$ 
                       &  $\frac{\cos^2\alpha}{1+\cos^2\alpha}$ 
                       & $\frac{\cos^2\alpha}{1+\cos^2\alpha}$\\
  \hline
    $\sin^2\theta_{12}$   
    & $\frac{1}{2}+ \frac{\sqrt{2}\cos\sigma \sin 2\alpha}{3+\cos 2\alpha}$  
    & $\frac{1}{2}- \frac{\sqrt{3}\cos\sigma \sin 2\alpha}{2(2+\cos 2\alpha)}$  
    & $\frac{2+4\cos\alpha \sin\alpha \cos\sigma}{6-3\sin^2\alpha}$ 
    & $-\frac{1}{4}+ \frac{4+\sqrt{6}\cos\sigma \sin 2\alpha}{6+2\cos 2\alpha}$\\
  \hline
  $\sin^2\delta_{CP}$  
  & $\frac{2(3+\cos2\alpha)^2\sin^2\sigma}{15+12\cos2\alpha+5\cos 4\alpha-8\cos 2\sigma \sin^2 2\alpha}$   
  & $\frac{4(2+\cos2\alpha)^2\sin^2\sigma}{15+16\cos2\alpha+5\cos 4\alpha-6\cos 2\sigma \sin^2 2\alpha}$ 
  &  \scalebox{0.6}{$\frac{(3+\cos 2\alpha)^2\sin^2\sigma}{8+6\cos 2\alpha+2\cos 4\alpha+ 2\sin 2\alpha(\cos\sigma-
  2\cos 2\sigma \sin 2\alpha) +3\cos\sigma \sin 4\alpha}$}  
  & \scalebox{0.6}{$\frac{3(3+\cos 2\alpha)^2 \sin^2\sigma}{4(2\cos^2\alpha +3\sin^2\alpha+\sqrt{6}\cos\sigma \sin 2\alpha)
  (8-2\cos^2\alpha-7\sin^2\alpha-\sqrt{6}\cos\sigma\sin2\alpha)}$}\\
    \hline
   $J_{CP}$  
                 & $\frac{1}{8\sqrt{2}}\sin 2\alpha \sin\sigma$   
                 & $-\frac{1}{6\sqrt{3}}\sin 2\alpha \sin\sigma$ 
                 & $\frac{1}{6}\sin\alpha \cos\alpha \sin\sigma$ 
                 & $\frac{\sqrt{3}}{16\sqrt{2}}\sin 2\alpha \sin\sigma$\\
    \hline
    \hline
   $U_{13}^l\cdot V_M$ &&&&\\
   \hline
   $\sin^2\theta_{13}$   
                           & $\frac{1}{2}\sin^2\gamma$   
                           & $\frac{1}{3}\sin^2\gamma$ 
                           & $\frac{1}{2}\sin^2\gamma$ 
                           & $\frac{1}{2}\sin^2\gamma$\\
  \hline
  $\sin^2\theta_{23}$   
                           & $\frac{1}{1+\cos^2\gamma}$  
                           & $\frac{4}{5+\cos 2\gamma}$ 
                           & $\frac{1}{1+\cos^2\gamma}$ 
                           & $\frac{1}{1+\cos^2\gamma}$\\
  \hline
    $\sin^2\theta_{12}$   
                           &  $\frac{1}{2}- \frac{\sqrt{2}\cos\sigma \sin 2\gamma}{3+\cos 2\gamma}$    
                           &  $\frac{1}{2} + \frac{\sqrt{6}\cos\sigma \sin 2\gamma}{5+\cos 2\gamma}$ 
                           & $\frac{2+4\cos\gamma\cos\sigma \sin\gamma}{6-3\sin^2\gamma}$ 
                           & $-\frac{1}{4}+ \frac{4+\sqrt{6}\cos\sigma \sin 2\gamma}{6+2\cos 2\gamma}$\\
  \hline
  $\sin^2\delta_{CP}$  
                       & $\frac{2(3+\cos2\gamma)^2\sin^2\sigma}{15+12\cos2\gamma+5\cos 4\gamma-8\cos 2\sigma \sin^2 2\gamma}$  
                       & $\frac{2(5+\cos2\gamma)^2\sin^2\sigma}{39+20\cos2\gamma+13\cos 4\gamma-24\cos 2\sigma \sin^2 2\gamma}$
                       & \scalebox{0.7}{$\frac{(3+\cos 2\gamma)^2 \sin^2\sigma}{4(1+\cos\sigma \sin 2\gamma)(6-2\cos^2\gamma -5\sin^2\gamma-2\cos\sigma \sin 2\gamma)}$} 
                       & \scalebox{0.6}{$\frac{3(3+\cos 2\gamma)^2 \sin^2\sigma}{4(2\cos^2\gamma +3\sin^2\gamma+\sqrt{6}\cos\sigma \sin 2\gamma)
                                 (8-2\cos^2\gamma-7\sin^2\gamma-\sqrt{6}\cos\sigma\sin2\gamma)}$}\\
    \hline
   $J_{CP}$  
                       & $\frac{1}{8\sqrt{2}}\sin 2\gamma \sin\sigma$  
                       & $-\frac{1}{6\sqrt{6}}\sin 2\gamma \sin\sigma$ 
                       &   $-\frac{1}{6}\sin\gamma\cos\gamma \sin\sigma$ 
                       & $-\frac{\sqrt{3}}{16\sqrt{2}}\sin 2\gamma \sin\sigma$\\
  \hline
\end{tabular}\captionof{table}{\it{Here $\alpha$, $\beta$ and $\gamma$ are rotation parameters while $\sigma$ is a phase parameter in 
correction matrix.}}\label{Table2}  
\end{center}
\end{landscape}

\begin{landscape}
\begin{center}
\renewcommand{\arraystretch}{1.8}
\begin{tabular}{ |p{1.4cm}||p{5.2cm}|p{5.2cm}|p{5.2cm}|p{5.5cm}|  }
 \hline
 \multicolumn{5}{|c|}{Relevant Formuale for $V_M \cdot U_{13}^r$ and $V_M \cdot U_{23}^r$ Mixing Scheme} \\
 \hline
 $V_M \cdot U_{13}^r$  & BM &DC&TBM & HG\\
 \hline
 $\sin^2\theta_{13}$   
                       & $\frac{1}{2}\sin^2\gamma$   
                       &  $\frac{1}{2}\sin^2\gamma$ 
                       &  $\frac{2}{3}\sin^2\gamma$ 
                       & $\frac{3}{4}\sin^2\gamma$\\
  
  \hline
    $\sin^2\theta_{23}$   
                           & $\frac{1}{2}- \frac{\sqrt{2}\cos\sigma \sin 2\gamma}{3+\cos 2\gamma}$  
                           & $\frac{4\cos^2\gamma+\sin^2\gamma-2\cos\sigma \sin 2\gamma}{6-3\sin^2\gamma}$  
                           & $\frac{1}{2}-\frac{\sqrt{3}\cos\sigma \sin 2\gamma}{2(2+\cos 2\gamma)}$
                           & $\frac{1}{2}+ \frac{2\cos\sigma \sin 2\gamma}{5+3\cos 2\gamma}$\\
                           
  \hline
  $\sin^2\theta_{12}$   
                       & $\frac{1}{1+\cos^2\gamma}$   
                       & $\frac{1}{1+\cos^2\gamma}$ 
                       &  $\frac{1}{2+\cos 2\gamma}$ 
                       & $\frac{1}{4-3\sin^2\gamma}$\\ 
  \hline
 $\sin^2\delta_{CP}$ 
                        & $\frac{2(3+\cos2\gamma)^2\sin^2\sigma}{15+12\cos2\gamma+5\cos 4\gamma-8\cos 2\sigma \sin^2 2\gamma}$  
                        & \scalebox{0.8}{$\frac{(3+\cos2\gamma)^2\sin^2\sigma}{4(4\cos^2\gamma+\sin^2\gamma-2\cos\sigma \sin 2\gamma)(1+\cos\sigma \sin 2\gamma)}$}
                        &  \scalebox{0.8}{$\frac{4(2+\cos 2\gamma)^2 \sin^2\sigma}{15+16 \cos 2\gamma +5 \cos 4\gamma-6\cos 2\sigma \sin^2 2\gamma}$}  
                        & \scalebox{0.8}{$\frac{2(5+3\cos 2\gamma)^2 \sin^2\sigma}{51+60\cos 2\gamma+ 17\cos 4\gamma -16\cos 2\sigma \sin^2 2\gamma}$}\\
    \hline
   $J_{CP}$  
                            & $\frac{1}{8\sqrt{2}}\sin 2\gamma \sin\sigma$   
                            & $\frac{1}{12}\sin 2\gamma \sin\sigma$ 
                            & $\frac{1}{6\sqrt{3}}\sin 2\gamma \sin\sigma$ 
                            & $-\frac{3}{32}\sin 2\gamma \sin\sigma$\\
    \hline
    \hline
   $V_M \cdot U_{23}^r$ &&&&\\
   \hline
   $\sin^2\theta_{13}$   & $\frac{1}{2}\sin^2\beta$   
                         & $\frac{1}{2}\sin^2\beta$ 
                         & $\frac{1}{3}\sin^2\beta$ 
                         & $\frac{1}{4}\sin^2\beta$\\
 
  \hline
    $\sin^2\theta_{23}$   
     &  $\frac{1}{2}+ \frac{\sqrt{2}\cos\sigma \sin 2\beta}{3+\cos 2\beta}$    
     & $\frac{4\cos^2\beta +\sin^2\beta + 2\cos\sigma \sin 2\beta}{6-3\sin^2\beta}$ 
     & $\frac{1}{2} + \frac{\sqrt{6}\cos\sigma \sin 2\beta}{5+\cos 2\beta}$ 
     & $\frac{1}{2} - \frac{2\sqrt{3}\cos\sigma \sin 2\beta}{7+\cos 2\beta}$\\
     
  \hline
  $\sin^2\theta_{12}$   
                         & $\frac{\cos^2\beta}{1+\cos^2\beta}$  
                         & $\frac{\cos^2\beta}{1+\cos^2\beta}$  
                         & $1-\frac{4}{5+\cos 2\beta}$ 
                         & $1-\frac{6}{7+\cos 2\beta}$\\
  \hline
  $\sin^2\delta_{CP}$  
   & $\frac{2(3+\cos2\beta)^2\sin^2\sigma}{15+12\cos2\beta+5\cos 4\beta-8\cos 2\sigma \sin^2 2\beta}$   
   & \scalebox{0.8}{$\frac{(3+\cos 2\beta)^2\sin^2\sigma}{4(1-\cos\sigma\sin 2\beta)(4\cos^2\beta +\sin^2\beta+2\cos\sigma \sin 2\beta)}$} 
   & \scalebox{0.8}{$\frac{2(5+\cos 2\beta)^2\sin^2\sigma}{39+20\cos 2\beta + 13\cos 4\beta -24 \cos 2\sigma \sin^2 2\beta}$} 
   & \scalebox{0.8}{$\frac{(7+\cos 2\beta)^2 \sin^2\sigma}{(7+\cos^2\beta)^2-2(7+\cos^2\beta(1+96\cos^2\sigma))\sin^2\beta +\sin^4\beta}$}\\
    \hline
   $J_{CP}$  
                 & $\frac{1}{8\sqrt{2}}\sin 2\beta \sin\sigma$  
                 & $\frac{1}{6}\sin\beta \cos\beta \sin\sigma$ 
                 & $\frac{1}{6\sqrt{6}}\sin2\beta \sin\sigma$  
                 & $-\frac{\sqrt{3}}{32}\sin 2\beta \sin\sigma$\\
  \hline
\end{tabular}\captionof{table}{\it{Here $\alpha$, $\beta$ and $\gamma$ are rotation parameters while $\sigma$ is a phase parameter in correction 
matrix.}}\label{Table3}  
\end{center}
\end{landscape}

\begin{center}
\begin{tabular}{ |p{2.2cm}||p{2.2cm}|p{2.2cm}|p{2.2cm}|p{2.2cm}|  }
 \hline
 \multicolumn{5}{|c|}{($\chi^2_{min}$,~{\text{Best fit level}}) for NH and IH from Mixing angles fitting} \\
 \hline
 Rotation-NH  & BM &DC&TBM & HG\\
 \hline
 $U_{12}^l$   & $(20.2,~3\sigma)$   & $(51.4,~\times)$ &   $(1.95,~1\sigma)$ & $(1.94,~1\sigma)$\\
  \hline
  $U_{13}^l$   & $(18.9,~3\sigma)$   & $(8.56,~\times)$ &   $(0.82,~1\sigma)$ & $(0.81,~1\sigma)$\\
  \hline
    $U_{13}^r$   & $(234.7,~\times)$   & $(234.7,~\times)$ &   $(7.35,~3\sigma)$ & $(12.9,~\times)$\\
  \hline
  $U_{23}^r$  & $(187.7,~\times)$   & $(187.7,~\times)$ &   $(1.19,~2\sigma)$ & $(40.8,~\times)$\\
    \hline
    \hline
   Rotation-IH &&&&\\
   \hline
    $U_{12}^l$   & $(30.2,~3\sigma)$   & $(70.4,~\times)$ &   $(11.0,~2\sigma)$ & $(11.0,~2\sigma)$\\
  \hline
  $U_{13}^l$   & $(23.4,~3\sigma)$   & $(36.8,~\times)$ &   $(5.0,~2\sigma)$ & $(5.0,~2\sigma)$\\
      \hline
    $U_{13}^r$   & $(237.5,~\times)$   & $(237.5,~\times)$ &   $(7.82,~3\sigma)$ & $(12.3,~\times)$\\
  \hline
  $U_{23}^r$  & $(189.3,~\times)$   & $(189.3,~\times)$ &   $(1.31,~2\sigma)$ & $(86.7,~\times)$\\
  \hline
\end{tabular}\captionof{table}{\it{Here `$\times$' refers to the case which is unable to fit mixing angles even 
at $3\sigma$ level.}}\label{Table4} 
\end{center}

\begin{landscape}
\begin{center}
\begin{tabular}{ |p{2.5cm}||p{4.5cm}|p{1.7cm}|p{4.5cm}|p{4.5cm}|  }
 \hline
 \multicolumn{5}{|c|}{Allowed Dirac CP Phase($\delta_{CP}$) range for NH and IH } \\
 \hline
 Rotation-NH  & BM &DC&TBM & HG\\
 \hline
 $U_{12}^l$   & $(-4.4^\circ \leq \delta_{CP} \leq 3.4^\circ)$   & $\times$ &   $(61.0^\circ \leq |\delta_{CP}| \leq 89.9^\circ)$ & $(39.0^\circ \leq |\delta_{CP}| \leq 78.7^\circ)$\\
  \hline
  $U_{13}^l$   & $(-3.6^\circ \leq  \delta_{CP} \leq  4.2^\circ)$   & $\times$ &   $(61.0^\circ \leq |\delta_{CP}| \leq 89.9^\circ)$ & $(39.0^\circ \leq |\delta_{CP}| \leq 78.7^\circ)$\\
  \hline
    $U_{13}^r$   & $\times$   & $\times$ &   $(-89.9^\circ \leq \delta_{CP} \leq 89.9^\circ) $ & $\times$\\
  \hline
  $U_{23}^r$  & $\times$   & $\times$ &   $(60.1^\circ \leq |\delta_{CP}| \leq 89.9^\circ)$ & $\times$\\
    \hline
    \hline
   Rotation-IH &&&&\\
   \hline
 $U_{12}^l$   & $(-4.7^\circ \leq  \delta_{CP} \leq  5.3^\circ)$   & $\times$ &    $(60.9^\circ<|\delta_{CP}|<89.9^\circ)$ & $(40.4^\circ \leq |\delta_{CP}| \leq 79.2^\circ)$\\
  \hline
  $U_{13}^l$   & $(-4.9^\circ \leq  \delta_{CP} \leq 5.5^\circ)$   & $\times$ &   $(60.9^\circ \leq |\delta_{CP}| \leq 89.9^\circ)$ & $(40.4^\circ \leq |\delta_{CP}| \leq 79.2^\circ)$\\
  \hline
    $U_{13}^r$   & $\times$   & $\times$ &                                $(-89.9^\circ \leq \delta_{CP} \leq 89.9^\circ) $ & $\times$\\
  \hline
  $U_{23}^r$  & $\times$   & $\times$ &                                   $(60.3^\circ \leq |\delta_{CP}|\leq 89.9^\circ)$ & $\times$\\
  \hline
\end{tabular}\captionof{table}{\it{Here `$\times$' refers to the case which is unable to fit mixing angles 
even at $3\sigma$ level.}}\label{Table5} 
\end{center}

\begin{center}
\begin{tabular}{ |p{2.6cm}||p{4.8cm}|p{1.5cm}|p{4.8cm}|p{4.8cm}|  }
 \hline
 \multicolumn{5}{|c|}{Allowed Jarkslog Invariant($J_{CP}$) range for NH and IH} \\
 \hline
 Rotation-NH  & BM &DC&TBM & HG\\
 \hline
 $U_{12}^l$   & $(-0.0027 \leq J_{CP} \leq 0.0021)$   & $\times$ &   $(0.026 \leq |J_{CP}| \leq 0.035)$ & $(0.020 \leq  |J_{CP}| \leq 0.032)$\\
  \hline
  $U_{13}^l$   & $(-0.0025 \leq  J_{CP} \leq 0.0026)$   & $\times$ &   $(0.026< |J_{CP}| < 0.035)$ & $(0.020< |J_{CP}| < 0.032)$\\
  \hline
    $U_{13}^r$   & $\times$   & $\times$ &  $(-0.035 \leq J_{CP} \leq 0.035)$ & $\times$\\
  \hline
  $U_{23}^r$  & $\times$   & $\times$ &   $(0.026 \leq |J_{CP}| \leq 0.035)$ & $\times$\\
    \hline
    \hline
   Rotation-IH &&&&\\
   \hline
    $U_{12}^l$   & $(-0.0029 \leq J_{CP} \leq 0.0033)$   & $\times$ &   $(0.026 \leq |J_{CP}| \leq 0.035)$     & $(0.021\leq |J_{CP}| \leq 0.032)$\\
  \hline
  $U_{13}^l$   & $(-0.0034 \leq J_{CP} \leq 0.0030)$   & $\times$ &   $(0.026 \leq |J_{CP}| \leq 0.035)$       & $(0.021\leq |J_{CP}| \leq 0.032)$\\
  \hline
    $U_{13}^r$   & $\times$   & $\times$ &  $(-0.036 \leq J_{CP} \leq 0.036)$ & $\times$\\
  \hline
  $U_{23}^r$  & $\times$   & $\times$ &   $(0.027 \leq |J_{CP}| \leq 0.035)$ & $\times$\\
  \hline
\end{tabular}\captionof{table}{\it{Here `$\times$' refers to the case which is unable to fit mixing angles even at $3\sigma$ level.}}\label{Table6} 
\end{center}
\end{landscape}

\end{document}